\documentclass[12pt]{article}
\usepackage{amsmath,graphicx}
\usepackage{amssymb, amsxtra} 
\usepackage{blkarray}
\usepackage{mathtools}

\usepackage{graphicx}
\usepackage{yfonts}
\usepackage{epsfig}
\usepackage{tikz}
\usetikzlibrary{arrows,snakes,decorations.pathmorphing}
\usepackage{supertabular}
\def\therefore{
\leavevmode
\lower0.1ex\hbox{$\bullet$}
\kern-0.2em\raise0.9ex\hbox{$\bullet$}
\kern-0.2em\lower0.2ex\hbox{$\bullet$}
\thinspace}
\title{  \vspace{1cm} 
A new way to unify all fermion and boson fields, including gravity
}
\author{Norma Susana  Manko\v c Bor\v stnik$^{1}$,
\\
$^1$Department of Physics, University of Ljubljana\\
SI-1000 Ljubljana, Slovenia\footnote{norma.mankoc@fmf.uni-lj.si}}
\date{}
\begin{document}
\maketitle

\begin{abstract}
The description of the internal spaces of fermion and boson fields with
"basis vectors", which are the superposition of odd and even products
of the operators $\gamma^a$, offers in $d=2(2n+1)$-dimensions,
such as $d=(13+1)$, a unified picture of all so far observed fermions
(quarks, leptons, antiquarks and antileptons that appear
in families) and bosons (gravitons, photons, weak bosons,
gluons and scalars), under the condition that all fields have non-zero
angular momenta only in the $d =(3+1)$, $SO(3,1)$, of ordinary 
space-time.
Bosons, which also carry the spatial index $\alpha$ (which is for
tensors and vectors $\mu =(0,1,2,3)$ and for scalars $\sigma \ge 5$)
appear in two orthogonal groups.
In any $d=2(2n+1)$-dimensional space the number of internal states
of fermions in all families and their Hermitian conjugate partners is
equal to the number of internal states of boson states.
The article presents general properties of massless fermion and boson
fields and their mutual interactions in this theory, which determine
the Lagrangian density of both fields and their interactions. It
particularly illustrates "basis vectors" and their properties in $d=(13+1)$
and $d=(5+1)$. The article presents new results and discusses open
problems in this theory.\\
\end{abstract} 
%

% 
% Keywords: second quantization of fermion and boson fields with 
% Clifford algebra; beyond the standard model; Lorentz symmetry
% in internal spaces; unique description of fermion and boson fields 
% 
 
\maketitle

% G
\section{Introduction}
\label{introduction}
The author, with collaborators, succeeded in demonstrating in a long
series of works~\cite{%norma92,norma95,
norma93,pikanorma2005,nh02,nd2017,%2020PartIPartII,
nh2021RPPNP,%{nh2017FDP},%
n2024Bled,gmdn2008,gn2009,n2014matterantimatter,%
gn2014} that the model, named the {\it spin-charge-family}
theory, offers an elegant description of the second-quantised fermion
fields, appearing in families, written as the tensor products of the
basis in ordinary space-time and the basis, named ``basis vectors'',
in internal spaces, presented as superpositions of odd products of
operators $\gamma^a$, arranged in nilpotents and projectors,
which are eigenvectors of the (chosen) Cartan subalgebra members~%
\cite{norma93,pikanorma2005,nh02,nd2017,nh2021RPPNP}.
% 2020PartIPartII,nh2021RPPNP}.

% G
Four years ago~\cite{%n2022IARD,
n2023NPB,n2023MDPI,%
n2024NPB,%n2025Bled
nh2025Bled} the author started to use 
an equivalent description for
boson fields, as so far used for fermion fields, recognising the
possibility  from 30 years ago~\cite{%norma92,norma95,%
norma93}: The internal space
of boson second quantised fields can be described by the ``basis
vectors'', presented as superpositions of even products of operators
$\gamma^a$, arranged in nilpotents and projectors, which are
eigenvectors of the Cartan subalgebra members.

{\it
The theory of massless fermions and bosons, with the non-zero
momentum only in $d=(3+1)$, describing internal spaces by an
odd number of nilpotents, the rest of the projectors (for fermions)
and an even number of nilpotents, the rest of the projectors (for
bosons, which also carry the ordinary space-time index $\alpha$),
determines all the properties of fermion and boson fields and of
their mutual interactions, realised by the algebraic
multiplication:}\\
{\bf a.\,\,\,} Explains the Dirac's postulates for the second quantised
fermion and boson fields.\\
{\bf b.\,\,\,} Determines Lorentz (and correspondingly the gauge)
symmetry of fermion and boson fields.\\
{\bf c.\,\,\,} Determines couplings among fermion and boson fields,
and consequently the Lagrange densities.\\
{\bf d.\,\,\,} Determines families of fermions, any family of which
includes in $d=2(2n +1)$-dimensional internal space fermions and
antifermions. Consequently, the vacuum does not have the negative
energy of the Dirac vacuum; it is just the quantum vacuum. \\
{\bf e.\,\,\,} Determines two orthogonal kinds of boson fields; One
applies to fermions from the left-hand side, the other from the
right-hand side. One transforms fermions within each family, the
other transforms a member of a family into the same member of
other families. Both are expressible as algebraic products of
fermion fields and their Hermitian conjugate
partners~\footnote{
Knowing the ``basic vectors'' of fermion fields, we
know also the ``basic vectors'' of the boson fields, although the
properties of fermion fields are very different from the properties of
the boson fields.
}.
\\
{\bf f.\,\,\,} Fermion states are algebraically orthogonal.\\
{\bf g.\,\,\,} Although the internal spaces of fermions and bosons
demonstrate so many different properties (anticommuting fermions
appear in families, and have half-integer spins and charges in the
fundamental representations,  commuting bosons appear in two
orthogonal groups, have no families, and have integer spins and
charges in adjoint representations), the simple algebraic
multiplication with the $\gamma^a$ relates both kinds of
``basis vectors''~(\cite{n2024NPB},~ App.~B).\\
{\bf h.\,\,\,} Properties presented from {\,\bf d.\,} to {\,\bf g.\,}
influence the Feynman diagrams, which should reproduce the
experimental data.\\
{\bf i.\,\,\,}  Second quantised fermion and boson fields are
described as a tensor product of bases in ordinary space-time and
of ``basis vectors'' describing the internal spaces of fields.
\\
% \vspace{2mm}

\noindent
{\bf  . \,\, } The analysis of the fermion and boson internal spaces
with respect to the subgroups  $SO(1,3), SU(2),$ $SU(2), SU(3),
U(1)$ of the group $SO(13,1)$, offers the description of the
observed families of quarks and leptons, appearing in families, and
of tensor (gravitons), vector (photons, weak bosons, gluons), and
scalar (Higgs) boson fields, explaining also other observed
properties of fermions and bosons (like the appearance of the dark
matter~\cite{gn2009}, the matter-antimatter asymmetry in the
universe~\cite{n2014matterantimatter}, and
several other predictions~\cite{gn2014,tdn2014IJMP,nh2021RPPNP}).
\\
{\bf . \,\, } The Pauli matrices for spins and charges in any even
$d$ for either fundamental or adjoint representations can easily be
found by applying the corresponding operators, $S^{ab}$ or
${\cal S}^{ab}$, on the ``basis vectors'' of fermions or of
bosons.\\
{\bf . \,\, } There are in internal space with $d=2(2n+1)$,
$2^{\frac{d}{2}-1}$ $\times \,2^{\frac{d}{2}-1}$ ``basis
vectors'' of fermion fields, arranged in an odd number of nilpotents,
and the same number of their Hermitian conjugate partners. And
there are twice $2^{\frac{d}{2}-1}$ $\times \,2^{\frac{d}{2}-1}$
``basis vectors'' with an even number of nilpotents describing
two orthogonal boson fields. \\
% 03.04.2026
{\bf . \,\, } In odd-dimensional spaces,  $d=(2n+1)$, the fermion
and boson fields have very peculiar properties: Half of the ``basis
vectors'', $2^{\frac{2n}{2}-1}$ $\times \,2^{\frac{2n}{2}-1}$,
have the properties of fields in the $2n$-dimensional space
(the anticommuting ``basis vectors'' appear in families and have
their Hermitian conjugate partners in a separate group, the
commuting ``basis vectors'' appear in two orthogonal groups).
Among the rest  of  the ``basis vectors'', that is, in
$2^{\frac{2n}{2}-1}$ $\times \, 2^{\frac{2n}{2}-1}$ cases,
the anticommuting appear in two orthogonal groups, and commuting
appear in families and have their Hermitian conjugate partners in
a separate group~\cite{n2023MDPI,n2023NPB,n2024NPB}.\\
{\bf . \,\, }
In this contribution, all fields, fermions and bosons
(tensors, vectors and scalars) are massless.
This contribution does not discuss the breaking of symmetries
and the appearance of
massive fermion fields. There are condensates which make several
scalar fields, as well as some of the fermion and vector boson
fields, massive, discussed in Ref.~\cite{nh2021RPPNP}.~\footnote{
The breaks of symmetries are expected to follow similarly to the
case when we describe one kind of boson fields with
$\omega^{ab}{}_{\alpha}$ and another kind  with
$\tilde{\omega}^{ab}{}_{\alpha}$, presented in
Ref.~\cite{nh2021RPPNP}.
}\\

\vspace{2mm}

There is a lot of literature using the Clifford algebra for many different purposes, 
and many of the contributions do have many a thing in common  with what this paper (and our previous papers, cited in this paper) is offering. However, it is not easy to comment on the differences, not even on the level when several authors ~\cite{dh1986,dhgs1984,mp,iabgav1981,iabgav1983} are trying to 
understand the Dirac second quantisation postulates~\cite{pd1930} in 
a new way, Refs.~\cite{dh1986,dhgs1984}, as also this paper does,
since we all use different models.

Let us try to review what this paper is proposing with respect to the Dirac
second quantisation postulates, built on the recognition that there are anticommuting fermions interacting with commuting bosons, governing our 
world, in order to comment possible similarities with other papers, cited 
in this one.

This paper presents the equivalent description of fermion and boson fields 
in a new way: The internal spaces of fermion and boson fields are described 
by superposition of odd products of operators $\gamma^{a}$ if they are 
fermions, and by the superposition of even products of operators 
$\gamma^{a}$ if they are bosons. For easier recognition, understanding, 
we take care that 
the superposition of odd and even products of operators $\gamma^{a}$ are products of nilpotents and projectors, so that the states are eigenvectors of 
the chosen Cartan subalgebra members of the Lorentz algebra. 
Correspondingly, fermions have an odd number of nilpotents, the rest are 
projectors, bosons have an even number of nilpotents besides projectors.

All the rest follows: {\bf \,\, i. } All the eigenstates with an odd number of 
nilpotents anticommute, and are mutually orthogonal; they are called odd
``basis vectors'' of $\psi$, $\psi *_A \psi=0$. {\,\bf \,\, ii. } All their Hermitian conjugate  partners, $\psi^{\dagger}$, are mutually orthogonal, 
$\psi^{\dagger} *_A \psi^{\dagger}=0$. {\,\bf \,\, iii. } Products of one
``basis vector'' and one of the Hermitian conjugate partners,
$\psi *_A \psi^{\dagger}$ are in general not zero, as well as $\psi^{\dagger}
*_A \psi$ are not. It can be seen that internal spaces of boson fields, 
even ``basis vectors'', of  all bosons, can be written as even products of 
odd ``basis vectors'' (of fermions $\psi$). Even more, they have properties 
of the internal spaces of two kinds of mutually orthogonal bosons. We call them
even ``basis vectors'' of ${}^I{\cal A}_{a}$ and ${}^{II}{\cal A}_{a}$.
One immediately sees that these two kinds of bosons must multiply 
fermions, one from the left-hand side, ${}^I{\cal A}_{a}$, $(\psi *_A \psi^{\dagger}) *_A \psi$,
the other from the right-hand side, ${}^{II}{\cal A}_{a}$,
$\psi *_A (\psi^{\dagger} *_A \psi)$. 
Refs .~\cite {mp,dh1986,dhgs1984,iabgav1981,iabgav1983,vvtps} 
observe that there are two kinds of boson fields.
(Multiplications of fermions and bosons are associative due to the properties 
of the odd and even``basis vectors''.) 
{\,\bf \,\, iv. } Since in our way of building second quantised fermion and 
boson fields,  fermions appear in families, there is the same number of 
the two kinds of the even ``basis vectors'' of boson fields as it is the 
number of odd ``basis vectors'' appearing in families and of their Hermitian conjugated partners. The authors of Refs.~\cite{abgngg,abgnggEPJC}
use triality to achieve families of fermions, Refs.~\cite{vvtps,rcrg1979} use
Clifford algebra, quaternions and octonions to reproduce the 
{\it standard model} groups.
{\,\bf \,\, v. } While in our way of building second quantised fermion 
and boson fields, all the internal degrees of freedom, spins and charges, 
of fermions and bosons (of all bosons with gravity included), are 
described by the odd and even ``basis vectors'' (spins in $SO(3,1)$ 
subgroup of $SO(13,1)$,
charges in $SU(2)\times SU(2)\times SU(3)\times U(1)$ of $SO(13,1)$), 
and correspondingly fermions anticommute and bosons 
commute~\cite{nh2021RPPNP,n2023NPB}, most of papers use Clifford 
algebra only to reproduce fermion fields~\cite{dlls2009}.

\vspace{2mm}

In Sect.~\ref{internalspaces} 
the internal spaces of fermion and boson fields are presented as
``basis vectors'' which are algebraic products, $\,*_A\,$, of an odd
number of nilpotents (for fermions) and an even number of nilpotents
(for bosons),  the rest are projectors. Nilpotents and projectors
are chosen to be eigenvectors of the Cartan subalgebra members
of the Lorentz algebra,  as demonstrated in
Subsect.~\ref{basisvectors}.

Subsect.~\ref{basisvectors} also presents algebraic relations among the
``basis vectors'' of fermion and two kinds of boson fields, what
determines the Lagrange densities of interacting fermion and boson
fields.

The creation operators, presented in Subsect.~\ref{creationoperators},
are tensor products, $\,*_T\,$, of the ``basis vectors'' and the basis in
ordinary space-time.

In Subsect.~\ref{statesactivein3+1}, the states active only in
$d=(3+1)$-dimensions of ordinary  space-time, while the internal
space is active in $d=2(2n+1)$ dimensions, are discussed. In
Subsubsect.~\ref{5+1and13+1} of this section, the concrete
algebraic relations among fermion and boson fields
for the two cases that  the internal space has $d=(5+1)$ and
$d=(13+1)$, are presented.

In Sect.~\ref{algebraicstructure}, the general algebraic structure of the
second quantised fermion and boson fields, following from the
properties presented in Sect.~\ref{internalspaces}, is presented.\\

In Sect.~\ref{conclusion}, we present shortly what we have learned
in the last three years.

In Subsect.~\ref{problems}, the problems which remain to be
solved in this theory, to find out whether the theory offers the
right description of the observed fermion and boson second
quantised fields which determine the history (and the future) of our
universe, are discussed.

\vspace{1mm}

In App.~\ref{13+1representation} ``basis vectors'' of one family
of quarks and leptons, and antiquarks and antileptons are presented.

In App.~\ref{grassmannclifford}, the relations among the Grassmann
algebra and the two kinds of the Clifford algebras are discussed.

In App.~\ref{usefulrelations}, the relations are presented that are needed in all
the sections.

In App.~\ref{oddevend} odd and even ``basis vectors'' for $d=(5+1)$
cases are presented in detail, meant to be used like an exercise for
$d=13+1)$.
\section{Internal spaces of second quantised fermion and boson fields}
\label{internalspaces}

We shall need all the requirements of algebraic relations among fermion 
and boson fields derived in Sect.~\ref{internalspaces} in the next section~\ref{genaralalgebraicstructure}, which generalizes these
requirements.

Sect.~\ref{internalspaces} overviews briefly (following several papers~%
\cite{n2024NPB} and the references therein) the description of
the internal spaces of the second-quantised fermion and boson
fields as algebraic products of nilpotents and projectors, which
are the superposition of odd and even products of $\gamma^a$'s.
However, the new recognitions, some of them presented
in~\cite{n2024NPB} and the others, which the author has later 
clarified, are presented.

As explained in  Sect.~\ref{grassmannclifford}, Eq.~(\ref{clifftheta1}),
the Grassmann algebra offers two kinds of operators, $\gamma^a$'s
and $\tilde{\gamma}^a$'s with the properties,
Eq.~(\ref{gammatildeantiher})
\begin{eqnarray}                             
\label{gammatildeantiher}
\{\gamma^{a}, \gamma^{b}\}_{+}&=&2 \eta^{a b}=
\{\tilde{\gamma}^{a},
\tilde{\gamma}^{b}\}_{+}\,, \nonumber\\
\{\gamma^{a}, \tilde{\gamma}^{b}\}_{+}&=&0\,,\quad
(a,b)=(0,1,2,3,5,\cdots,d)\,, \nonumber\\
(\gamma^{a})^{\dagger} &=& \eta^{aa}\, \gamma^{a}\, ,
\quad (\tilde{\gamma}^{a})^{\dagger} = \eta^{a a}\,
\tilde{\gamma}^{a}\,.
\end{eqnarray}
We use one of the two kinds, $\gamma^a$'s, to generate the
``basis vectors'' describing internal spaces of fermions and
bosons. They are arranged in products of nilpotents and
projectors.
\begin{small}
\begin{eqnarray}
\label{nilproj}
\stackrel{ab}{(k)}:&=&\frac{1}{2}(\gamma^a +
\frac{\eta^{aa}}{ik} \gamma^b)\,, \;\;\; (\stackrel{ab}{(k)})^2=0\, ,
\nonumber \\
\stackrel{ab}{[k]}:&=&
\frac{1}{2}(1+ \frac{i}{k} \gamma^a \gamma^b)\,, \;\;\;
(\stackrel{ab}{[k]})^2=
\stackrel{ab}{[k]}\,,
\end{eqnarray}
\end{small}
so that each nilpotent and each projector is the eigenstate of
one of the Cartan (chosen) subalgebra members of the Lorentz
algebra
\begin{small}
\begin{eqnarray}
% &&{\cal {\bf S}}^{03}, {\cal {\bf S}}^{12}, {\cal {\bf S}}^{56}, \cdots,
% {\cal {\bf S}}^{d-1 \;d}\,, \nonumber\\
&&S^{03}, S^{12}, S^{56}, \cdots, S^{d-1 \;d}\,, \nonumber\\
&&\tilde{S}^{03}, \tilde{S}^{12}, \tilde{S}^{56}, \cdots,  \tilde{S}^{d-1\; d}\,,
\nonumber\\
&&{\cal {\bf S}}^{ab} = S^{ab} +\tilde{S}^{ab}\,,
%\nonumber\\ &&(= i \, (\theta^{a} \frac{\partial}{\partial \theta_{b}} -
% \theta^{b} \frac{\partial}{\partial \theta_{a}}))\,.
\label{cartangrasscliff}
\end{eqnarray}
\end{small}
where $ S^{ab}= \frac{i}{4} \{\gamma^a, \gamma^b\}_+$, while
$\tilde{S}^{ab}=  \frac{i}{4} \{\tilde{\gamma}^a, \tilde{\gamma}^b\}_+$
are used to determine additional quantum numbers, in the case of
fermions are called the family quantum numbers.

Being eigenstates of both operators, of $S^{ab}$ and $\tilde{S}^{ab}$,
nilpotents and projectors carry both quantum numbers $S^{ab}$ and
$\tilde{S}^{ab}$
%
%{calsab}
\begin{small}
\begin{eqnarray}
\label{calsab}
S^{ab} \,\stackrel{ab}{(k)} = \frac{k}{2} \,\stackrel{ab}{(k)}\,,\quad && \quad
\tilde{S}^{ab}\,\stackrel{ab}{(k)} = \frac{k}{2} \,\stackrel{ab}{(k)}\,,\nonumber\\
S^{ab}\,\stackrel{ab}{[k]} = \frac{k}{2} \,\stackrel{ab}{[k]}\,,\quad && \quad
\tilde{S}^{ab} \,\stackrel{ab}{[k]} = - \frac{k}{2} \,\,\stackrel{ab}{[k]}\,,
\end{eqnarray}
\end{small}
with $k^2=\eta^{aa} \eta^{bb}$.\\

In even-dimensional spaces, the states in internal spaces are defined by
the ``basis vectors'' which are products of  $\frac{d}{2}$ nilpotents and
projectors, and are the eigenstates of all the Cartan subalgebra
members.

{\it Fermions are products of an odd number of nilpotents}
(at least one), {\it the rest are projectors; Bosons are products of an
even number of nilpotents} (or none), {\it the rest are projectors}.
We call them odd and even ``basis vectors''.\\

The odd ``basis vectors'' have the eigenvalues of the Cartan subalgebra
members, Eq.~(\ref{cartangrasscliff}, \ref{calsab}), 
either of $S^{ab}$ or $\tilde{S}^{ab}$,
equal to half integer, $\pm \frac{i}{2}$ or $\pm \frac{1}{2}$.

The even ``basis vectors'' have the eigenvalues of the Cartan subalgebra
members, Eq.~(\ref{cartangrasscliff}, \ref{calsab}), ${\cal S}^{ab}=$
$S^{ab}+\tilde{S}^{ab}$, equal to $\pm i$ or $\pm 1$ or zero.

% *********\\04.04.2026\\
%
\subsection{``Basis vectors'' describing internal spaces of fermion and 
boson fields}
\label{basisvectors}

The odd products of nilpotents (at least one, the rest are projectors), 
called {\it odd ``basis vectors''}, differ essentially from the 
even products of nilpotents (none or at least two), called {\it even 
``basis vectors''} (the rest are projectors). 
Either odd or even ``basis vectors'' are chosen to be eigenvectors 
of all the Cartan subalgebra members, Eq.~(\ref{cartangrasscliff}).
\\

The algebraic product of any two members of the odd or even
``basis vectors'' can easily be calculated when taking into account the
relations following from Eq.~(\ref{gammatildeantiher}). 
%}
The most needed relations are presented in Eq.~(\ref{usefulrel}). %\footnote{
\begin{small}
\begin{eqnarray}
\label{usefulrel}
%{\rm it \; then} &&\; {\rm follows}\nonumber\\
\gamma^a \stackrel{ab}{(k)}&=& \eta^{aa}\stackrel{ab}{[-k]},\; \quad
\gamma^b \stackrel{ab}{(k)}= -ik \stackrel{ab}{[-k]}, \; \quad % \nonumber\\
\gamma^a \stackrel{ab}{[k]}= \stackrel{ab}{(-k)},\;\quad \;\;
\gamma^b \stackrel{ab}{[k]}= -ik \eta^{aa} \stackrel{ab}{(-k)}\,,\nonumber\\
\tilde{\gamma^a} \stackrel{ab}{(k)} &=& - i\eta^{aa}\stackrel{ab}{[k]},\quad
\tilde{\gamma^b} \stackrel{ab}{(k)} = - k \stackrel{ab}{[k]}, \;\qquad \,%\nonumber\\
\tilde{\gamma^a} \stackrel{ab}{[k]} = \;\;i\stackrel{ab}{(k)},\; \quad
\tilde{\gamma^b} \stackrel{ab}{[k]} = -k \eta^{aa} \stackrel{ab}{(k)}\,,
\nonumber\\ %
\stackrel{ab}{(k)}\stackrel{ab}{(-k)}& =& \eta^{aa} \stackrel{ab}{[k]}\,,\quad \;%\qquad \qquad
\stackrel{ab}{(-k)}\stackrel{ab}{(k)} = \eta^{aa} \stackrel{ab}{[-k]}\,,\quad\;
\stackrel{ab}{(k)}\stackrel{ab}{[k]} =0\,,\quad \quad\,
\stackrel{ab}{(k)}\stackrel{ab}{[-k]} =
\stackrel{ab}{(k)}\,,\quad %\qquad \qquad
\nonumber\\ %%
\stackrel{ab}{(-k)}\stackrel{ab}{[k]} &=& \stackrel{ab}{(-k)}\,,\quad \quad\quad
\stackrel{ab}{[k]}\stackrel{ab}{(k)}= \stackrel{ab}{(k)}\,,
%\stackrel{ab}{[k]}\stackrel{ab}{(k)}= \stackrel{ab}{(k)}\,, \quad \quad \quad
% \stackrel{ab}{(k)}\stackrel{ab}{[-k]} = \stackrel{ab}{(k)}\,,
\quad \quad \quad \;
\stackrel{ab}{[k]}\stackrel{ab}{(-k)} =0\,,\quad \quad\,
\stackrel{ab}{[k]}\stackrel{ab}{[-k]} =0\,,\quad
\nonumber\\
\stackrel{ab}{(k)}^{\dagger} &=& \eta^{aa}\stackrel{ab}{(-k)}\,,\quad
(\stackrel{ab}{(k)})^2 =0\,, \quad \stackrel{ab}{(k)}\stackrel{ab}{(-k)}
=\eta^{aa}\stackrel{ab}{[k]}\,,\nonumber\\
% \stackrel{ab}{[k]}:&=&
% \frac{1}{2}(1+ \frac{i}{k} \gamma^a \gamma^b)\,,\quad \;\,
\stackrel{ab}{[k]}^{\dagger} &=& \,\stackrel{ab}{[k]}\,, \quad \quad \quad \quad
(\stackrel{ab}{[k]})^2 = \stackrel{ab}{[k]}\,,
\quad \stackrel{ab}{[k]}\stackrel{ab}{[-k]}=0\,.
 \end{eqnarray}
 \end{small} 
More algebraic relations can be found in App.~\ref{usefulrelations}.

{\it The odd ``basis vectors''}, named  ${\hat b}^{m \dagger}_{f}$, $m$
determines the family member quantum number, $f$ determines the  
quantum number of a family, appear in $2^{\frac{d}{2}-1}$ irreducible representations - families, all with the same properties with respect to 
$S^{ab}$, while distinguishing with respect $\tilde{S}^{ab}$. Each family
has $2^{\frac{d}{2}-1}$ members. Their Hermitian conjugate partners
$({\hat b}^{m \dagger}_{f})^{\dagger}=$${\hat b}^{m }_{f}$,
appearing in a separate group, have $2^{\frac{d}{2}-1} \times
2^{\frac{d}{2}-1}$ members.  As already written, the odd ``basis 
vectors'' have the eigenvalues of the Cartan subalgebra members,
Eq.~(\ref{cartangrasscliff}), either of $S^{ab}$ or $\tilde{S}^{ab}$
half integer, $\pm \frac{i}{2}$ or $\pm \frac{1}{2}$.

The algebraic product of any two members of the odd ``basis vectors'' 
are equal to zero. The same is true for any two members of the 
Hermitian conjugated partner.
\begin{eqnarray}
\hat{b}^{m \dagger}_f *_{A} \hat{b}^{m `\dagger }_{f `}&=& 0\,,
\quad \hat{b}^{m}_f *_{A} \hat{b}^{m `}_{f `}= 0\,, \nonumber\\
% \hat{b}^{m}_f *_{A} \hat{b}^{m `\dagger}_{f `}&=&
% \delta^{m m'} \delta_{f f `}\,,\nonumber\\ 
\forall m,m', && f,f `\,. \nonumber\\
\label{orthogonalodd}
\end{eqnarray}
Choosing the vacuum state equal to
\begin{eqnarray}
\label{vaccliffodd}
|\psi_{oc}>= \sum_{f=1}^{2^{\frac{d}{2}-1}}\,\hat{b}^{m}_{f}{}_{*_A}
\hat{b}^{m \dagger}_{f} \,|\,1\,>\,,
\end{eqnarray}
for one of the members $m$, which can be anyone of the odd irreducible
representations $f$, it follows that the odd ``basis vectors'' obey the relations
\begin{eqnarray}
\label{almostDirac}
\hat{b}^{m}_{f} {}_{*_{A}}|\psi_{oc}>&=& 0.\, |\psi_{oc}>\,,\nonumber\\
\hat{b}^{m \dagger}_{f}{}_{*_{A}}|\psi_{oc}>&=&  |\psi^m_{f}>\,,\nonumber\\
\{\hat{b}^{m}_{f}, \hat{b}^{m'}_{f `}\}_{*_{A}+}|\psi_{oc}>&=&
 0.\,|\psi_{oc}>\,, \nonumber\\
\{\hat{b}^{m \dagger}_{f}, \hat{b}^{m' \dagger}_{f  `}\}_{*_{A}+}|\psi_{oc}>
&=& 0. \,|\psi_{oc}>\,,\nonumber\\
\{\hat{b}^{m}_{f}, \hat{b}^{m' \dagger}_{f `}\}_{*_{A}+}|\psi_{oc}>
&=& \delta^{m m'} \,\delta_{f f `}|\psi_{oc}>\,,
\end{eqnarray}\\
as postulated by Dirac for the second quantised fermion fields.
Here the odd ``basis vectors'' anti-commute, since 
the odd products of $\gamma^a$'s anti-commute.\\

The odd ``basis vectors''
and their Hermitian conjugate partners are normalised as follows 
 \begin{eqnarray}
<\psi_{oc}|(\hat{b}^{m \dagger}_{f})^{\dagger}\,*_A\,
 \hat{b}^{m' \dagger}_{f `}|\psi_{oc}> =\delta^{m m'}\delta_{f f `}
<\psi_{oc}|\psi_{oc}>\,, 
\end{eqnarray}
the vacuum state $<\psi_{oc}|\psi_{oc}>$ is normalised to identity.~\footnote{
Let us, as an exercise,  present the odd ``basis vectors'' and their Hermitian conjugate partners
for $d=(3+1)$, taken from Ref.~\cite{n2023MDPI}. The odd ``basis vectors'' 
appear in two families, each family has 
two members.
\begin{small}
\begin{eqnarray}
\label{3+1oddb}
\begin{array} {ccrr}
f=1&f=2&&\\
\tilde{S}^{03}=\frac{i}{2}, \tilde{S}^{12}=-\frac{1}{2}&
\;\;\tilde{S}^{03}=-\frac{i}{2}, \tilde{S}^{12}=\frac{1}{2}\;\;\; &S^{03}\, &S^{12}\\
\hat{b}^{ 1 \dagger}_{1}=\stackrel{03}{(+i)}\stackrel{12}{[+]}&
\hat{b}^{ 1 \dagger}_{2}=\stackrel{03}{[+i]}\stackrel{12}{(+)}&\frac{i}{2}&
\frac{1}{2}\\
\hat{b}^{ 2 \dagger}_{1}=\stackrel{03}{[-i]}\stackrel{12}{(-)}&
\hat{b}^{ 2 \dagger}_{2}=\stackrel{03}{(-i)}\stackrel{12}{[-]}&-\frac{i}{2}&
-\frac{1}{2}\,.
\end{array}
\end{eqnarray}
\end{small}
Their Hermitian conjugate partners have the properties
\begin{small}
\begin{eqnarray}
\label{3+1oddHb}
\begin{array} {ccrr}
S^{03}=- \frac{i}{2}, S^{12}=\frac{1}{2}&
\;\;S^{03}=\frac{i}{2}, S^{12}=-\frac{1}{2}\;\;&\tilde{S}^{03} &\tilde{S}^{12}\\
\hat{b}^{ 1 }_{1}=\stackrel{03}{(-i)}\stackrel{12}{[+]}&
\hat{b}^{ 1 }_{2}=\stackrel{03}{[+i]}\stackrel{12}{(-)}&-\frac{i}{2}&
-\frac{1}{2}\\
\hat{b}^{ 2 }_{1}=\stackrel{03}{[-i]}\stackrel{12}{(+)}&
\hat{b}^{ 2 }_{2}=\stackrel{03}{(+i)}\stackrel{12}{[-]}&\frac{i}{2}&
\frac{1}{2}\,.
\end{array}
\end{eqnarray}
\end{small}
The vacuum state $|\psi_{oc}>$, Eq.~(\ref{vaccliffodd}), is equal to:
$|\psi_{oc}>= \frac{1}{\sqrt{2}} (\stackrel{03}{[-i]}\stackrel{12}{[+]}
+\stackrel{03}{[+i]}\stackrel{12}{[-]} )$.
}\\

{\it The even ``basis vectors''}, named ${}^{I}\hat{\cal A}^{m \dagger}_{f}$ 
and ${}^{II}\hat{\cal A}^{m \dagger}_{f}$, appear in two orthogonal groups
\begin{eqnarray}
\label{AIAIIorth}
{}^{I}{\hat{\cal A}}^{m \dagger}_{f} *_A {}^{II}{\hat{\cal A}}^{m \dagger}_{f}
&=&0={}^{II}{\hat{\cal A}}^{m \dagger}._{f} *_A
{}^{I}{\hat{\cal A}}^{m \dagger}_{f}\,.
\end{eqnarray}
%.  
Each group has $2^{\frac{d}{2}-1}\times 2^{\frac{d}{2}-1}$ members 
with the Hermitian conjugate partners within the group.

The even ``basis vectors'' have the eigenvalues of the Cartan subalgebra 
members, Eq.~(\ref{cartangrasscliff}), ${\cal S}^{ab}=$$S^{ab}+ 
\tilde{S}^{ab}$, equal to $\pm i$ or $\pm 1$ or zero.

The algebraic products, $ \,*_A\,$, of two members of each of these two groups
 have the property
\begin{eqnarray}
%\begin{displaymath}
\label{ruleAAI}
{}^{i}{\hat{\cal A}}^{m \dagger}_{f} \,*_A\, {}^{i}{\hat{\cal A}}^{m' \dagger}_{f `}
\rightarrow \left \{ \begin{array} {r}
{}^{i}{\hat{\cal A}}^{m \dagger}_{f `}\,, i=(I,II) \\
% {\rm for \;chosen \,} m,f,f ' %= 1,2,\dots, 2^{\frac{d}{2}-1} \,{\rm and} \,
% f `=1,2,\dots 2^{\frac{d}{2}-1}
% {\rm there is only\; one} \; % m' \, {\rm }
% \forall f `=1,2,\dots 2^{\frac{d}{2}-1}\,,\\
{\rm or \,zero}\,.
\end{array} \right.
%\end{displaymath}
%i=(I,II)\,.
\end{eqnarray}
For a chosen ($m, f, f `$), there is (out of $2^{\frac{d}{2}-1}$) only one $m'$  
giving a non-zero contribution.~\footnote{
\begin{small}
Let us, for an exercise, present the $2^{\frac{4}{2}-1}\times 2^{\frac{4}{2}-1} $  
''basis vectors'' for $d=(3+1)$,
the members of the group ${}^{I}{\bf {\cal A}}^{m \dagger}_{f}$, taken 
from Ref.~\cite{n2023MDPI}
\end{small}
\begin{small}
\begin{eqnarray}
\label{3+1evenAI}
\begin{array} {crrcrr}
&{\bf {\cal S}}^{03}&{\bf {\cal S}}^{12}&&{\bf {\cal S}}^{03}&{\bf {\cal S}}^{12}\\
{}^{I}{\bf {\cal A}}^{1 \dagger}_{1}= \stackrel{03}{[+i]}\stackrel{12}{[+]}&0&0&\,,
{}^{I}{\bf {\cal A}}^{1 \dagger}_{2}= \stackrel{03}{(+i)}\stackrel{12}{(+)}&i&1\\
{}^{I}{\bf {\cal A}}^{2 \dagger}_{1}= \stackrel{03}{(-i)}\stackrel{12}{(-)}&-i&-1&\,,
{}^{I}{\bf {\cal A}}^{2 \dagger}_{2}= \stackrel{03}{[-i]}\stackrel{12}{[-]}&0&0\,,
\end{array}
\end{eqnarray}
\end{small}
\begin{small}
and  $2^{\frac{4}{2}-1}\times 2^{\frac{4}{2}-1} $  even ''basis vectors'' 
${}^{II}{\bf {\cal A}}^{m \dagger}_{f}$, $m=(1,2), f=(1,2)$,
\end{small}
\begin{small}
\begin{eqnarray}
\label{3+1evenAII}
\begin{array} {crrcrr}
&{\bf {\cal S}}^{03}&{\bf {\cal S}}^{12}&&{\bf {\cal S}}^{03}&{\bf {\cal S}}^{12}\\
{}^{II}{\bf {\cal A}}^{1 \dagger}_{1}= \stackrel{03}{[+i]}\stackrel{12}{[-]}&0&0&\,,
{}^{II}{\bf {\cal A}}^{1 \dagger}_{2}= \stackrel{03}{(+i)}\stackrel{12}{(-)}&i&-1\\
{}^{II}{\bf {\cal A}}^{2 \dagger}_{1}= \stackrel{03}{(-i)}\stackrel{12}{(+)}&-i&1&\,,
{}^{II}{\bf {\cal A}}^{2 \dagger}_{2}= \stackrel{03}{[-i]}\stackrel{12}{[+]}&0&0\,.
\end{array}
\end{eqnarray}
\end{small}
One can easily check the above relations from Eqs.~(\ref{3+1oddb} - \ref{AIIbdaggerb})
if taking into  account Eq.~\ref{gammatildeantiher}, or Eq.(\ref{usefulrel}).
}
%. 

To be able to propose the action for fermion and boson second quantized fields,
we need to know the algebraic application, $*_{A}$, of boson fields on fermion
fields and fermion fields on boson fields.\\

{\it The algebraic application}, $*_{A}$, {\it of the even ``basis vectors''}
${}^{I}{\hat{\cal A}}^{m \dagger}_{f }$ {\it on the odd ``basis vectors''}
$ \hat{b}^{m' \dagger}_{f `} $, we call this left multiplication, gives
\begin{eqnarray}
%\begin{displaymath}
\label{calIAb1234gen}
{}^{I}{\hat{\cal A}}^{m \dagger}_{f } \,*_A \, \hat{b}^{m' \dagger }_{f `}
\rightarrow \left \{ \begin{array} {r} \hat{b }^{m \dagger}_{f `}\,, \\
{\rm or \,zero}\,.
\end{array} \right.
%\end{displaymath}
\end{eqnarray}
Eq.~(\ref{calIAb1234gen}) demonstrates that
${}^{I}{\hat{\cal A}}^{m \dagger}_{f}$,
applying on $\hat{b}^{m' \dagger }_{f `} $, transforms the odd
``basis vector'' into another odd ``basis vector'' of the same family,
transferring to the odd ``basis vector'' integer spins or gives zero.\\

{\it We find for the second group of boson fields}, 
${}^{II}{\hat{\cal A}}^{m \dagger}_{f }$,
\begin{eqnarray}
%\begin{displaymath}
\label{calbIIA1234gen}
\hat{b}^{m \dagger }_{f } *_{A} {}^{II}{\hat{\cal A}}^{m' \dagger}_{f `} \,
\rightarrow \left \{ \begin{array} {r} \hat{b }^{m \dagger}_{f ``}\,, \\
{\rm or \,zero}\,,
\end{array} \right.
%\end{displaymath}
\end{eqnarray}
demonstrating that the application of the odd ``basis vector'' 
$\hat{b}^{m \dagger }_{f }$ on 
$ {}^{II}{\hat{\cal A}}^{m' \dagger}_{f `}$ leads to another 
odd ``basis vector'' $\hat{b }^{m \dagger}_{f ``}$ belonging 
to the same family member $m$ of a different family $f ``$.
We call this the right multiplication of 
${}^{II}{\hat{\cal A}}^{m \dagger}_{f }$ on the odd ``basis vector'' 
$\hat{b}^{m \dagger }_{f }$.%

The rest of possibilities give zero.
\begin{eqnarray}
%\begin{displaymath}
\label{calbIA1234gen}
\hat{b}^{m \dagger }_{f } *_{A} {}^{I}{\hat{\cal A}}^{m' \dagger}_{f `} = 0\,, \quad
{}^{II}{\hat{\cal A}}^{m \dagger}_{f } \,*_A \, \hat{b}^{m' \dagger }_{f `}= 0\,,\;\;
\forall (m, m`, f, f `)\,.
%\end{displaymath}
\end{eqnarray}
\vspace{2mm}
{\it Let us add that the internal spaces of boson second quantized fields can be written
as the algebraic products of the odd ``basis vectors'' and their Hermitian conjugate
partners: $\hat{b}^{m \dagger}_{f}$ and $(\hat{b}^{m'' \dagger}_{f `})^{\dagger}$.}
\begin{eqnarray}
\label{AIbbdagger}
{}^{I}{\hat{\cal A}}^{m \dagger}_{f}&=&\hat{b}^{m' \dagger}_{f `} *_A 
(\hat{b}^{m'' \dagger}_{f `})^{\dagger}\,,
\end{eqnarray}
\begin{eqnarray}
\label{AIIbdaggerb}
 {}^{II}{\hat{\cal A}}^{m \dagger}_{f}&=&
(\hat{b}^{m' \dagger}_{f `})^{\dagger} *_A 
\hat{b}^{m' \dagger}_{f `'}\,. 
\end{eqnarray}
Family members $\hat{b}^{m' \dagger}_{f `}$ of any family $f '$ 
generates in the algebraic product $\hat{b}^{m' \dagger}_{f `}\, *_A\,$
($\hat{b}^{m'' \dagger}_{f `})^{\dagger}$ the same  
$2^{\frac{d}{2}-1}\times$ $2^{\frac{d}{2}-1}$ 
even ``basis vectors'' ${}^{I}{\hat{\cal A}}^{m \dagger}_{f}$, 
each family member $m'$ generates 
in $(\hat{b}^{m' \dagger}_{f `})^{\dagger} *_A$ 
$\hat{b}^{m' \dagger}_{f `'}$ the same  $2^{\frac{d}{2}-1}\times$ 
$2^{\frac{d}{2}-1}$ even ``basis vectors''
${}^{II}{\hat{\cal A}}^{m \dagger}_{f}$~\footnote{
It follows that ${}^{I}{\hat{\cal A}}^{m \dagger}_{f}$, expressed by
$\hat{b}^{m' \dagger}_{f `} *_A$ ($\hat{b}^{m'' \dagger}_{f `})^{\dagger}$, 
applying on $\hat{b}^{m''' \dagger}_{f `''}$, obey Eq.~(\ref{calIAb1234gen}), 
and $\hat{b}^{m''' \dagger}_{f ``'}$ applying on 
${}^{II}\hat{\cal A}^{m \dagger}_{f}$, expressed by 
$(\hat{b}^{m' \dagger}_{f `})^{\dagger}$ $ *_A$ 
$\hat{b}^{m' \dagger}_{f ``}$, obey Eq.~(\ref{calbIIA1234gen}).}.

\subsection{Fermions and bosons creation operators}
\label{creationoperators}
The creation operators for either fermions or bosons must be defined
as the tensor products, $*_{T}$, of both contributions, the ``basis
vectors'' describing the internal space of fermions or bosons and the
basis in ordinary space-time in the momentum or coordinate
representation~\footnote{ 
To the boson second quantised fields, we need to add the space
index $\alpha$. Since our space-time is flat, we shall use $a$ instead
of $\alpha$, and $n$ instead of $\mu$ and for the scalar index, 
$a\ge 5$, (our fermions and bosons have non-zero momentum only 
in $d=(3+1)$ of ordinary space-time) index $s$.
}.

\vspace{2mm}

Let us start with the definition of the single particle states in 
ordinary space-time in momentum representation, briefly 
overviewing Refs.~\cite{n2024NPB}, (\cite{nh2021RPPNP}, 
Subsect.~3.3 and App. J). 
\begin{eqnarray}
\label{creatorp}
|\vec{p}>&=& \hat{b}^{\dagger}_{\vec{p}} \,|\,0_{p}\,>\,,\quad
<\vec{p}\,| = <\,0_{p}\,|\,\hat{b}_{\vec{p}}\,, \nonumber\\
<\vec{p}\,|\,\vec{p}'>&=&\delta(\vec{p}-\vec{p}')=
<\,0_{p}\,|\hat{b}_{\vec{p}}\; \hat{b}^{\dagger}_{\vec{p}'} |\,0_{p}\,>\,,
\nonumber\\
<\,0_{p}\,| \hat{b}_{\vec{p'}}\, \hat{b}^{\dagger}_{\vec{p}}\,|\,0_{p}\, > 
&=&\delta(\vec{p'}-\vec{p})\,,
\end{eqnarray}
with $<\,0_{p}\, |\,0_{p}\,>=1$. The operator 
$\hat{b}^{\dagger}_{\vec{p}}$ pushes a  single particle state 
with zero momentum by an amount $\vec{p}$.
Taking into account that $\{\hat{p}^i\,, \hat{p}^j \}_{-}=0$ and 
$\{\hat{x}^k\,, \hat{x}^l \}_{-}=0$, while
$\{\hat{p}^i\,, \hat{x}^j \}_{-}=i \eta^{ij}$, it follows
\begin{small}
\begin{eqnarray}
\label{eigenvalue10}
<\vec{p}\,| \,\vec{x}>&=&<0_{\vec{p}}\,|\,\hat{b}_{\vec{p}}\;
\hat{b}^{\dagger}_{\vec{x}}
|0_{\vec{x}}\,>=(<0_{\vec{x}}\,|\,\hat{b}_{\vec{x}}\;
\hat{b}^{\dagger}_{\vec{p}} \,\,
|0_{\vec{p}}\,>)^{\dagger}\, \nonumber\\
<0_{\vec{p}}\,|\{\hat{b}^{\dagger}_{\vec{p}}\,, \,
\hat{b}^{\dagger}_{\vec{p}\,'}\}_{-}|0_{\vec{p}}\,>&=&0\,,\qquad
<0_{\vec{p}}\,|\{\hat{b}_{\vec{p}}\,, \,\hat{b}_{\vec{p}\,'}\}_{-}|0_{\vec{p}}\,>=0\,,\qquad
<0_{\vec{p}}\,|\{\hat{b}_{\vec{p}}\,, \,\hat{b}^{\dagger}_{\vec{p}\,'}\}_{-}|0_{\vec{p}}\,>=0\,,
\nonumber\\
<0_{\vec{x}}\,|\{\hat{b}^{\dagger}_{\vec{x}}\,, \,\hat{b}^{\dagger}_{\vec{x}\,'}\}_{-}|0_{\vec{x}}\,>&=&0\,,
\qquad
<0_{\vec{x}}\,|\{\hat{b}_{\vec{x}}\,, \,\hat{b}_{\vec{x}\,'}\}_{-}|0_{\vec{x}}\,>=0\,,\qquad
<0_{\vec{x}}\,|\{\hat{b}_{\vec{x}}\,, \,\hat{b}^{\dagger}_{\vec{x}\,'}\}_{-}|0_{\vec{x}}\,>=0\,,
\nonumber\\
%{\rm while}&&\nonumber\\
<0_{\vec{p}}\,|\{\hat{b}_{\vec{p}}\,, \,\hat{b}^{\dagger}_{\vec{x}}\}_{-}|0_{\vec{x}}\,>&=&
e^{i \vec{p} \cdot \vec{x}} \frac{1}{\sqrt{(2 \pi)^{d-1}}}\,,\quad
<0_{\vec{x}}\,|\{\hat{b}_{\vec{x}}\,, \,\hat{b}^{\dagger}_{\vec{p}}\}_{-}|0_{\vec{p}}\,>=
e^{-i \vec{p} \cdot \vec{x}} \frac{1}{\sqrt{(2 \pi)^{d-1}}}\,.
\end{eqnarray}
\end{small}
The momentum basis is continuously infinite, while the internal space 
of either fermion or boson fields has a finite number of ``basis vectors'' -  
twice $2^{\frac{d}{2}-1}\times 2^{\frac{d}{2}-1}$ for fermions and twice
$2^{\frac{d}{2}-1}\times 2^{\frac{d}{2}-1}$ for bosons, provided that the
internal space includes $d$-dimensions. We assume that fermions and bosons 
are active only $d=(3+1)$.\\

The creation operator for a free massless fermion field of the energy
$p^0 =|\vec{p}|$, belonging to the family $f$ and to a superposition of
family members $m$ applying on the vacuum state
($|\psi_{oc}>\,*_{T}\, |0_{\vec{p}}>$) 
can be written as~(we follow \cite{nh2021RPPNP}, Subsect.3.3.2, and
the references therein)
\begin{small}
\begin{eqnarray}
\label{wholespacefermions}
{\bf \hat{b}}^{s \dagger}_{f} (\vec{p}) \,&=& \,
\sum_{m} c^{sm}{}_f (\vec{p})\, \,\hat{b}^{\dagger}_{\vec{p}}\,*_{T}\,
\hat{b}^{m \dagger}_{f} \,.
\end{eqnarray}
\end{small}

The vacuum state for fermions, $|\psi_{oc}>\,*_{T}\, |0_{\vec{p}}> $, 
includes both spaces, the internal part, Eq.(\ref{vaccliffodd}), and the momentum
part, Eq.~(\ref{creatorp}). 
The creation operators in the coordinate representation can be written as
$\hat{\bf b}^{s \dagger}_{f }(\vec{x},x^0)=
\sum_{m} \,\hat{b}^{ m \dagger}_{f} \, *_{T}\, \int_{- \infty}^{+ \infty} \,
\frac{d^{d-1}p}{(\sqrt{2 \pi})^{d-1}} \, c^{s m }{}_{f}\;
(\vec{p}) \; \hat{b}^{\dagger}_{\vec{p}}\;
e^{-i (p^0 x^0 - \vec{p}\cdot \vec{x})}$,
\cite{n2023NPB}, (\cite{nh2021RPPNP}, subsect. 3.3.2. and the 
references therein).
                                                
The creation operators, $ \hat{\bf b}^{s\dagger}_{f }(\vec{p}) $, and their
Hermitian conjugate partners annihilation operators,
$(\hat{\bf b}^{s \dagger}_{f }(\vec{p}))^{\dagger}$
$=\hat{\bf b}^{s}_{f }(\vec{p})$, creating and annihilating the 
single fermion states, respectively, fulfil when applying the vacuum 
state, ($|\psi_{oc}> *_{T} |0_{\vec{p}}>$), the anti-commutation 
relations for the second quantized fermions, postulated by Dirac 
(Ref.~\cite{nh2021RPPNP}, Subsect.~3.3.1, 
Sect.~5). The anticommuting properties of the creation operators for 
fermions are determined by  the odd ``basis vectors'', the basis in 
ordinary space-time, namely, commute
~\footnote{                                              
\begin{small}
\begin{eqnarray}
<0_{\vec{p}}\,|
\{ \hat{\bf b}^{s' }_{f `}(\vec{p'})\,,\,
\hat{\bf b}^{s \dagger}_{f }(\vec{p}) \}_{+} \,|\psi_{oc}> |0_{\vec{p}}>&=&
\delta^{s s'} \delta_{f f'}\,\delta(\vec{p}' - \vec{p})\,\cdot |\psi_{oc}>
\,,\nonumber\\
\{ \hat{\bf b}^{s' }_{f `}(\vec{p'})\,,\,
\hat{\bf b}^{s}_{f }(\vec{p}) \}_{+} \,|\psi_{oc}> |0_{\vec{p}}>&=&0\, \cdot \,
|\psi_{oc}> |0_{\vec{p}}>
\,,\nonumber\\
\{ \hat{\bf b}^{s' \dagger}_{f '}(\vec{p'})\,,\,
\hat{\bf b}^{s \dagger}_{f }(\vec{p}) \}_{+}\, |\psi_{oc}> |0_{\vec{p}}>&=&
0\, \cdot \,|\psi_{oc}> |0_{\vec{p}}>
\,,\nonumber\\
\hat{\bf b}^{s \dagger}_{f }(\vec{p}) \,|\psi_{oc}> |0_{\vec{p}}>&=&
|\psi^{s}_{f}(\vec{p})>\,,\nonumber\\
\hat{\bf b}^{s}_{f }(\vec{p}) \, |\psi_{oc}> |0_{\vec{p}}>&=&0\, \cdot\,
\,|\psi_{oc}> |0_{\vec{p}}>\,, \nonumber\\
|p^0| &=&|\vec{p}|\,.
\label{Weylpp'comrel}
\end{eqnarray}
\end{small}
 }.
 
\vspace{2mm}

The creation operator for a free massless boson field of the energy
$p^0 =|\vec{p}|$,  with the ``basis vectors'' belonging to one of the 
two groups, ${}^{i}{\hat{\cal A}}^{m \dagger}_{f }, i=(I,II)$,
applying on the vacuum state, $|\,1\,>\,*_{T}\, |0_{\vec{p}}>$, 
must carry the space index $a$, describing the $a$ component 
of the boson field in the ordinary space%
~\footnote{
The vacuum state for bosons is chosen to be identity.
}.
We, therefore, add the space index $a$, 
as well as the dependence on the momentum~\cite{n2024NPB}
\begin{eqnarray}
\label{wholespacebosons}
{\bf {}^{i}{\hat{\cal A}}^{m \dagger}_{f a}} (\vec{p}) \,&=&
{}^{i}{\cal C}^{ m}{}_{f a} (\vec{p})\,*_{T}\,
{}^{i}{\hat{\cal A}}^{m \dagger}_{f} \, \,, i=(I,II)\,,
\end{eqnarray}
with ${}^{i}{\cal C}^{ m}{}_{f a} (\vec{p})=
{}^{i}{\cal C}^{ m}{}_{f a}\,\hat{b}^{\dagger}_{\vec{p}}$, with
$\hat{b}^{\dagger}_{\vec{p}}$ defined in Eqs.~(\ref{creatorp}, 
\ref{eigenvalue10})~\footnote{
In the general case, the energy eigenstates of bosons are in a superposition of
${\bf {}^{i}{\hat{\cal A}}^{m \dagger}_{f}}$, for either $i=I$ or $i=II$.
}.

The creation operators for boson fields in the coordinate representation 
one finds using Eqs.~(\ref{creatorp}, \ref{eigenvalue10}),\,
% \begin{eqnarray}
${\bf {}^{i}{\hat{\cal A}}^{m \dagger}_{f a}}
(\vec{x}, x^0) = {}^{i}{\hat{\cal A}}^{m \dagger}_{f} \,*_{T}\,
\int_{- \infty}^{+ \infty} \,
\frac{d^{d-1}p}{(\sqrt{2 \pi})^{d-1}} \,{}^{i}{\cal C}^{ m}{}_{f a}\,
\hat{b}^{\dagger}_{\vec{p}}\,
e^{-i (p^0 x^0- \varepsilon \vec{p}\cdot \vec{x})}|_{p^0=|\vec{p}|}\,,i=(I,II)\,.$\\

Assuming that the internal space has $d=(13+1)$, while fermions and bosons 
have non-zero momenta only in $d=(3+1)$ of the ordinary space-time, both 
creation operators manifest (after analysing $SO(13,1)$ with respect to the 
subgroups $SO(3,1)$, $SU(2) \times SU(2)$, $SU(3)$ and $U(1)$ of the 
Lorentz group $SO(13,1)$) the properties of fermions and bosons              %Eq.~(\ref{ruleAAI}), 
 as assumed by the {\it standard model} before the electroweak phase 
transitions: 
Clifford odd creation operators defining the fermion fields manifest all the
properties of quarks and leptons and antiquarks and antileptons, appearing in
 families, and the Clifford even boson creation operators, 
 ${\bf {}^{I}{\hat{\cal A}}^{m \dagger}_{f a}}$, defining the  boson fields
manifest for $a$ equal to $n =(0,1,2,3)$ all the properties of gauge fields
(photons, weak bosons, and gluons - predicting the second weak fields and 
explaining the gravitons). 

For $a$ equal to $s \ge 5$, the Clifford even boson creation operators, 
${\bf {}^{II}{\hat{\cal A}}^{m \dagger}_{f s}}$, manifest properties of the
scalar Higgs, causing after the electroweak phase transitions masses of quarks and 
leptons and antiquarks and antileptons, and some of the gauge fields.

The assumption that the internal spaces of fermion  
and boson fields are describable by the 
odd and even ``basis vectors",  respectively, leads to the conclusion that the internal 
spaces of all the boson fields - gravitons (the gauge fields of the spins $SO(3,1)$), 
photons (the gauge fields of $U(1)$), weak bosons (the gauge fields of one of the 
$SU(2)$) and gluons (the gauge fields of $SU(3)$ (together with $U(1)$ origin in 
$SO(6)$)  - must also be  described by the even ``basis vectors'', all must carry 
the index $a=$ $n=(0,1,2,3)$.\\

{\it Both groups of even ``basis vectors'' manifest as the gauge fields
of the corresponding fermion fields: One concerning the family members
quantum numbers,  determined by $S^{ab}$, the other concerning the 
family quantum numbers, determined by $\tilde{S}^{ab}$.}\\

Let us point out that although it looks like that this theory postulates 
two kinds of boson fields, not yet observed so far, this is not the case:
All the theories so far postulate the families of fermions and the scalar 
fields giving masses to fermions and weak bosons in addition to the 
internal spaces of fermions and bosons. In our case, the families are 
present without being postulated. Our boson fields of the second 
kind have, in theories so far, realization in Higgs. 

The proposed description of the internal spaces offers families of 
fermions, scalar fields and gauge fields:
${}^{I}{\hat{\cal A}}^{m \dagger}_{f }$, transferring the integer 
quantum numbers to the odd ``basis vectors'', 
$\hat{b}^{m \dagger}_{f}$, changes the family members' quantum 
numbers, leaving the family quantum numbers unchanged, manifesting
the properties of the gauge fields; The 
second group, ${}^{II}{\hat{\cal A}}^{m \dagger}_{f }$,
transferring the integer quantum numbers to the ``basis vector''
$\hat{b}^{m \dagger}_{f}$, changes the family quantum
numbers leaving the family members quantum numbers unchanged,
manifesting properties of the scalar fields, which give masses to 
quarks and leptons, and to the weak bosons.

The propose description of the internal spaces of fermions and bosoms
predicts  more families than observed so far, and more gauge fields.
We expect that the breaks of the starting symmetry will show, why 
we have not observed them~\cite{nh2021RPPNP,n2023NPB}(yet).
\subsection{States of fermions and bosons active only in $d=(3+1)$}
\label{statesactivein3+1}

We take the states of fermion and boson fields to have non-zero
momentum only in $d=(3+1)$. This refers to the Poincaré group (with 
the infinitesimal generators $M^{ab}(= L^{ab} +{\cal S}^{ab}), 
p^c$) applying only in $d=(3+1)$, while in the internal space, the 
Lorentz group (with the infinitesimal generators ${\cal S}^{ab}$, 
${\cal S}^{ab}=S^{ab}+\tilde{S}^{ab}$) 
applies to the whole internal  space $d=2(2n+1)$. We discuss in 
this section the algebraic relations among fermion and boson 
fields~(\ref{internalspaces}, \ref{basisvectors}) in the case that  
the internal space has $d=(5+1)$ and  $d=(13+1)$, 
Subsect.~\ref{5+1and13+1}.

The odd and even ``basis vectors'' are presented in the case that
$d=(5+1)$ in App.~\ref{oddevend} in 
Tables~(\ref{Table Clifffourplet.},%
~\ref{transverseCliff basis5+1even I.},%
~\ref{S120Cliff basis5+1even I.},%
~\ref{transverseCliff basis5+1even II.},~\ref{S120Cliff basis5+1even II.}).
This toy model is discussed as an exercise.

In Table~\ref{Table so13+1.} the odd  ``basis vectors'' are
presented in the case that $d=(13+1)$ for one family of fermions -
quarks and leptons and antiquarks and antileptons -  as products
of an odd number of nilpotents (at least one, up to seven).
The  ``basis vectors'' are eigenstates of all the Cartan subalgebra
memebers, Eq.~(\ref{cartangrasscliff}), of the Lorentz algebra.

The creation and annihilation operators are for odd and even
``basis vectors'' the tensor products, $\,*_{T}\,$, of the basis in
ordinary space-time in  $d=(3+1)$, and the ``basis vectors'' in
internal space, with $d=(5+1)$ or $d=(13+1)$: For anti-commuting
creation operators  
we have ${\bf \hat{b}}^{s \dagger}_{f} (\vec{p}) \,= \,
\sum_{m} c^{sm}{}_f (\vec{p}) \,\hat{b}^{\dagger}_{\vec{p}}\,*_{T}\,
\hat{b}^{m \dagger}_{f}$, Eq.~(\ref{wholespacefermions}).

For the commuting creation operators  with the ``basis vectors''
belonging to one of the two groups,
${}^{i}{\hat{\cal A}}^{m \dagger}_{f }, i=(I,II)$, carrying the
space index $a$, we have
${\bf {}^{i}{\hat{\cal A}}^{m \dagger}_{f a}} (\vec{p}) \,=
{}^{i}{\cal C}^{ m}{}_{f a} (\vec{p})\,*_{T}\,
{}^{i}{\hat{\cal A}}^{m \dagger}_{f} \, \,, i=(I,II)$,
Eq.~(\ref{wholespacebosons}).
\subsubsection{Internal spaces of fermions and bosons in $d=(5+1)$ 
and $d=(13+1)$}
\label{5+1and13+1}
{\bf a.} $\,\,$  Let us start with {\it the toy model for electrons,
positrons, photons and gravitons in the case that the internal  space is 
$d=(5+1)$, while fields have non-zero momenta in $d=(3+1)$}.

We follow here to some extent a similar part in the
Ref.~(\cite{n2024NPB}, and the references therein). This toy model
is to show the reader, in a simple model, what the new description
of the internal spaces of fermion and boson fields offers.

In Table~\ref{Table Clifffourplet.} the odd ``basis vectors'',
$\hat{b}^{m \dagger}_{f}$, appearing in four ($2^{\frac{d=6}{2}-1}$)
families, each family having four ($2^{\frac{d=6}{2}-1}$) family 
members, are presented in the first group, as products of an odd number 
of nilpotents (one or three) and the remaining projectors.
Their Hermitian conjugate partners are presented in the second
group, again with $16$ members.

The even basis vectors appear in the third and the fourth group.

Table~\ref{Table Clifffourplet.} presents the eigenvalues of all
Cartan subalgebra members, Eq.~(\ref{cartangrasscliff}); $S^{ab}$
for a family members, and $\tilde{S}^{ab}$ for a family. 
${\cal S}^{ab}=(S^{ab}+\tilde{S}^{ab})$ determine the Cartan 
eigenvalues of the even ``basis vectors'', presenting internal spaces 
of boson fields.

The reader can check the relations among ``basis vectors'' of fermions 
and bosons appearing in Eqs.~(\ref{orthogonalodd} -- \ref{AIIbdaggerb}),
for the case that  the internal space has $d=(5+1)$, by taking into account
Eqs.~(\ref{usefulrel}, \ref{graficcliff0}, \ref{graficfollow1}).

The corresponding creation and annihilation operators for free massless
fermion fields (${\bf \hat{b}}^{m\dagger}_{f} (\vec{p}) \,= \,
\hat{b}^{\dagger}_{\vec{p}}\,*_{T}\,
\hat{b}^{m \dagger}_{f}$), and for
free massless boson fields (${\bf {}^{i}{\hat{\cal A}}^{m \dagger}_{f  a}}, 
i=(I,II)$, carrying the space index $a$, 
${\bf {}^{i}{\hat{\cal A}}^{m \dagger}_{f a}} (\vec{p}) \,=
{}^{i}{\cal C}^{ m}{}_{f a} (\vec{p})\,*_{T}\,
{}^{i}{\hat{\cal A}}^{m \dagger}_{f} \, \,, i=(I,II)$), can be found in 
Eqs.~(\ref{wholespacefermions},~\ref{wholespacebosons}).

Let us call the first $\hat{b}^{m \dagger}_{f}$  of the ``basis vectors'' in
Table~\ref{Table Clifffourplet.}, $\hat{b}^{1 \dagger}_{1}=$
$\stackrel{03}{(+i)}\stackrel{12}{[+]}\stackrel{56}{[+]}$, the ``basis vector''
of the ``electron'',
and the third ``basis vector'' $\hat{b}^{3 \dagger}_{1}=$
$\stackrel{03}{[-i]}\stackrel{12}{[+]}\stackrel{56}{(-)}$, both belong to the 
first family, the ``basis vector''
of the ``positron'', choosing the quantum numbers of the ``electron''  equal to 
($S^{03} =\frac{i}{2}$, $S^{12}=\frac{1}{2}$ and $S^{56} =\frac{1}{2}$), and
of the ``positron'' equal to ($S^{03}
=-\frac{i}{2}$, $S^{12}=\frac{1}{2}$ and $S^{56} =\frac{1}{2}$).
One can transform the ``electron'' to the ``positron'' by $S^{05}$.

The ``basis vectors'' of the ``positron'' and ``electron'' have fractional
charges and both appear in four families, reachable from the first one by 
the application of $\tilde{S}^{ab}$.

For example, one generates the second family by applying 
$\tilde{S}^{05}$ on the first family.

The corresponding ``photon'' field, its ``basis vector'' indeed,
describing the internal space of the ``photon'', must be a product of projectors 
only, since the photon does not change the charge of the positron or electron.

There is only one even ``basis vector'', that applying to the
``basis vector'' of the ``electron'' gives a non-zero contribution, 
 the ``basis vector'' ${}^{I}{\hat{\cal A}}^{1 \dagger}_{3}=$ 
$\stackrel{03}{[+i]}\stackrel{12}{[+]} \stackrel{56}{[+]}$. It is  
presented  in Table~\ref{S120Cliff basis5+1even I.}.
 
There is also only one even ``basis vector'', which, applying to
the ``basis vector'' of the ``positron'', gives a non-zero contribution. 
Both even ``basis vectors'' have the properties of photons.
 It is presented  in Table~\ref{S120Cliff basis5+1even I.}.\\

\begin{small}
\begin{eqnarray}
\label{phelpo}
&&{}^{I}{\hat{\cal A}}^{1 \dagger}_{3 ph}
(\equiv \stackrel{03}{[+i]}\stackrel{12}{[+]} \stackrel{56}{[+]})\,
\,*_A\,\hat{b}^{1 \dagger}_{f} (\equiv \stackrel{03}{(+i)}\stackrel{12}{[+]} 
\stackrel{56}{[+]})\,\rightarrow \hat{b}^{1 \dagger}_{f}\,,
\,\,\nonumber\\
&&{}^{I}{\hat{\cal A}}^{3 \dagger}_{2 ph}
(\equiv \stackrel{03}{[-i]}\stackrel{12}{[+]} \stackrel{56}{[-]})\,
\,*_A\,\hat{b}^{3 \dagger}_{f} (\equiv \stackrel{03}{[-i]}\stackrel{12}{[+]} 
\stackrel{56}{(-)})\,\rightarrow \hat{b}^{3 \dagger}_{f}\,.
\end{eqnarray}
\end{small}
The same ``photon'' makes the same transformations on the corresponding
``electron''  (or ``positron'') of all the families.
Obviously, the Cartan subalgebra quantum numbers, Eq.~(\ref{cartangrasscliff}), 
($S^{ab}+\tilde{S}^{ab}$), applying on any member of the ``photon'' is equal 
to zero: ($S^{03}+\tilde{S}^{03} =0$, $S^{12}+\tilde{S}^{12} =0$ and 
$S^{56} +\tilde{S}^{56}=0$) of  either 
${}^{I}{\hat{\cal A}}^{1 \dagger}_{3 ph}$ or  
${}^{I}{\hat{\cal A}}^{3 \dagger}_{2 ph}$,  are zero, since the projectors have
properties that $S^{ab}$$=
-\tilde{S}^{ab}$, Eq.~(\ref{calsab}).\\

Let us check the relation of Eq.~(\ref{AIbbdagger}), using Eqs.~(\ref{usefulrel},~\ref{usefulrel0}).
$${}^{I}{\hat{\cal A}}^{1 \dagger}_{3}
 (\equiv \stackrel{03}{[+i]}\stackrel{12}{[+]} \stackrel{56}{[+]})\,
 =\hat{b}^{1 \dagger}_{1} 
 (\equiv \stackrel{03}{(+i)}\stackrel{12}{[+]} \stackrel{56}{[+]})\, *_A \,
(\hat{b}^{1 \dagger}_{1})^{\dagger}
(\equiv (\stackrel{03}{(+i)}\stackrel{12}{[+]} \stackrel{56}{[+]})^{\dagger})\,.$$
$${}^{I}{\hat{\cal A}}^{3 \dagger}_{2}
 (\equiv \stackrel{03}{[-i]}\stackrel{12}{[+]} \stackrel{56}{[-]})\,
 =\hat{b}^{3 \dagger}_{1} 
 (\equiv \stackrel{03}{[-i]}\stackrel{12}{[+]} \stackrel{56}{(-)})\, *_A \,
(\hat{b}^{3 \dagger}_{1})^{\dagger}
(\equiv (\stackrel{03}{[-i]}\stackrel{12}{[+]} \stackrel{56}{(-)})^{\dagger})\,.$$

We demonstrated on one example, that knowing the odd ``basis vectors'' we can
reproduce all the even ``basis vectors'', ${}^{I}{\hat{\cal A}}^{m \dagger}_{f}$.
\footnote{ 
In Ref.~\cite{n2024NPB} the relations among even ``basis vectors'', and the odd 
``basis vectors'' are presented in Tables~(2,3,4,5). Tables~(2,3) relate 
${}^{I}{\hat{\cal A}}^{m \dagger}_{f}$  and odd ``basis vectors'', while
Tables~(4,5) relate ${}^{II}{\hat{\cal A}}^{m \dagger}_{f}$  and odd ``basis 
vectors''.
} 
Tables~(\ref{transverseCliff basis5+1even II.} and \ref{S120Cliff basis5+1even II.} relate the odd ``basis vectors'' and their Hermitian 
conjugated partners.\\

We can repeat all the relations obtained for 
${}^{I}{\hat{\cal A}}^{m \dagger}_{f}$ in this subsection also for 
${}^{II}{\hat{\cal A}}^{m \dagger}_{f}$. Kipping in mind Eq.~(\ref{AIIbdaggerb}),
we easily see the essential difference between 
${}^{I}{\hat{\cal A}}^{m \dagger}_{f}$ and 
${}^{II}{\hat{\cal A}}^{m \dagger}_{f}$. While 
${}^{I}{\hat{\cal A}}^{m \dagger}_{f}$ transform family members  of  odd 
``basis vectors'' among themselves, keeping family quantum number unchanged, 
transform ${}^{II}{\hat{\cal A}}^{m \dagger}_{f}$  a particular family member 
to the same family member of all the families, changing the family quantum 
numbers. \footnote{ 
Let us demonstrate the right application of  
${}^{II}{\hat{\cal A}}^{m \dagger}_{f}$ on $\hat{b}^{m' \dagger}_{f `} $
for the case $d=(5+1)$.

The same ${}^{II}{\hat{\cal A}}^{m \dagger}_{f}$, applying from the 
right-hand side on $\hat{b}^{m' \dagger}_{f `} $, transforms any family 
member of one family to the same family member of another family. 
$$\hat{b}^{1 \dagger}_{1}\,(\equiv \stackrel{03}{(+i)}\stackrel{12}{[+]} \stackrel{56}{[+]}) \, *_A \,{}^{II}{\hat{\cal A}}^{1 \dagger}_{4}
 (\equiv \stackrel{03}{(-i)}\stackrel{12}{(+)} \stackrel{56}{[+]})\,
 \rightarrow\hat{b}^{1 \dagger}_{3} 
 (\equiv \stackrel{03}{[+i]}\stackrel{12}{(+)} \stackrel{56}{[+]})\, ,
$$
$$\hat{b}^{1 \dagger}_{1}\,(\equiv \stackrel{03}{(+i)}\stackrel{12}{[+]} \stackrel{56}{[+]}) \, *_A \,{}^{II}{\hat{\cal A}}^{1 \dagger}_{1}
 (\equiv \stackrel{03}{[-i]}\stackrel{12}{(+)} \stackrel{56}{(+)})\,
 \rightarrow \hat{b}^{1 \dagger}_{4} 
 (\equiv \stackrel{03}{(+i)}\stackrel{12}{(+)} \stackrel{56}{(+)})\, ,
$$
$$\hat{b}^{2 \dagger}_{1}\,(\equiv \stackrel{03}{[-i]}\stackrel{12}{(-)} \stackrel{56}{[+]}) \, *_A \,{}^{II}{\hat{\cal A}}^{1 \dagger}_{4}
 (\equiv \stackrel{03}{(-i)}\stackrel{12}{(+)} \stackrel{56}{[+]})\,
 \rightarrow\hat{b}^{2 \dagger}_{3} 
 (\equiv \stackrel{03}{(-i)}\stackrel{12}{[-]} \stackrel{56}{[+]})\, ,
$$
$$\hat{b}^{2 \dagger}_{1}\,(\equiv \stackrel{03}{[-i]}\stackrel{12}{(-)} \stackrel{56}{[+]}) \, *_A \,{}^{II}{\hat{\cal A}}^{1 \dagger}_{1}
 (\equiv \stackrel{03}{[-i]}\stackrel{12}{(+)} \stackrel{56}{(+)})\,
 \rightarrow \hat{b}^{2 \dagger}_{4} 
 (\equiv \stackrel{03}{[-i]}\stackrel{12}{[-]} \stackrel{56}{(+)})\, .
$$
}
Correspondingly, we see in the second kind of bosons, when carrying 
the scalar space-time index $a\ge 5$, a scalar boson - a kind of Higgs. \\

Let us point out that the even ``basis vectors'',  determining the creation and
annihilation operators in a tensor product with the basis in ordinary space-time,
determine spins and charges of boson fields. Having non zero momentum only
in $d=(3+1)$, they carry space index $a =n=(0,1,2,3)$. They behave
in the case that internal space has $(5+1)$ dimensions as a ``photon'', as we
just discussed. Our ``photon'' can exchange the momentum in ordinary
space-time with ``electron'' or ``positron'', but can not influence any internal
property, like there are the spins, $S^{03} $ and $S^{12}$, or the charge
$ S^{56}$.

Let us find out what represents the even ``basis vectors'',
${}^{I}{\hat{\cal A}}^{1 \dagger}_{4}$, with two nilpotents in the
$SO(3,1)$  subgroup of the group $SO(5,1)$. The two spins, $S^{03} $ and
$S^{12}$, enables the creation operators, which are the tensor product of
the basis in ordinary space-time and the even ``basis vectors'' with two
nilpotents, Eq.~(\ref{wholespacebosons}), to form ``gravitons'', presented in
Table~\ref{transverseCliff basis5+1even I.}.
We presents two gravitons, ${\bf {}^{I}{\hat{\cal A}}^{1 \dagger}_{4 n}}%
(\vec{p}) $ and ${\bf {}^{I}{\hat{\cal A}}^{2 \dagger}_{3n}} (\vec{p})$,
\begin{small}
$$
{\bf {}^{I}{\hat{\cal A}}^{1 \dagger}_{4 n}} (\vec{p}) \,=
{}^{I}{\cal C}^{ 1}{}_{4  n} (\vec{p})\,*_{T}\,
{}^{I}{\hat{\cal A}}^{1 \dagger}_{4}
(\equiv \hat{b}^{1 \dagger}_{1}\,*_A\,(\hat{b}^{2 \dagger}_{1})^{\dagger}=
 \stackrel{03}{(+i)}\stackrel{12}{(+)} \stackrel{56}{[+]})\,,
$$
$$
{\bf {}^{I}{\hat{\cal A}}^{2 \dagger}_{3n}} (\vec{p}) \,=
{}^{I}{\cal C}^{ 2}{}_{3 n} (\vec{p})\,*_{T}\,
{}^{I} \hat{{\cal A}}^{2 \dagger}_{3} (\equiv \stackrel{03}{(- i)}
\stackrel{12}{(-)} \stackrel{56}{[+]}=(\hat{b}^{2 \dagger}_{1}
\,*_A\,(\hat{b}^{1 \dagger}_{1})^{\dagger})\,,
$$
with the basis vectors ${}^{I}{\hat{\cal A}}^{1 \dagger}_{4}
(\equiv \stackrel{03}{(+i)}\stackrel{12}{(+)} \stackrel{56}{[+]})$, the first one, 
and with %the basis vectors 
${}^{I}{\hat{\cal A}}^{2 \dagger}_{3}
(\equiv \stackrel{03}{(-i)} \stackrel{12}{(-)} \stackrel{56}{[+]})$, the second one,
which change the  spins, $S^{03}$ and $S^{12}$, of fermions.
When a boson ${\bf {}^{I}{\hat{\cal A}}^{1 \dagger}_{4 n}} (\vec{p}) $ scatters
on a ``electron'' with the spin down, ${\bf \hat{b}}^{2 \dagger}_{1} (\vec{p}) 
(\equiv \hat{b}^{\dagger}_{\vec{p}}\,*_{T}\, \hat{b}^{2 \dagger}_{1}$, 
Eq.~(\ref{wholespacefermions}),
changes its spin from $\downarrow$ to $\uparrow$, and transfers the momentum  to the
``electron''. This  boson ${\bf {}^{I}{\hat{\cal A}}^{1 \dagger}_{4 n}} (\vec{p}) $,
transferring the integer spin to the ``electron'' in addition to momentum of the space-time,
is obviously ``graviton'' with ${\cal S}^{03}=i$ and  ${\cal S}^{12}=1$, changing the
quantum numbers $S^{03}=-\frac{i}{2}$ and  $S^{12}=-\frac{1}{2}$ of
${\bf \hat{b}}^{2 \dagger}_{1} (\vec{p})$ to $S^{03}=\frac{i}{2}$ and
$S^{12}=\frac{1}{2}$ of ${\bf \hat{b}}^{1 \dagger}_{1} (\vec{p})$.\\
\end{small}

Let us check for two cases, how do the ``basis vectors'' of ``gravitons'' behave when ``gravitons'' scatter.

\begin{small}
\begin{eqnarray}
\label{grgrgr}
{}^{I}{\hat{\cal A}}^{2 \dagger}_{3 gr}\,(\equiv \stackrel{03}{(-i)}\stackrel{12}{(-)} \stackrel{56}{[+]}) \,*_A\,
{}^{I}{\hat{\cal A}}^{1 \dagger}_{4 gr}\, (\equiv \stackrel{03}{(+i)}
\stackrel{12}{(+)} \stackrel{56}{[+]})\, \rightarrow \,
{}^{I}{\hat{\cal A}}^{2 \dagger}_{4 ph}\, (\equiv \stackrel{03}{[-i]}\stackrel{12}{[-]} \stackrel{56}{[+]})\,\,,
\nonumber\\
{}^{I}{\hat{\cal A}}^{1 \dagger}_{4 gr}\,
 (\equiv \stackrel{03}{(+i)} \stackrel{12}{(+)} \stackrel{56}{[+]})\,*_A\,
{}^{I}{\hat{\cal A}}^{2 \dagger}_{3 gr}\, (\equiv \stackrel{03}{(-i)}
\stackrel{12}{(-)} \stackrel{56}{[+]})\, \rightarrow \,
{}^{I}{\hat{\cal A}}^{1 \dagger}_{3 ph}(\equiv \stackrel{03}{[+i]}\stackrel{12}{[+]} \stackrel{56}{[+]})\,\,.
\end{eqnarray}
\end{small}
\begin{small}
There are also even ``basis vectors'' of the kind ${}^{I}{\hat{\cal A}}^{m \dagger}_{f}$ 
which change spin and charges, changing for example 
``positrons'' into ``electron''~\footnote{ 
The corresponding bosons  transform ``positrons'' into ``electron'',
$\,{}^{I}{\hat{\cal A}}^{2 \dagger}_{1}\,(\equiv \stackrel{03}{(-i)}\stackrel{12}{[-]} 
\stackrel{56}{(+)}) \,*_A\, \hat{b}^{4 \dagger}_{1}\, (\equiv \stackrel{03}{(+i)}
\stackrel{12}{(-)} \stackrel{56}{(-)}) \rightarrow \hat{b}^{2 \dagger}_{1}\, 
(\equiv \stackrel{03}{[-i]}\stackrel{12}{(-)} \stackrel{56}{[+]})\,$. 
}, changing at the same time the handedness.

\vspace{4mm}

Looking at the even ``basis vector'' in this toy model, there are
one fourth of ${}^{I}{\hat{\cal A}}^{m \dagger}_f$, which are 
``photons'' (like ${}^{I}{\hat{\cal A}}^{1 \dagger}_{3}$ 
and ${}^{I}{\hat{\cal A}}^{2 \dagger}_4$, 
not able to change the quantum numbers of the ``electrons'' and 
``positrons'', presented in Table~\ref{Table Clifffourplet.}) 
or  ``gravitons'' (like ${}^{I}{\hat{\cal A}}^{2 \dagger}_3$ and 
${}^{I}{\hat{\cal A}}^{1 \dagger}_4$, which change the spin of 
``electrons'' and ``positrons'').

The rest eight ${}^{I}{\hat{\cal A}}^{m \dagger}_f$ relate ``electrons'' 
and ``positrons''.
\end{small} % 
\begin{small} 
As we already said, repeating the relations for 
${}^{I}{\hat{\cal A}}^{m \dagger}_{f}$, Eq.~(\ref{phelpo}, \ref{grgrgr}), 
also for ${}^{II}{\hat{\cal A}}^{m \dagger}_{f}$, we shall not get 
``photons''  or ``gravitons'', which both transform family members  of  odd 
``basis vectors'' among themselves, keeping the family quantum number 
unchanged. Carrying the space index equal to $(5,6)$, the scalar bosons 
of the second kind, ${}^{II}{\hat{\cal A}}^{m \dagger}_{f}$, (``photons'' 
and ``gravitons'') cause, as a kind of ``Higgs'', after breaking symmetries 
in this toy model, the masses of fermion fields.
\end{small}\\

{\bf b.} $\,\,$  {\it The realistic case, which offers the ``basis vectors'' 
for all the so far observed fermion and boson fields, requires for internal 
space $d=(13+1)$, and for the space-time, in which fermions and bosons 
have non zero momenta, $d=(3+1)$,} at least at observable energies.

In Table~\ref{Table so13+1.}, App.~\ref{13+1representation}, the 
$2^{\frac{14}{2}-1}$ odd ``basis vectors'' present one irreducible 
representation, one family, of quarks and leptons and antiquarks and 
antileptons (both appearing in the same family), analysed with respect 
to the subgroups $SO(3,1), SU(2)_I, SU(2)_{II},$ $ SU(3),$ $ U(1)$ 
of the group $SO(13,1)$. One can notice in Table~\ref{Table so13+1.}, 
that the content of the subgroup $SO(7,1)$ (including subgroups 
$SO(3,1), SU(2)_I, SU(2)_{II}$) are identical for quarks and leptons, 
and identical for antiquarks and antileptons; due to two $SU(2)$ subgroups 
$SU(2)_I, SU(2)_{II}$, first representing the weak charge, postulated 
by the {\it standard model}, the second $SU(2)_{II}$ group members 
are not (yet) observed at low energies. Quarks and leptons, and 
antiquarks and antileptons distinguish only in the $SU(3)\times U(1)$ 
part of the group $SO(13,1)$.  

From the first member, the odd ``basis vector'' $u_R^{c1}$  in 
Table~\ref{Table so13+1.}, follow the rest odd ``basis vectors'' by the 
application of the infinitesimal generators of the Lorentz group $S^{ab}$, 
as well as by the application of ${}^{I}{\hat{\cal A}}^{m \dagger}_{f}$).
All the first members of the other families follow from the one presented in 
Table~\ref{Table so13+1.} by applying on $u_R^{c1}$ by $\tilde{S}^{ab}$, 
as well as by the application of ${}^{II}{\hat{\cal A}}^{m \dagger}_{f}$.

The corresponding creation and annihilation operators  of fermions are 
tensor products of a ``basis vector'' and the basis in ordinary space-time, 
for example, ${\bf u_R^{c1}}(\vec{p}) = u_R^{c1} \,*_T\, 
\hat{b}_{\vec{p}}$.\\

The even ``basis vectors'' can be obtained, according 
to Eqs.~(\ref{AIbbdagger}, \ref{AIIbdaggerb}), as the algebraic products 
of the odd ``basis vectors'' and their Hermitian conjugate partners; In a 
tensor product with the basis in ordinary space-time, and with the space index $a=n (=0,1,2,3)$ added,
 ${\bf {}^{I}{\hat{\cal A}}^{m \dagger}_{f a}} (\vec{p}) =$
 $       {}^{I}{\cal C}^{ m}{}_{f a} (\vec{p})\,*_{T}\,$
$ {}^{I}{\hat{\cal A}}^{m \dagger}_{f}$.

${\bf {}^{I}{\hat{\cal A}}^{m \dagger}_{f a}} (\vec{p})$ manifest the 
properties of the tensor ($a=n$), vector ($a=n$) and scalar ($a=s \ge 5$)
gauge fields, observed so far.

 In a tensor product with the basis in ordinary space-time, and with the space 
index $a=s \ge 5$ added, 
${\bf {}^{I}{\hat{\cal A}}^{m \dagger}_{f a}} (\vec{p}) $ manifest the 
properties of the scalar fields, like the Higgs and other scalar fields, bringing 
masses to quarks and leptons and antiquarks and antileptons and to weak 
bosons, for example.

Let us look in Table~\ref{Table so13+1.} for $e^{- \dagger}_L$,  $29^{th} $
 line. The photon  ${}^{I}{\hat{\cal A}}^{\dagger}_{ph\, e^{- \dagger}_L
 \rightarrow e^{- \dagger}_L}$ interacts with $e^{- \dagger}_L$ as  
 follows
\begin{small}
\begin{eqnarray}
\label{phAee}
&&{}^{I}{\hat{\cal A}}^{ \dagger}_{ph \, e^{- \dagger}_L
 \rightarrow e^{- \dagger}_L}
(\equiv \stackrel{03}{[-i]}\stackrel{12}{[+]} \stackrel{56}{[-]}
 \stackrel{7 8}{[+]}\stackrel{9\,10}{[+]} \stackrel{11\,12}{[+]}
\stackrel{13\,14}{[+]})\,*_A\,     e^{- \dagger}_{L},
(\equiv \stackrel{03}{[-i]}\stackrel{12}{[+]} 
\stackrel{56}{(-)}\stackrel{78}{(+)}\stackrel{9\,10}{(+)} 
\stackrel{11\,12}{(+)} \stackrel{13\,14}{(+)})\,\rightarrow 
\nonumber\\
&& e^{- \dagger}_{L}
(\equiv \stackrel{03}{[-i]}\stackrel{12}{[+]} \stackrel{56}{(-)}
 \stackrel{7 8}{(+)}\stackrel{9\,10}{(+)} \stackrel{11\,12}{(+)}
\stackrel{13\,14}{(+)f})\,,
 \quad \quad \quad
{}^{I}{\hat{\cal A}}^{ \dagger}_{ph \, e^{- \dagger}_L
 \rightarrow e^{- \dagger}_L} =  e^{- \dagger}_L,
\,*_A\, ( e^{- \dagger}_L)^{\dagger}\,,
%\nonumber\\
\end{eqnarray}
 \end{small}

Let us look for the weak boson, transforming $ e^{- \dagger}_L$ from the
$29^{th}$ line into $ \nu^{\dagger}_L$ from the $31^{st}$ line.

It follows
\begin{small}
\begin{eqnarray}
\label{w1enuL}
&&{}^{I}{\hat{\cal A}}^{ \dagger}_{w1\, e_{L}
\rightarrow \nu_{L}}
(\equiv \stackrel{03}{[-i]}\stackrel{12}{[+]} \stackrel{56}{(+)}
 \stackrel{7 8}{(-)}\stackrel{9\,10}{[+]} \stackrel{11\,12}{[+]}
\stackrel{13\,14}{[+]})\,*_A\,  e^{- \dagger}_{L}
(\equiv \stackrel{03}{[-i]}\stackrel{12}{[+]} 
\stackrel{56}{(-)}\stackrel{78}{(+)}\stackrel{9\,10}{(+)} 
\stackrel{11\,12}{(+)} \stackrel{13\,14}{(+)})\,\rightarrow 
\nonumber\\
&&\nu^{\dagger}_{L},
(\equiv \stackrel{03}{[-i]}\stackrel{12}{[+]} 
\stackrel{56}{[+]}\stackrel{78}{[-]}\stackrel{9\,10}{(+)} 
\stackrel{11\,12}{(+)} \stackrel{13\,14}{(+)})\,, 
\quad \quad \quad {}^{I}{\hat{\cal A}}^{ \dagger}_{w1\, e_{L}
\rightarrow \nu_{L}}= \nu^{\dagger}_{L}\, *_A\,
(e^{-\dagger}_{L})^{\dagger}\,.%\nonumber\\
\end{eqnarray}
\end{small}
Knowing ``basis vectors'' of fermions, we can find ``basis vectors''  
of all bosons fields. Only few of them are among the so far observed
boson fields.~\footnote{
We discuss in Refs.~\cite{nh2021RPPNP,n2024NPB}, that breaks 
of symmetries make that not all possibilities of the predicted vector 
and scalar boson fields are observable at low energies. Also among 
the observed fermions and antifermions the predicted right-handed
neutrinos and left-handed antineutrinos wait to be observed. Our 
model predicts more families than observed, like the fourth family 
to the observed three,~\cite{gmdn2008,gn2009} and the dark
matter~\cite{n2014matterantimatter,gn2014} as the (almost stable)
fifth family belonging to the upper four family.
}

However, studying all the boson fields might help to recognise
why and how the properties of fermions and bosons change
with breaking symmetries, if this theory describing the internal 
spaces of fermion and boson fields with odd and even  ``basis 
vectors'' is what our universe obeys.

Demonstrating so many simple and elegant descriptions of the 
second quantized fields, explaining the assumptions of other 
theories, makes us hop that the theory might be what the 
universe obeys.\\

Since the graviton in this theory is understood in an 
equivalent way as all the gauge fields observed so far, let 
us at the end of this section, try to analyse the ``basis vectors'' 
of the gravitons if the internal space has $d=(13+1)$.

We must take into account that the ``gravitons'' do have the spin
and handedness (non-zero ${\cal S}^{03}$ and $ {\cal S}^{12}$, 
which means that this part must be presented by two nilpotents,
$\stackrel{03}{(\pm i)}\stackrel{12}{(\pm)} $) in $d=(3+1)$, 
and do not have weak, colour and $U(1)$ charges (what means 
that all the rest must be projectors), and have, as all the vector 
gauge fields, the space index $n=(0,1,2,3)$.

We can then easily find the ``basis vector'' of the graviton, 
${}^{I}{\hat{\cal A}}^{ \dagger}_{gr\, u^{c1 \dagger}_{R}
\rightarrow u^{c1 \dagger}_{R}}$, which applying on 
$u^{c1 \dagger}_{R}$ with spin up, appearing 
in the first line of the table~\ref{Table so13+1.}, transforms it into 
$u^{c1 \dagger}_{R}$ with spin down, appearing in the second 
line of the table~\ref{Table so13+1.}.
%
 % G
\begin{small}
\begin{eqnarray}
\label{grAqeqe}
&&{}^{I}{\hat{\cal A}}^{ \dagger}_{gr \,u^{c1 \dagger}_{R\uparrow}
 \rightarrow u^{c1 \dagger}_{R\downarrow}}
(\equiv \stackrel{03}{(-i)}\stackrel{12}{(-)} \stackrel{56}{[+]}
 \stackrel{7 8}{[+]}\stackrel{9\,10}{[+]} \stackrel{11\,12}{[-]}
\stackrel{13\,14}{[-]})\,*_A\, u^{c1 \dagger}_{R \uparrow} 
(\equiv \stackrel{03}{(+i)}\stackrel{12}{[+]} 
\stackrel{56}{[+]}\stackrel{78}{(+)}\stackrel{9\,10}{(+)} 
\stackrel{11\,12}{[-]} \stackrel{13\,14}{[-]})\,\rightarrow 
\nonumber\\
&&u^{c1 \dagger}_{R\downarrow},
(\equiv \stackrel{03}{[-i]}\stackrel{12}{(-)} 
\stackrel{56}{[+]}\stackrel{78}{(+)}\stackrel{9\,10}{(+)} 
\stackrel{11\,12}{[-]} \stackrel{13\,14}{[-]})\,,
 \quad \quad \quad
{}^{I}{\hat{\cal A}}^{ \dagger}_{gr\, u^{c1 \dagger}_{R\uparrow}
 \rightarrow u^{c1 \dagger}_{R\downarrow}} = 
u^{c1 \dagger}_{R\downarrow}\,*_A\, (u^{c1 \dagger}_{R\uparrow})^{\dagger}\,.
\end{eqnarray}
\end{small}

Let us look at the ``scattering'' (that is the algebraic application, $\,*_A\,$) of the 
graviton with the ``basis vector'' 
${}^{I}{\hat{\cal A}}^{ \dagger}_{gr \,u^{c1 \dagger}_{R\uparrow}
\rightarrow u^{c1 \dagger}_{R\downarrow}}$ with the graviton with  the
``basis vector'' \\
${}^{I}{\hat{\cal A}}^{ \dagger}_{gr \,u^{c1 \dagger}_{R\downarrow}
 \rightarrow u^{c1 \dagger}_{R\uparrow}}\,
(\equiv \stackrel{03}{(+i)}\stackrel{12}{(+)} \stackrel{56}{[+]}
 \stackrel{7 8}{[+]}\stackrel{9\,10}{[+]} \stackrel{11\,12}{[-]}
\stackrel{13\,14}{[-]})$, that is
 $${}^{I}{\hat{\cal A}}^{ \dagger}_{gr \,u^{c1 \dagger}_{R\uparrow}
 \rightarrow u^{c1 \dagger}_{R\downarrow}}
(\equiv \stackrel{03}{(-i)}\stackrel{12}{(-)} \stackrel{56}{[+]}
 \stackrel{7 8}{[+]}\stackrel{9\,10}{[+]} \stackrel{11\,12}{[-]}
\stackrel{13\,14}{[-]})\,*_A\, 
{}^{I}{\hat{\cal A}}^{ \dagger}_{gr \,u^{c1 \dagger}_{R\downarrow}
 \rightarrow u^{c1 \dagger}_{R\uparrow}}
(\equiv \stackrel{03}{(+i)}\stackrel{12}{(+)} \stackrel{56}{[+]}
 \stackrel{7 8}{[+]}\stackrel{9\,10}{[+]} \stackrel{11\,12}{[-]}
\stackrel{13\,14}{[-]}) \rightarrow$$
$$ (\equiv \stackrel{03}{[-i]}\stackrel{12}{[-]} \stackrel{56}{[+]}
 \stackrel{7 8}{[+]}\stackrel{9\,10}{[+]} \stackrel{11\,12}{[-]}
\stackrel{13\,14}{[-]})= u^{c1 \dagger}_{R\downarrow} %,
\,*_A\, (u^{c1 \dagger}_{R\downarrow})^{\dagger}=\,
{}^{I}{\hat{\cal A}}^{ \dagger}_{ph \,u^{c1 \dagger}_{R\downarrow}
 \rightarrow u^{c1 \dagger}_{R\downarrow}}\,,
 $$
to recognize how easily one finds the internal space of bosons.\\

The creation operators for gravitons must carry the space index 
$n=(0,1,2,3)$, like: 
${\bf {}^{I}{\hat{\cal A}}^{ \dagger}_{gr \,u^{c1 \dagger}_{R\uparrow}
 \rightarrow u^{c1 \dagger}_{R\downarrow} n}} (\vec{p})$.

% G down
\section{General algebraic structure of fermion and boson second
quantised fields}
\label{genaralalgebraicstructure}

We demonstrated in Sect.~\ref{internalspaces} the relations among
second quantised fermion and boson fields when treating internal spaces
of fermions and bosons in a unique way: fermions as a superposition of
odd products of operators $\gamma^{a}$, bosons as a superposition of
even products of operators $\gamma^{a}$  carrying in addition the 
ordinary space-time index $a$, named odd and even ``basis
vectors'', respectively. Although the corresponding states of the creation
and annihilation operators have, in the case that the internal space has
$d=(13+1)$ while in the ordinary space-time fermions and bosons are
active only in $d=(3+1)$, the desired properties as assumed by the
other theories, yet there are differences, which might lead to different
predictions and conclusions.

Let us point out the main differences:\\
{\bf \,\, a} All ``basis vectors'' of fermions are mutual orthogonal and
so are mutually orthogonal their Hermitian conjugate partners,
and correspondingly their creation and annihilation operators.\\
{\bf \,\, b} Fermions appear in families, bosons appear in two orthogonal
groups.\\
{\bf \,\, c} Bosons multiply fermions either from the left-hand side (when
causing transformations within one family) or from the right-hand side
(when causing transformations among families).\\

In this section, we generalize Sect.~\ref{internalspaces}, by generalising 
structure of massless fermion and boson second quantised fields with 
the non-zero momentum only in$d=(3+1)$ of the ordinary space-time, 
presented in Refs.~\cite{%n2022IARD,
n2023NPB,n2023MDPI,n2024NPB,%n2025Bled,
nh2025Bled}, and explained and discussed in 
Sect.~\ref{internalspaces} of this paper.

 The decision that the internal spaces of the second
quantised fermion and boson fields are described by a superposition of
odd numbers of operators $\gamma^a$'s for fermions and even
numbers of $\gamma^a$'s for bosons, bosons carry the ordinary
space-time index $a$, determines all the properties of the fermion
and boson fields, as well as their mutual interactions, which are
realised by the algebraic multiplication.

This decision explains Dirac's postulates for the second quantised
fermion and boson fields, determines the statistics of fermions and bosons,
determines families of fermions, each of which includes fermions and
antifermions~\footnote{
Consequently, the vacuum is not the negative-energy Dirac vacuum;
It is just the quantum vacuum.
},
determines the Lorentz (and correspondingly gauge) symmetry of
fermion and boson fields, Subsect.~\ref{internallorentzC}, determines
couplings among fermion and boson fields,
Subsect.~\ref{covariantderivative}, and consequently the Lagrange
densities, Subsects.~(\ref{fermionlagrange},~\ref{bosonlagrange}),
determines two orthogonal kinds of boson fields,
Subsects.~(\ref{structurebosons},~\ref{bosonlagrange}), determines
orthogonality of fermion states, what all influences Feynman
diagrams~\ref{feynmanrules} (which should reproduce
the experimental data).

\subsection{Algebraic structure}
\label{algebraicstructure}

The theory offers fermion fields, they are tensor products of ``basis 
vectors'' and basis in ordinary space-time, Eq.~(\ref{wholespacefermions}),
\begin{equation} %\[
\label{psigen}
\psi(x) \in Cl^{\mathrm{odd}}(d-1,1) ,
\end{equation}
% \]
as Clifford odd $2^{\frac{d}{2}-1}$ members in $2^{\frac{d}{2}-1}$ 
irreducible representations, families, together with the same number of their 
Clifford odd Hermitian conjugate partners. They satisfy the relations of  
Eq.~(\ref{orthogonalodd})~\footnote{
The oddness of ``basis vectors'' determines the orthogonality.
}
\begin{equation} % \[
\label{orthogonaloddC}
\psi *_A \psi = 0,
\qquad
\psi^\dagger *_A \psi^\dagger = 0,
\qquad
\psi^\dagger *_A \psi \ne 0 ,
\end{equation} %\]
where $*_A$ denotes an associative algebraic product.

The theory also contains two orthogonal bosonic fields, which are
tensor products of ``basis vectors'' and basis in ordinary space-time,
Eq.~(\ref{wholespacebosons}), each with
$2^{\frac{d}{2}-1}\times 2^{\frac{d}{2}-1}$ members,
\[
{}^{I}\!\hat{\mathcal A}_a(x), 
\qquad
{}^{II}\!\hat{\mathcal A}_a(x),
\]
with
\[
{}^{I}\!\hat{\mathcal A}_a,\,
{}^{II}\!\hat{\mathcal A}_a 
\in Cl^{\mathrm{even}}(d-1,1) ,
\]
$a$ is the space-time index~\footnote{ 
The space-time index $a$ is in this theory equal to $n=(0,1,2,3)$
for vectors and tensors, and $a\ge 5$ for scalars.
}.
They satisfy the orthogonality condition, Eq.~(\ref{AIAIIorth}),
\begin{equation} %
\label{AIAIIorthC}
{}^{I}\!\hat{\mathcal A}_a *_A {}^{II}\!\hat{\mathcal A}_b
=
{}^{II}\!\hat{\mathcal A}_a *_A {}^{I}\!\hat{\mathcal A}_b
=0 .
\end{equation} % 

The chosen algebraic structure leads to the relations of Eq.~(\ref{ruleAAI}),
\begin{eqnarray}
\label{ruleAAIC}
{}^{i}\!\hat{\mathcal A}_a *_A {}^{i}\!\hat{\mathcal A}_a
\rightarrow 
{}^{i}\!\hat{\mathcal A}_a, 
\qquad i=(I,II),
\end{eqnarray}
and of Eq.~(\ref{calIAb1234gen}),
\begin{eqnarray}
\label{calIAb1234genC}
{}^{I}\!\hat{\mathcal A}_a *_A \psi
\rightarrow \psi ,
\end{eqnarray}
transforming fermion fields within an irreducible representation - family, 
and to Eq.~(\ref{calbIIA1234gen}),
\begin{eqnarray}
\label{calbIIA1234genC}
\psi *_A {}^{II}\!\hat{\mathcal A}_a
\rightarrow
\psi ,
\end{eqnarray}
transforming fermion fields among irreducible representations - families. 
\subsection{Internal Lorentz symmetry}
\label{internallorentzC}

Local internal Lorentz transformations in the internal space of fermion 
and boson fields are generated by, Ref.~\cite{n2023NPB},
\[
{\cal S}^{ab} = (S^{ab} + \tilde S^{ab}) .
\]

The corresponding local transformation reads
\[
\Lambda (x) =
e^{i \omega_{ab}(x) {\cal S}^{ab}},
\qquad
\Lambda^{\dagger} (x) *_A \Lambda (x)=1 ,
\]
provided that $\omega_{0i}^*=-\omega_{0i}$ and $\omega_{ij}^*=
\omega_{ij}$.
%%%%%%%%%%%%%%%%%%%%%%

The fermion field transforms as
\[
\psi \longrightarrow \Lambda *_A \psi .
\] 

Translations in ordinary space-time are generated by operators
$p_a$. Rotations in
both spaces, internal and ordinary space-time, are generated by
$M^{ab}= ({L}^{ab} + S^{ab} + \tilde S^{ab})$~\footnote{
${L}^{ab}$ gives non-zero contributions only in $d={3+1}$. Only
in $d={3+1}$ are the rotations in internal space and in ordinary
space-time related.
}.

\subsection{Covariant derivative}
\label{covariantderivative}

The covariant derivative corresponding to the algebraic properties of the 
fermion and boson fields acting on $\psi$ is defined by, 
Eqs.~(\ref{calIAb1234gen},~\ref{calIAb1234genC},~\ref{calbIIA1234gen},%
~\ref{calbIIA1234genC}),  
\begin{equation} 
\label{covderC}
D_a \psi
=
p_a \psi
-
{}^{I}\!\hat{\mathcal A}_a *_A \psi
-
\psi *_A {}^{II}\!\hat{\mathcal A}_a .
\end{equation} 

Under a local Lorentz transformation in internal space (defined by 
$( S^{ab}+ \tilde{S}^{ab})$, concerning the internal and space-time 
Lorentz transformations one needs to take $M^{ab}=(L^{ab} + S^{ab}+
\tilde{S}^{ab})$.)
\[
p_a \longrightarrow
\Lambda^{\dagger} *_A p_a *_A \Lambda
=
p_a + (p_a \Lambda^{\dagger}) *_A \Lambda .
\]

Lorentz covariance of $D_a \psi$ requires
\begin{equation}
\label{covLC}
D_a *_A \,(\Lambda *_A \psi ) \longrightarrow 
 (\Lambda^{\dagger} *_A  D_a  *_A  \Lambda )*_A  \psi  ,
\end{equation}
as we shall see.
% G up

%
\subsection{Fermion Lagrangian and local Lorentz transformations
 in internal space}
\label{fermionlagrange}

Following the properties of the fermion and boson fields, repeated 
above in Sects.~(\ref{internalspaces},~\ref{genaralalgebraicstructure}
up to this subsection) we are able to consider the fermionic 
Lagrangian density in a Dirac way, and with our covariant derivative 
$D_a$
\begin{equation}
\label{ferLC}
\mathcal L_F (\psi)
=
\frac{1}{2}
\left[
\psi^\dagger *_A \gamma^0 \gamma^a *_A D_a \psi
+
(D_a \psi)^\dagger *_A \gamma^0 \gamma^a *_A \psi
\right].
\end{equation}

The covariant derivative acting on the fermion operator $\psi$ is
\begin{equation}
\label{covderLC}
D_a *_A \psi
=
(p_a - {}^{I}{\cal A}_a) *_A  \psi
-
\psi\, *_A {}^{II}{\cal A}_a .
\end{equation}

We consider a local Lorentz transformation in internal space
\begin{equation} %\[
\label{lambdaDirac}
\psi \rightarrow \psi' = \Lambda \psi , % \] 
\end{equation}
with
\begin{equation}
\Lambda(x)=e^{i\omega_{ab}(x)\mathcal S^{ab}}\, .\nonumber
%\exp\!\left(\frac{i}{2}\omega_{ab}(x)\mathcal S^{ab}\right).\] %
\end{equation}
Using this in the Lagrange density it follows~\footnote{
\begin{small}
Looking at transformations in the first order in the way
%%%%%%%%%%%%%%%%%%%%%%
\begin{eqnarray}
\label{lambdaDiracC}
  \frac{1}{2} \{(\gamma^0 \gamma^a p_a \,  (1+ i \omega_{ij} S^{ij}  
+ i \omega_{0i} S^{0i}) \psi') ^{\dagger}) (1+ i \omega_{ij} S^{ij}+ 
i \omega_{0i} S^{0i}) \psi' +  \nonumber \\
 ((1+ i \omega_{ij} S^{ij}  +i \omega_{0i} S^{0i}) \psi')^{\dagger}
\,\gamma^0 \gamma^a p_a \, (1+ i \omega_{ij} S^{ij}  + 
i \omega_{0i} S^{0i}) \psi'   \}\,,\nonumber\\
=\frac{1}{2} \{ (p_a \psi') ^{\dagger}  \gamma^0 \gamma^a \psi' +  (\psi')^{\dagger}  
 \gamma^0 \gamma^a p_a  \psi' \}\,,\quad  \quad \quad \quad 
 \quad \quad 
 \end{eqnarray}
 after taking into account that  $ \omega_{ij}^{*}= \omega_{ij}$, 
while $\omega_{0i}^{*}= - \omega_{0i}$, and that $(S^{ij})^{\dagger}
= S^{ij}, $  $(S^{0i})^{\dagger}= - S^{0i}$\,.
\end{small}
}
\begin{equation}
\mathcal L_F(\Lambda \psi)
=
\frac{1}{2}
\left[
(\Lambda \psi)^\dagger *_A \gamma^0 \gamma^a *_A D_a *_A (\Lambda \psi)
+
(D_a *_A \Lambda \psi)^\dagger *_A \gamma^0 \gamma^a *_A \Lambda \psi
\right] ,  \nonumber% \] % 
\end{equation}
which leads to
% \[  % 
\begin{equation}
\mathcal L_F(\Lambda \psi)
=
\frac{1}{2}
\left[
(\psi)^{\dagger } \Lambda^{\dagger}*_A \gamma^0 \gamma^a *_A 
\Lambda \Lambda^{\dagger}*_A D_a *_A \Lambda \psi
+
(\Lambda \Lambda^{\dagger}*_A D_a *_A  \Lambda \psi)^\dagger *_A 
\gamma^0 \gamma^a *_A \Lambda \psi
\right] . 
\nonumber %  
\end{equation}
Taking into account that $ \Lambda^{\dagger}*_A \gamma^0 \gamma^a 
*_A \Lambda = \gamma^{`0} \gamma^{`a}$ 
is the Lorentz transformed object, and $ \Lambda^{\dagger} *_A  D_a 
*_A  \Lambda =D'_a$ is the Lorentz
transformed object, while $ (\Lambda *_A D'_a \psi)^{\dagger}=(D'_a 
*_A \psi)^{\dagger} \Lambda^{\dagger}$,
the new, transformed,  fermion Lagrange density is 
\begin{equation}
\label{LTferLC}
\mathcal L_F(\Lambda \psi)
=
\frac{1}{2}
\left[
\psi^\dagger *_A \gamma^{`0} \gamma^{`a} *_A D'_a  *_A \psi
+
(D'_a \psi)^\dagger *_A \gamma^{`0} \gamma^{`a}*_A \psi
\right] .
\end{equation}
Expanding the derivative term yields
\begin{equation}
p_a(\Lambda\psi)
=
(p_a \Lambda)\psi
+
\Lambda(p_a\psi) . \nonumber % 
\end{equation}

The new covariant derivative is, if we call $-  \Lambda^{\dagger} *_A \,
\, {}^{I}{\cal A}_a *_A \Lambda + p_a \Lambda= - {}^{I}{\cal A'}_a$
 and $- \Lambda^{\dagger} *_A \,
\, {}^{II}{\cal A}_a  *_A\Lambda= - {}^{II}{\cal A'}_a$.
\begin{equation}
\label{newcovC}
 \Lambda^{\dagger}D_a (\Lambda \psi)
= D'_a \psi=\{%(p_a \Lambda)
 - {}^{I}{\cal A'}_a \} *_A  \psi - \psi *_A \,\,{}^{II}{\cal A'}_a .
\end{equation}

This determines the transformation laws of the bosonic fields
\begin{eqnarray}
\label{Ltransbosons}
{}^{I}{\cal A}'_a
&=&
\Lambda^{\dagger} {}^{I}{\cal A}_a \Lambda
+
\Lambda^{\dagger} (p_a\Lambda)\, ,  \nonumber
\\
{}^{II}{\cal A}'_a
&=&
\Lambda^{\dagger} {}^{II}{\cal A}_a \Lambda\, .
 %\nonumber
\end{eqnarray}
With these transformation rules one obtains
\begin{equation}
\label{eqfermionC}
\gamma^0 \gamma^a D_a (\Lambda \psi)
=
\gamma^{`0} \gamma^{`a} *_A D'_a  *_A \psi ,
\end{equation}
and the fermion Lagrange density preserves its form under 
local Lorentz transformations as we promised in 
Subsect.~\ref{covariantderivative}.~\footnote{
The gauge transformations of boson fields coincide with (some of) 
the Lorentz transformations concerning only the internal space with
$d >3$ if the boson fields carry the space index $n=(0,1,2,3)$.
}

We could as well assume the quadratic form of the Lagrange density for 
fermions, again with  the covariant derivative $(D_a  \psi$ presented in 
Eq.~(\ref{newcovC})
\begin{equation}
\label{quadLfermionC}
{\cal L}_F = (D_a  \psi)^{\dagger}\, *_A \,  (D^a  \psi)\,.
\end{equation}
 This Lagrange density, obviously Hermitian, transforming under 
Lorentz rotations as required by Eqs.~(\ref{newcovC},~\ref{Ltransbosons}),
 is waiting to be studied.

\subsubsection{Equation of motion for fermion fields}
\label{equationfermions}

Varying the fermion Lagrangian density with respect to the Hermitian
conjugate field $\psi^\dagger$ gives

\begin{equation} 
\delta {\cal L}_F
=
\frac12
\left[
\delta\psi^\dagger *_A \gamma^0\gamma^a *_A D_a\psi
+
(D_a\delta\psi)^\dagger *_A \gamma^0\gamma^a *_A \psi
\right]\,. \nonumber
\end{equation} 
After integration by parts one obtains the equation of motion

\begin{equation}
\label{eqfermion}
\gamma^0\gamma^a *_A D_a \psi =0 \,.
\end{equation}

Using the definition of the covariant derivative
\begin{equation} 
D_a\psi
=
p_a\psi
-
{}^{I}\!\hat{\mathcal A}_a *_A \psi
-
\psi *_A {}^{II}\!\hat{\mathcal A}_a \,, \nonumber
\end{equation} 
the explicit form of the fermion equation becomes
\begin{equation}
\label{expliciteqfermion}
\gamma^0\gamma^a
\left(
p_a\psi
-
{}^{I}\!\hat{\mathcal A}_a *_A \psi
-
\psi *_A {}^{II}\!\hat{\mathcal A}_a
\right)
=0 \,.
\end{equation}

We see that the theory, assuming that internal spaces of fermion 
and boson fields are described by the odd number of nilpotents 
(fermions) and the even number of nilpotents (bosons), the rest 
of projectors, fermions can interact only by exchanging bosons,
either with left or right multiplication.

\subsection{Algebraic structure of boson fields}
\label{structurebosons}
Boson fields in the present approach carry a space-time index
\begin{equation} 
a = 0,1,2,3,5,6,\dots ,d ,
\nonumber
\end{equation} 
~\footnote{Here the indices $(0,1,2,3)$ correspond to the usual 
$(3+1)$-dimensional space-time when representing vectors and 
tensors (only in this space-time bosons, as well as fermions, are 
active), while $a\ge 5$ determine scalars.
The internal space is active in all $d=2(2n+1)$.
}

For $a=0,1,2,3$ the fields
\[
{\cal A}_a(x)
\]
transform as vectors in space-time. For $a\ge 5$ the corresponding
fields appear as scalars from the $(3+1)$--dimensional point of view.
Boson fields therefore appear either as space-time vectors or as
space-time scalars.

The internal structure of bosons can be expressed in terms of fermion
operators $\psi$ and $\psi^\dagger$, Eqs.~(\ref{AIbbdagger},%
~\ref{AIIbdaggerb}). The algebra satisfies the
nilpotent relations, Eq.~(\ref{orthogonalodd}),
\[
\psi *_A \psi =0 ,
\qquad
\psi^\dagger *_A \psi^\dagger =0 ,
\]
while mixed products do not vanish. The internal structure of bosonic
operators can be expressed through the bilinears,  
Eqs.~(\ref{AIbbdagger},~\ref{AIIbdaggerb}),
\[
\psi *_A \psi^\dagger ,
\qquad
\psi^\dagger  *_A \psi .
\]
The algebra distinguishes two types of bosons, Eqs.~(\ref{AIbbdagger},~\ref{AIIbdaggerb}),
\[
{}^{I}{\cal A}_a \sim \psi *_A \psi^\dagger ,
\qquad
{}^{II}{\cal A}_a \sim \psi^\dagger  *_A \psi .
\]

Both transform under Lorentz transformation as presented in 
Eq.~(\ref{Ltransbosons}).~\footnote{
 Let us see what Eq.~(\ref{Ltransbosons}) does say about the  transformation
of bosons ${}^{II}{\cal A}_a \sim \psi^\dagger  *_A \psi$ and 
${}^{I}{\cal A}_a \sim \psi  *_A \psi^{\dagger}$
\begin{small}
\begin{eqnarray}
\label{AIAIItransLC}
{}^{I}{\cal A}_a&=& \psi *_A \psi^{\dagger}\, ,\quad
{}^{II}{\cal A}_a = \psi^{\dagger} *_A \psi\, , \nonumber\\
{}^{I}{\cal A'}_a &=& (\Lambda^{\dagger} \psi) *_A (\Lambda^{\dagger} \psi)^{\dagger}\, =
\Lambda^{\dagger} \psi *_A \psi^{\dagger} \Lambda= {}^{I}{\cal A'}_a \nonumber\\
{}^{II}{\cal A'}_a &=& (\Lambda^{\dagger} \psi)^{\dagger} *_A (\Lambda^{\dagger} \psi)\, =
\psi^{\dagger} \Lambda *_A  \Lambda^{\dagger} \psi = {}^{II}{\cal A'}_a \nonumber\\
{\rm leading} & & {\rm \,to} \nonumber\\
{}^{I}{\cal A'}_a &=& \Lambda^{\dagger} \psi *_A \psi^{\dagger} \Lambda \, ,
\quad {}^{II}{\cal A'}_a = \psi^{\dagger} *_A \psi\,.
 %\nonumber
\end{eqnarray}
\end{small}
While ${}^{I}{\cal A}_a$ transform as we use to see transformation 
properties of vector gauge fields, the second kind does not transform
at all.
}
\\

Because of the nilpotent relations - their mutual algebraic products are
zero - these two sectors do not interact directly. They can interact only 
in the presence of a fermion. One finds schematically~\footnote{
Let us demonstrate the relation in Eq.~(\ref{Apsi}) on the case that 
$d=(5+1)$, with the fermion and boson ``basis vectors'' presented in
Table~\ref{Table Clifffourplet.}. Let the ``basis vector'' of 
${}^{I}{\cal A}_a$ be equal to ${}^{I}\hat{\cal A}^{1 \dagger}_{4}
(\equiv \stackrel{03}{(+i)}\stackrel{12}{(+)} \stackrel{56}{[+]}))=
{\hat b}^{1 \dagger}_1 (\equiv \stackrel{03}{(+i)}\stackrel{12}{[+]}
\stackrel{56}{[+]})\,*_A\, ({\hat b}^{2 \dagger}_1)^{\dagger}
(\equiv (\stackrel{03}{[-i]}\stackrel{12}{(-)} 
\stackrel{56}{[+]})^{\dagger})$, and let it applies on a fermion $\psi$ 
with the ``basis vector'' equal to ${\hat b}^{2 \dagger}_1(\equiv
\stackrel{03}{[-i]}\stackrel{12}{(-)}\stackrel{56}{[+]}$. $\psi$ with
the ``basis vector'' $ {\hat b}^{2 \dagger}_1$ scatters  to 
${\hat b}^{1 \dagger}_1$ and ${}^{II}{\cal A}_a$  with the ``basis 
vector'' equal to ${}^{II}\hat{\cal A}^{1 \dagger}_{3}(\equiv 
\stackrel{03}{[-i]}\stackrel{12}{[+]}\stackrel{56}{[+]}$). $\psi$ and
${}^{II}{\cal A}_a$ either share the momentum and energy, or
${}^{II}{\cal A}_a$ remains momentumless in the vacuum, obeying
Eq.~(\ref{calIAb1234gen}).
}
\begin{equation} 
\label{Apsi}
{}^{I}{\cal A}_a *_A \psi
=
(\psi *_A \psi^\dagger) *_A \psi
=
\psi *_A (\psi^\dagger *_A \psi)
\sim
\psi\,*_A\,\,{}^{II}{\cal A}_a .
\end{equation} 
Thus bosons of type I and type II are connected only through fermions, what 
must be recognized in Feynman diagrams, which must, of course,
reproduce the experimental data. 

When ${}^{I}{\cal A}_a *_A \psi \rightarrow \psi$, it is indeed 
${}^{I}{\cal A}_a *_A \psi \rightarrow \psi *_A \,\,{}^{II}{\cal A}_a$.
In this case
${}^{II}{\cal A}_a$ can either remain without energy and momentum,
taken by $\psi$, or both - ${}^{II}{\cal A}_a$  and $\psi$ - share energy 
and momentum~\footnote{
\begin{small}
When applying either  ${}^{I}{\cal A}_a(x) $ on $\psi $ or $\psi $ on  
${}^{II}{\cal A}_a(x) $, we always get two fields back.  In the first case, 
when ${}^{I}{\cal A}_a(x) $ apply on $\psi $ we get (another) $\psi$ and
${}^{II}{\cal A}_a(x) $. The boson field may  stay in the vacuum without 
energy, or appear as well as a propagating field. In the second case when
$\psi$ applies on ${}^{II}{\cal A}_a(x) $ we get
$\psi *_A (\psi^\dagger *_A \psi)= (\psi *_A \psi^\dagger) *_A \psi$, 
that is ${}^{I}{\cal A}_a(x) $ and another $\psi$. The boson fields may  
stay in the vacuum without energy, or appear as well as a propagating 
field.
This will be very important when looking at the Feynman diagrams.
\end{small}
}.

In general a boson field can be written as an expansion over Clifford
algebra elements,  Eq.~(\ref{wholespacebosons}),
\begin{equation} % 
\label{wholespacebosonsC}
{\cal A}_a(x)=\sum_k A_a^{(k)}(x)\,\Gamma_k ,
\end{equation} % 
where $\Gamma_k$ denote elements of the Clifford algebra built from
nilpotents and projectors. 
A boson field is a tensor product, $\,*_T\,$, of a field in the ordinary 
space-time in the coordinate representation and of one of the 
two kinds of Clifford even  ``basis vectors'': ${\cal C}^{ m}{}_{f a} (x)$ 
$\,*_T\,{\hat{\cal A}}^{m \dagger}_{f} $, Eq.~(\ref{wholespacebosons}).
Here $\Gamma_k $ replaces ${\hat{\cal A}}^{m \dagger}_{f}$ and
$A_a^{(k)}(x)$ replaces ${\cal C}^{ m}{}_{f a} (x)$.

 The space-time index $a$ determines whether
the field behaves as a vector or scalar in space-time, while the
Clifford algebra structure determines the physical properties such as
spin and charges.\\

\subsection{Bosonic Lagrange density} % Correct
\label{bosonlagrange}

Since the algebra distinguishes two kinds of non interacting bosons,
\[
{}^{I}{\cal A}_a ,
\qquad
{}^{II}{\cal A}_a ,
\]
the bosonic Lagrange density separates naturally into two parts,
\[
{\cal L}_B = {\cal L}_B^{I} + {\cal L}_B^{II} .
\]
Taking into account Eqs.~(\ref{ruleAAI},~\ref{covderC}), and requiring
that each kind contains the kinetic term for the corresponding boson
fields, and the self interacting term following 
Eqs.~(\ref{ruleAAI},~\ref{covderC}), we end up with the field strengths
\begin{eqnarray} 
\label{Fab}
\,{}^{I}F_{ab} &=&
-i\, (p_a\,{}^{I}{\cal A}_b - 
p_b\,{}^{I}{\cal A}_a -
\{{}^{I}{\cal A}_a,{}^{I}{\cal A}_b\}_{-}) ,
\nonumber\\ 
\,{}^{II}F_{ab} &=&
-i (p_a\,{}^{II}{\cal A}_b -
p_b\,{}^{II}{\cal A}_a -
\{{}^{II}{\cal A}_a,{}^{II}{\cal A}_b\}_{-} ,
\end{eqnarray} 
and correspondingly with the bosonic Lagrange density 
\begin{eqnarray} 
\label{bosLC}
{\cal L}_B =
\frac14\,{}^{I}F_{ab}\,{}^{I}F^{ab} + 
\frac14\,{}^{II}F_{ab}\,{}^{II}F^{ab} .
\end{eqnarray} %\]
Because the two bosonic sectors correspond to different Clifford
structures, orthogonal to each other, they do not interact in the 
bosonic part of the Lagrangian.

The interaction between the two sectors occurs only through fermions.
The fermion interaction term can schematically be, taken from Eq.~(\ref{ferLC})
into account, written as
\begin{eqnarray} 
\label{bosfermintLC} 
{\cal L}_{int} &=& \frac{1}{2}\,
\{(\psi^{\dagger}\,\gamma^0\gamma^a\,\,{}^{I}{\cal A}_a\,\psi
+
\psi^{\dagger}\,\gamma^0\gamma^a\,\psi \,\,{}^{II}{\cal A}_a +
\nonumber\\
& &[\,{}^{I}{\cal A}_a\,\psi\,+ \psi\,\,{}^{II}{\cal A}_a]^{\dagger}\,\gamma^0\gamma^a\,
\,\psi\}\,.
\end{eqnarray} 

In this way bosons of type I and type II meet only in the presence
of fermionic fields.

\subsection{Feynman Rules and interaction structure}
\label{feynmanrules}

The algebraic structure discussed above have important 
consequences for the interaction vertices and the corresponding 
Feynman rules.

The proposed description of the internal space of bosonic operators 
with the even number of nilpotents,  leads to the relations among 
fermion operators and boson operators
\begin{equation} 
\psi *_A \psi^\dagger \,,
\qquad
\psi^\dagger *_A \psi\, ,\nonumber
\end{equation} 
describing two kinds of bosons,
\begin{equation} % \[
{}^{I}{\cal A}_a \,,
\qquad
{}^{II}{\cal A}_a \,.\nonumber
\end{equation} % \]
%
% \end{document}
Because of the nilpotent relations
\begin{equation} 
% \label{bosfermLC} %\[
\psi *_A \psi =0\, ,
\qquad
\psi^\dagger *_A \psi^\dagger =0 \,,\nonumber
\end{equation} %\]
the two kinds of bosons do not interact directly. They can interact only 
through fermionic operators. Schematically one finds relations of the form
\[
{}^{I}{\cal A}_a *_A \psi
=
(\psi *_A \psi^\dagger) *_A \psi
=
\psi *_A (\psi^\dagger *_A \psi)
\sim
\psi\,\,*_A {}^{II}{\cal A}_a .
\]

The above implies that bosons of type I and type II can be connected only
through fermion lines. Correspondingly, it is the fermionic Lagrange density
which is responsible for the interaction among fermions and bosons.
\[
{\cal L}_{int}
=
\psi^{\dagger}\,\,*_A \gamma^0\,\gamma^a\,({}^{I}{\cal A}_a\,*_A\,\psi
+
\psi\,\,*_A \,\, {}^{II}{\cal A}) .
\]
%

% G down
\section{Conclusion}
\label{conclusion}

The proposed theory, built on the assumption that the internal spaces of
fermion and boson fields are describable by odd (for fermions) and even
(for bosons) products of operators $\gamma^{a}$, offers the unique 
description of spins and charges of fermion and boson second quantised 
fields,  as we can noticed in Tables~(\ref{Table so13+1.} for the 
observed quarks and leptons,~\ref{Table Clifffourplet.}~for a toy model), 
as well as in the covariant derivatives of the Lagrange densities  for 
fermion and boson fields in interaction, Eqs.~(\ref{ferLC},~\ref{covderLC},~\ref{bosLC},~
\ref{bosfermintLC}). % \\

The theory offers the explanation for the appearance of all the so far observed 
fermions (quarks and leptons and antiquarks and antileptons; as well as for 
their families) and bosons ($SO(3,1)$ gravitons~\cite{n2023NPB,n2024NPB}~\footnote{
Let us point out that the graviton in this theory is explained in an 
equivalent way as all the gauge fields observed so far.
}, 
two $SU(2)$ weak fields~\cite{nh2021RPPNP,n2023NPB,n2024NPB}, 
$SU(3)$ gluons~\cite{nh2021RPPNP,n2023NPB,n2024NPB}, $U(1)$ photons
and new predicted fermions (right-handed neutrinos and left-handed 
antineutrinos~(\cite{nh2021RPPNP} and references therein), the fourth 
family to the observed three~\cite{gmdn2008,gn2014}, a new 
group of four families of quarks and leptons, the stable one at low 
energies explaining the Dark matter~\cite{gn2009}).                     

Both fields, fermions and bosons, are assumed to be massless and appear
in a flat space-time. 

Without breaking symmetries, there would also exist boson fields
carrying more than one charge at the same time, like the weak and
colour charge, or the spin, weak charge and colour charge, which we have
not (yet) observed.

There is the break of symmetries which make 
fermion fields and some of the boson fields massive, as studied in 
Ref.~\cite{nh2021RPPNP}\footnote{
The breaks of symmetries were studied when the boson fields were
described by $S^{ab}\omega_{a b \alpha}$ and
$\tilde{S}^{ab}\tilde{\omega}_{a b \alpha}$,~(\cite{nh2021RPPNP},
Subsect. 6.2 and references therein), instead of by
${}^{I}{\hat{\cal A}}^{m \dagger}_{f \alpha}$ and
${}^{II}{\hat{\cal A}}^{m \dagger}_{f \alpha}$.
}.

The Lorentz transformations in internal space, presented in
Subsect.~\ref{internallorentzC}, manifest the gauge transformations of
other theories when we treat the internal space with $d\ge 5$. Our
description of the internal spaces of fermion and boson fields requires 
that all bosons  are treated equivalently, with gravitons (with 
the space index $a=(0,1,2,3)$) and scalars included (scalars have the 
space index $a\ge 5$ while they can have all the internal space 
properties as vectors and tensors).

The choice that the internal spaces of fermions and bosons are described
by superposition of odd and even products of operators $\gamma^{a}$, 
respectively, explains the Dirac's postulates of the second quantized 
fermion and boson fields.

We assume the Dirac-like action for fermion fields, but with the 
covariant derivative required by our description of the internal 
spaces of fermion and boson fields, presented in
Eqs.~(\ref{ferLC},~\ref{covderLC}).  
The fermion Lagrangian density, preserving its form under the local 
Lorentz transformations as shown in Subsect.~\ref{fermionlagrange}, 
explains the gauge invariance of the usual theories if our internal 
space concerns $d\ge 5$~\footnote{
There are gravitons with two nilpotents in $SO(3,1)$, the rest of 
projectors, Eq.~(\ref{grAqeqe}), suggesting that gravitons should be 
treated according to the usual gauge fields.   
}.   
There is no negative energy Dirac sea for fermions. Fermions have only
ordinary quantum vacuum. Fermions and anti-fermions appear 
in the same family. Fermions and bosons appear in the quantum vacuum.

The quadratic  Lagrange density for fermion fields, preserving its form 
under local Lorentz transformations, presented in Eq.~(\ref{quadLfermionC}), remains to be studied.

Also the boson  Lagrangian density preserves its form under the local Lorentz
transformations, as shown in Eq.~(\ref{AIAIItransLC}).

The proposed description of the internal spaces of fermion and boson
fields requires two kinds of mutually non-interacting boson fields,
Subsect.~\ref{structurebosons}, ${}^{I}{\cal A}_a$ and
${}^{II}{\cal A}_a$,  ${}^{I}{\cal A}_a\,*_A\, {}^{II}{\cal A}_a=0$.
Their Lorentz transformation properties are presented in
Eqs.~(\ref{Ltransbosons},~\ref{AIAIItransLC}).

The Lagrange density for bosons, generated as in usual theories, is
correspondingly, the sum of two independent Lagrange densities,
presented in Subsect.~\ref{bosonlagrange},
Eqs.~(\ref{Fab},~\ref{bosLC}), for ${}^{I}{\cal A}_a$ and
${}^{II}{\cal A}_a$, which interact only through fermions.

The boson fields either manifest the left
multiplication to fermions, ${}^{I}{\cal A}_a$, or the right one,
${}^{II}{\cal A}_a$, Eq.~\ref{covderLC}.

Their mutual interaction term in the  Lagrange densities, presented
in Eq.~(\ref{bosfermintLC}),
are determined by our description of internal spaces of fermion and 
boson fields, requiring two kinds of bosons.

% G up   

We arrange in any $d=2(2n +1)$ dimensional internal space, the fermion
and boson states to be eigenvectors of all the members of the Cartan 
subalgebra, Eq.~(\ref{cartangrasscliff}), we call these eigenstates the
``basis vectors''. The ``basis vectors'' for fermion fields have an odd
number of nilpotents, and for the boson fields, an even number of
nilpotents, the rest are projectors, Eq.(\ref{calsab}).

The fermion ``basis vectors'' appear % $d=(13+1)$,
in $2^{\frac{d}{2}-1}$ families, each family having $2^{\frac{d}{2}-1}$
members; and there are  $2^{\frac{d}{2}-1}\times$$ 2^{\frac{d}{2}-1}$ of
their Hermitian conjugate partners, appearing in a separate group.

The boson ``basis vectors'' appear in two orthogonal groups, each with
$2^{\frac{d}{2}-1}\times$$ 2^{\frac{d}{2}-1}$ members and have their
Hermitian conjugate partners within the same group.

The ``basis vectors'' for bosons are expressible as the algebraic products
of fermion ``basis vectors'' and their Hermitian conjugate partners,
Eqs.~(\ref{AIbbdagger}, \ref{AIIbdaggerb}).

The second quantised fermion fields are tensor products of the ``basis
vectors'' and basis in ordinary space time, Eq.~(\ref{wholespacefermions}).

The second quantised boson fields are tensor products of the ``basis
vectors'' and basis in ordinary space time, and carry the space-time index,  Eq.~(\ref{wholespacebosons}). \\

The properties of fermion and boson fields, presented in this contributions,
like:\\
{\bf .}  the algebraic product of two fermions is zero
\[
\psi *_A \psi =0 ,
\qquad
\psi^\dagger *_A \psi^\dagger =0 ,
\]
{\bf .} the algebra distinguishes two types of bosons, Eqs.~(\ref{AIbbdagger},~\ref{AIIbdaggerb}),
\[
{}^{I}{\cal A}_a \sim \psi *_A \psi^\dagger ,
\qquad
{}^{II}{\cal A}_a \sim \psi^\dagger  *_A \psi ,
\]
{\bf .} bosons can interact only in the presence of a fermion
\begin{equation}
\label{Apsi}
{}^{I}{\cal A}_a *_A \psi
=
(\psi *_A \psi^\dagger) *_A \psi
=
\psi *_A (\psi^\dagger *_A \psi)
\sim
\psi\,*_A\,\,{}^{II}{\cal A}_a , \nonumber
\end{equation} 
influence the Feynman diagrams.
The Feynman rules are studied in Ref.~\cite{nh2025Bled}.\\

Demonstrating so many simple and elegant descriptions of the
second quantised fields, explaining the assumptions of other
theories, makes us hope that the theory might be what the
universe obeys.\\

\subsection{What should we understand }
\label{problems}
If our proposed description of the internal degrees of freedom of the 
second quantised fermion and boson fields is what nature uses in the 
case that the space-time is flat, and all the second quantised fields 
are massless, while all the fields are active only in $d=(3+1)$ in 
ordinary space-time, we are able not only to explain the {\it standard 
model} assumptions before the electroweak break, but also find the 
solution for all the open questions connected with elementary fermion 
and boson fields.

There are two kinds of boson fields in this theory. One kind describes
the observed gauge fields and the graviton field on an equal level,
unifying all the boson fields. This kind, causing transformations
within members of each family of fermions and antifermions,
predicts the existence of the second kind of the weak force and
correspondingly, the existence of the right-handed neutrino and the
left-handed anti-neutrino, as seen in Table~\ref{Table so13+1.}.

The second kind of boson fields transforms a family member of one
family to the same family member of all the families, offering the
explanation for the appearance of the Higgs scalars and predicting new
scalar fields~\cite{n2014matterantimatter}.

The internal spaces of all the observed boson fields are describable with
``basis vectors'' having only two nilpotents (gravitons in $SO(3,1)$ part of
$SO(13,1)$, two kinds of weak bosons in  $SO(4)$ part of $SO(13,1)$,
gluons in $SU(3)$ part of  $SO(6)$, part of $SO(13,1)$), and the rest of
projectors, photons with only projectors in all $SO(13,1)$.

There are, however, many more boson fields. Their ``basis vectors'' have
more than two nilpotents (they can have four or six nilpotents), carrying
correspondingly, more than one kind of the so-far-observed charges.
\\
% **********
There are many more families in this theory than the observed three.
The theory predicts that the three observed families are
the members of the group of four families~\cite{gn2014}. %\cite{gn2014Bled}.
The theory
predicts also the second group of four families, contributing to the dark
matter~\cite{gn2009}.% nm2015}.

To be able to explain why ``nature has decided'' to break symmetries,
we should know the properties this theory has with respect to: \\
a. \,\, The renormalisability and anomalies in even and odd dimensional
spaces.\\
b. \,\,  How does the second kind of bosons contribute to the breaking 
of symmetries, while the first kind of the boson ``basis vectors'' seems 
to mainly determine the properties of all the observed boson fields, with 
the gravity included. \\
c.  \,\, The differences in odd, $d=(2n+1)$, and even, $d=(2(2n+1))$,
dimensional spaces. While in even dimensional spaces, $d=2(2n+1)$,
the odd ``basis vectors'' anticommute and have their Hermitian
conjugated ``basis vectors'' in a separate group, and the even ``basis
vectors'' commute and appear in two orthogonal groups, have the ``basis
vectors'' in $d=2(2n+1)+1$ strange properties; half of the odd and even
``basis vectors'' behave like in $d=2(2n+1)$, in the second half, the 
anticommutng odd ``basis vectors'' appear in two orthogonal groups, 
while the commuting even ``basis vectors'' appear in families and have the Hermitian conjugate partners in a separate group~\cite{n2023MDPI}.\\
d. \,\, The differences in even dimensional internal spaces, when
$d=2(2n+1)$ and $d=4n$. While in $d=2(2n+1)$ the ``basis vectors''
for fermions and antifermions appear in the same family, in $d=4n$
the ``basis vectors'' of a family do not include antifermions.
Correspondingly, the vacuum in $d=2(2n+1)$ is just the quantum
vacuum, while in $d=4n$ the Dirac sea with the negative energies must
be invented.\\ 
%****02.04.2026\\
e.  \,\, How to present and interpret the Feynman diagrams in this theory
in comparison with the Feynman diagrams so far presented and 
interpreted~\cite{nh2025Bled}.\\ 
f. \,\, It might be needed to extend the second quantised fermion and
boson fields to strings, with the first step already done in Ref.~\cite{nh2025Bled}.\\

\appendix

\section{One family representation of odd ``basis vectors'' in $d=(13+1)$
}
\label{13+1representation}  

This appendix, is following similar appendices in Refs.~\cite{nh2021RPPNP,%
n2023MDPI,n2023NPB}.

One irreducible representation, one family, of the odd ``basis vectors'' 
describing the internal spaces of fermions in $d=(13+1)$, analysed  with 
respect to the subgroups $SO(3,1)\times SU(2) \times SU(2) \times 
SU(3)\times U(1)$, is presented. One family contains the ``basis vectors'' 
of quarks and leptons and antiquarks and antileptons, with the quantum 
numbers assumed by the {\it standard model} before the electroweak 
break, with the right-handed neutrinos and left-handed antineutrinos included. 
There are two $SU(2)$ subgroups, $SU(2)_I$ and $SU(2)_{II}$, of $SO(4)$
with the hypercharge of the {\it standard model} $Y=\tau^{23} + \tau^4$, 
Eqs.~(\ref{so42} - \ref{YQY'Q'andtilde}). The reader can see that the ``basis 
vectors'' of quarks distinguish from the ``basis vectors'' of leptons, and the 
``basis vectors'' of antiquarks distinguish from the ``basis vectors'' of antileptons only in the part $SU(3)\times U(1)$ of  $SO(13,1)$.

The generators $S^{ab}$ of the Lorentz transformations in the internal 
space of fermions with $d=(13+ 1)$, analysed with respect to the 
subgroups $SO(3,1)\times SU(2) \times SU(2) \times SU(3)\times U(1)$,
are presented as
\begin{small}
\begin{eqnarray}
\label{so1+3}
&&\vec{N}_{\pm}(= \vec{N}_{(L,R)}): = \,\frac{1}{2} (S^{23}\pm i S^{01},
S^{31}\pm i S^{02}, S^{12}\pm i S^{03} )\,,%\nonumber\\
% \nonumber\\ &&N^{\pm}_{+}      = N^{1}_{+} \pm i \,N^{2}_{+} = 
%  - \stackrel{03}{(\mp i)} \stackrel{12}{(\pm )}\,, \quad N^{\pm}_{-}= 
% N^{1}_{-} \pm  i\,N^{2}_{-} =    \stackrel{03}{(\pm i)} \stackrel{12}{(\pm )}\,.
%
% &&\vec{\tilde{N}}_{\pm}(=\vec{\tilde{N}}_{(L,R)}): =
% \,\frac{1}{2} (\tilde{S}^{23}\pm i \tilde{S}^{01}\,,
% \tilde{S}^{31}\pm i \tilde{S}^{02}, \tilde{S}^{12}\pm i \tilde{S}^{03} )\,,
\end{eqnarray}
 \begin{eqnarray}
 \label{so42}
 \vec{\tau}^{1}:&=&\frac{1}{2} (S^{58}-  S^{67}, \,S^{57} + S^{68}, \,S^{56}-  S^{78} )\,,%\nonumber\\
 \quad
 \vec{\tau}^{2}:= \frac{1}{2} (S^{58}+  S^{67}, \,S^{57} - S^{68}, \,S^{56}+  S^{78} )\,,
%  \nonumber\\
% \tau^{1\pm}         &=& (\mp)\, \stackrel{56}{(\pm )} \stackrel{78}{(\mp )} \,, \qquad   
% \qquad \qquad  \qquad \qquad \;\;
% \tau^{2\mp}=            (\mp)\, \stackrel{56}{(\mp )} \stackrel{78}{(\mp )} \,,
% \vec{\tilde{\tau}}^{1}:&=&\frac{1}{2} (\tilde{S}^{58}-  \tilde{S}^{67}, \,\tilde{S}^{57} + 
% \tilde{S}^{68}, \,\tilde{S}^{56}-  \tilde{S}^{78} )\,, \quad 
% \vec{\tilde{\tau}}^{2}:=\frac{1}{2} (\tilde{S}^{58}+  \tilde{S}^{67}, \,\tilde{S}^{57} - 
% \tilde{S}^{68}, \,\tilde{S}^{56}+  \tilde{S}^{78} ),\,\,\;\;
 \end{eqnarray}
%
% and the generators of $SU(3)$ and $U(1)$ subgroups of $SO(6)$ $\subset SO(13,1)$, 
% describing  the colour charge and the ''fermion'' charge  of fermions as well as the corresponding 
% family quantum number $\tilde{\tau}^4$
% in the {\it spin-charge-family } theory
%
 \begin{eqnarray}
 \label{so64}
 \vec{\tau}^{3}: = &&\frac{1}{2} \,\{  S^{9\;12} - S^{10\;11} \,,
  S^{9\;11} + S^{10\;12} ,\, S^{9\;10} - S^{11\;12}\, ,  %\nonumber\\
  S^{9\;14} -  S^{10\;13} ,\,  \nonumber\\
  && S^{9\;13} + S^{10\;14} \,,  S^{11\;14} -  S^{12\;13}\,, %\nonumber\\
  S^{11\;13} +  S^{12\;14} ,\,  \frac{1}{\sqrt{3}} ( S^{9\;10} + S^{11\;12} - 
 2 S^{13\;14})\}\,,\nonumber\\
 \tau^{4}: = &&-\frac{1}{3}(S^{9\;10} + S^{11\;12} + S^{13\;14})\,,\;\;%\nonumber\\
%  \tilde{\tau}^{4}: = &&-\frac{1}{3}(\tilde{S}^{9\;10} + \tilde{S}^{11\;12} + \tilde{S}^{13\;14})\,,
 \end{eqnarray}
 \begin{eqnarray}
 \label{YQY'Q'andtilde}
 Y:= \tau^{4} + \tau^{23}\,,\;\; Q: =  \tau^{13} + Y\,,\;\; 
 % Y':= -\tau^{4}\tan^2\vartheta_2 + \tau^{23}\,, %\;\;
% \;\; Q':= -Y \tan^2\vartheta_1 + \tau^{13} \,,
&&%\nonumber\\
%  \tilde{Y}:= \tilde{\tau}^{4} + \tilde{\tau}^{23}\,,\,\;\tilde{Q}:= 
%  \tilde{Y} + \tilde{\tau}^{13}\,,\;\;
%   \tilde{Y'}:= -\tilde{\tau}^{4} 
%  \tan^2 \vartheta_2 + \tilde{\tau}^{23}\,,\;
%  \;\; \tilde{Q'}= -\tilde{Y} \tan^2 \vartheta_1 
%  + \tilde{\tau}^{13}\,. &&\,
 % \,,%\nonumber\\
%%  %{\cal Y}:&=& {\cal \tau}^{4} + {\cal \tau}^{23}\,,\;\; {\cal Y'}:=
%%% -{\cal \tau}^{4}\tan^2 \vartheta_2 + 
 %% %{\cal \tau}^{23}\,,\;\;
 %% %{\cal Q}: =  {\cal \tau}^{13} + {\cal Y}\,,\;\; {\cal Q'}:=
 %%% -{\cal Y} \tan^2 \vartheta_1 + %\cal \tau}^{13}\,
 %% %\nonumber\\
  \end{eqnarray}
 \end{small}
The (chosen) Cartan subalgebra operators, determining the commuting 
operators in the above equations, is presented in Eq.~(\ref{cartangrasscliff}). \\

The corresponding relations for $\tilde{S}^{ab}$, determining the family 
quantum numbers, follow if we replace in above equations $S^{ab}$ by 
$\tilde{S}^{ab}$. \\

The  hypercharge $Y$ and the electromagnetic charge $Q$ relate to the {\it 
standard model} quantum numbers. \\

 For fermions, the operator of handedness $\Gamma^d$ is determined as follows:
  \begin{small}
\begin{eqnarray}
\label{Gamma}
%\begin{displaymath}
 \Gamma^{(d)}= \prod_a (\sqrt{\eta^{aa}} \gamma^a)  \cdot \left \{ \begin{array}{l l}
 (i)^{\frac{d}{2}} \,, &\rm{ for\, d \,even}\,,\\
 (i)^{\frac{d-1}{2}}\,,&\rm{for \, d \,odd}\,.
  \end{array} \right.
% \end{displaymath}
 \end{eqnarray}
 \end{small}

All the families (all the irreducible representations) follow from this one
by applying, let say, on the first member, $u_R^{c1}$, all possible 
$\tilde{S}^{ab}$, Eq.~(\ref{graficfollow1}). 
Let us start with $\tilde{S}^{01}$ which transforms
$ u_{R, f=1}^{c1}$$(\equiv \stackrel{03}{(+i)}\,\stackrel{12}{[+]}|
\stackrel{56}{[+]}\,\stackrel{78}{(+)}||\stackrel{9 \;10}{(+)}\;
\stackrel{11\;12}{[-]}\;\;\stackrel{13\;14}{[-]})$ of this first family 
to $ u_{R, f=2}^{c1}$$(\equiv \stackrel{03}{[+i]}\,\stackrel{12}{(+)}|
\stackrel{56}{[+]}\,\stackrel{78}{(+)}||\stackrel{9 \;10}{(+)}\;
\stackrel{11\;12}{[-]}\;\;\stackrel{13\;14}{[-]})$. From the first family 
member of the second family all the members of the second family follow
by the application of $S^{ab}$. There are obviously, the same number 
of families as there is the number of the family members.\\

% G
%
The even ``basis vectors'', analysed with respect to the same subgroups,
($SO(3,1)\times SU(2) \times SU(2) \times SU(3)\times U(1)$) of the 
$SO(13,1)$ group, offer the description of the internal spaces of the 
corresponding tensor, vector and scalar gauge fields, appearing in the 
{\it standard model} before the electroweak break~\cite{n2021SQ,nh2025Bled,%
n2023DOI}; 
as explained in Sect.~\ref{5+1and13+1}. There are breaks of symmetries
which make the very limited number of families observed at observable 
energies.

The even ``basis vectors'' are expressible as products of the odd
``basis vectors'' and their Hermitian conjugate partners, as presented
in Eqs.~(\ref{AIbbdagger}, \ref{AIIbdaggerb}).

\bottomcaption{\label{Table so13+1.}%
%\tiny{
%\begin{tiny}
\begin{small}
The left-handed ($\Gamma^{(13,1)} = -1$, Eq.~(\ref{Gamma})) irreducible representation representing the ``basis vectors'' of one family of spinors --- 
the product of the odd number of nilpotents and of projectors, the eigenvectors 
of all the Cartan subalgebra members of the Lorentz algebra of the group 
$SO(13,1)$~\cite{n2014matterantimatter,nh02}, manifesting the
colour charged quarks and antiquarks and the colourless leptons and antileptons ---
is presented.
It contains the left-handed ($\Gamma^{(3,1)}=-1$) weak ($SU(2)_{I}$) charged
($\tau^{13}=\pm \frac{1}{2}$), %Eq.~(\ref{so42})),
and $SU(2)_{II}$ chargeless ($\tau^{23}=0$) %, %Eq.~(\ref{so42}))
quarks and leptons, and the right-handed
($\Gamma^{(3,1)}=1$) weak ($SU(2)_{I}$) chargeless and $SU(2)_{II}$ charged
($\tau^{23}=\pm \frac{1}{2}$) quarks and leptons, both with the spin $ S^{12}$ up
and down ($\pm \frac{1}{2}$, respectively).
Quarks distinguish from leptons only in the $SU(3) \times U(1)$ part: Quarks are triplets
of three colours ($c^i$ $= (\tau^{33}, \tau^{38})$ $ = [(\frac{1}{2},\frac{1}{2\sqrt{3}}),
(-\frac{1}{2},\frac{1}{2\sqrt{3}}), (0,-\frac{1}{\sqrt{3}}) $, %Eq.~(\ref{so64})),
carrying the "fermion charge" ($\tau^{4}=\frac{1}{6}$). %, Eq.~(\ref{so64})).
The colourless leptons carry the "fermion charge" ($\tau^{4}=-\frac{1}{2}$).
The same multiplet - family - contains also the left-handed weak ($SU(2)_{I}$) chargeless and $SU(2)_{II}$ charged antiquarks and antileptons and the 
right-handed weak
($SU(2)_{I}$) charged and $SU(2)_{II}$ chargeless antiquarks and antileptons.
Antiquarks distinguish from antileptons again only in the $SU(3) \times U(1)$ part:
Antiquarks are anti-triplets carrying the "fermion charge" ($\tau^{4}=-\frac{1}{6}$).
The anti-colourless antileptons carry the "fermion charge" ($\tau^{4}=\frac{1}{2}$).
$Y=(\tau^{23} + \tau^{4})$ is the hyper charge, the electromagnetic charge
is $Q=(\tau^{13} + Y$). One can calculate, taking into account 
Eq.~(\ref{graficfollow1}), also the family quantum numbers
of the presented family: 
$\tilde{S}^{03}=\frac{i}{2}, \tilde{S}^{12}=-\frac{1}{2}, \tilde{S}^{56}=
-\frac{1}{2}, \tilde{S}^{7 8}=\frac{1}{2}, \tilde{S}^{9\,10}=\frac{1}{2},
\tilde{S}^{11\,12}=\frac{1}{2}, \tilde{S}^{13\,14}=\frac{1}{2}$.
%
%The vacuum state, on which the nilpotents and projectors operate, is presented in 
\end{small}
%\end{tiny}
}

%\end{document}
%*}
%***************************************
\tablehead{\hline
i&$$&$|^a\psi_i>$&$\Gamma^{(3,1)}$&$ S^{12}$&
$\tau^{13}$&$\tau^{23}$&$\tau^{33}$&$\tau^{38}$&$\tau^{4}$&$Y$&$Q$\\
\hline
&& ${\rm (Anti)octet},\,\Gamma^{(7,1)} = (-1)\,1\,, \,\Gamma^{(6)} = (1)\,-1$&&&&&&&&& \\
&& ${\rm of \;(anti) quarks \;and \;(anti)leptons}$&&&&&&&&&\\
\hline\hline}
\tabletail{\hline \multicolumn{12}{r}{\emph{Continued on next page}}\\}
\tablelasttail{\hline}
%\begin{table}
%\begin{small}
% **
\begin{tiny}
%\begin{center}
%*
%\tiny{
\begin{supertabular}{|r|c||c||c|c||c|c||c|c|c||r|r|}
1&$ u_{R}^{c1}$&$ \stackrel{03}{(+i)}\,\stackrel{12}{[+]}|
\stackrel{56}{[+]}\,\stackrel{78}{(+)}
||\stackrel{9 \;10}{(+)}\;\;\stackrel{11\;12}{[-]}\;\;\stackrel{13\;14}{[-]} $ &1&$\frac{1}{2}$&0&
$\frac{1}{2}$&$\frac{1}{2}$&$\frac{1}{2\,\sqrt{3}}$&$\frac{1}{6}$&$\frac{2}{3}$&$\frac{2}{3}$\\
\hline
2&$u_{R}^{c1}$&$\stackrel{03}{[-i]}\,\stackrel{12}{(-)}|\stackrel{56}{[+]}\,\stackrel{78}{(+)}
||\stackrel{9 \;10}{(+)}\;\;\stackrel{11\;12}{[-]}\;\;\stackrel{13\;14}{[-]}$&1&$-\frac{1}{2}$&0&
$\frac{1}{2}$&$\frac{1}{2}$&$\frac{1}{2\,\sqrt{3}}$&$\frac{1}{6}$&$\frac{2}{3}$&$\frac{2}{3}$\\
\hline
3&$d_{R}^{c1}$&$\stackrel{03}{(+i)}\,\stackrel{12}{[+]}|\stackrel{56}{(-)}\,\stackrel{78}{[-]}
||\stackrel{9 \;10}{(+)}\;\;\stackrel{11\;12}{[-]}\;\;\stackrel{13\;14}{[-]}$&1&$\frac{1}{2}$&0&
$-\frac{1}{2}$&$\frac{1}{2}$&$\frac{1}{2\,\sqrt{3}}$&$\frac{1}{6}$&$-\frac{1}{3}$&$-\frac{1}{3}$\\
\hline
4&$ d_{R}^{c1} $&$\stackrel{03}{[-i]}\,\stackrel{12}{(-)}|
\stackrel{56}{(-)}\,\stackrel{78}{[-]}
||\stackrel{9 \;10}{(+)}\;\;\stackrel{11\;12}{[-]}\;\;\stackrel{13\;14}{[-]} $&1&$-\frac{1}{2}$&0&
$-\frac{1}{2}$&$\frac{1}{2}$&$\frac{1}{2\,\sqrt{3}}$&$\frac{1}{6}$&$-\frac{1}{3}$&$-\frac{1}{3}$\\
\hline
5&$d_{L}^{c1}$&$\stackrel{03}{[-i]}\,\stackrel{12}{[+]}|\stackrel{56}{(-)}\,\stackrel{78}{(+)}
||\stackrel{9 \;10}{(+)}\;\;\stackrel{11\;12}{[-]}\;\;\stackrel{13\;14}{[-]}$&-1&$\frac{1}{2}$&
$-\frac{1}{2}$&0&$\frac{1}{2}$&$\frac{1}{2\,\sqrt{3}}$&$\frac{1}{6}$&$\frac{1}{6}$&$-\frac{1}{3}$\\
\hline
6&$d_{L}^{c1} $&$ - \stackrel{03}{(+i)}\,\stackrel{12}{(-)}|\stackrel{56}{(-)}\,\stackrel{78}{(+)}
||\stackrel{9 \;10}{(+)}\;\;\stackrel{11\;12}{[-]}\;\;\stackrel{13\;14}{[-]} $&-1&$-\frac{1}{2}$&
$-\frac{1}{2}$&0&$\frac{1}{2}$&$\frac{1}{2\,\sqrt{3}}$&$\frac{1}{6}$&$\frac{1}{6}$&$-\frac{1}{3}$\\
\hline
7&$ u_{L}^{c1}$&$ - \stackrel{03}{[-i]}\,\stackrel{12}{[+]}|\stackrel{56}{[+]}\,\stackrel{78}{[-]}
||\stackrel{9 \;10}{(+)}\;\;\stackrel{11\;12}{[-]}\;\;\stackrel{13\;14}{[-]}$ &-1&$\frac{1}{2}$&
$\frac{1}{2}$&0 &$\frac{1}{2}$&$\frac{1}{2\,\sqrt{3}}$&$\frac{1}{6}$&$\frac{1}{6}$&$\frac{2}{3}$\\
\hline
8&$u_{L}^{c1}$&$\stackrel{03}{(+i)}\,\stackrel{12}{(-)}|\stackrel{56}{[+]}\,\stackrel{78}{[-]}
||\stackrel{9 \;10}{(+)}\;\;\stackrel{11\;12}{[-]}\;\;\stackrel{13\;14}{[-]}$&-1&$-\frac{1}{2}$&
$\frac{1}{2}$&0&$\frac{1}{2}$&$\frac{1}{2\,\sqrt{3}}$&$\frac{1}{6}$&$\frac{1}{6}$&$\frac{2}{3}$\\
\hline\hline
\shrinkheight{0.25\textheight}
9&$ u_{R}^{c2}$&$ \stackrel{03}{(+i)}\,\stackrel{12}{[+]}|
\stackrel{56}{[+]}\,\stackrel{78}{(+)}
||\stackrel{9 \;10}{[-]}\;\;\stackrel{11\;12}{(+)}\;\;\stackrel{13\;14}{[-]} $ &1&$\frac{1}{2}$&0&
$\frac{1}{2}$&$-\frac{1}{2}$&$\frac{1}{2\,\sqrt{3}}$&$\frac{1}{6}$&$\frac{2}{3}$&$\frac{2}{3}$\\
\hline
10&$u_{R}^{c2}$&$\stackrel{03}{[-i]}\,\stackrel{12}{(-)}|\stackrel{56}{[+]}\,\stackrel{78}{(+)}
||\stackrel{9 \;10}{[-]}\;\;\stackrel{11\;12}{(+)}\;\;\stackrel{13\;14}{[-]}$&1&$-\frac{1}{2}$&0&
$\frac{1}{2}$&$-\frac{1}{2}$&$\frac{1}{2\,\sqrt{3}}$&$\frac{1}{6}$&$\frac{2}{3}$&$\frac{2}{3}$\\
\hline
%$\cdots$&&&&&&&&&&&\\
%\hline
11&$d_{R}^{c2}$&$\stackrel{03}{(+i)}\,\stackrel{12}{[+]}|\stackrel{56}{(-)}\,\stackrel{78}{[-]}
||\stackrel{9 \;10}{[-]}\;\;\stackrel{11\;12}{(+)}\;\;\stackrel{13\;14}{[-]}$
&1&$\frac{1}{2}$&0&
$-\frac{1}{2}$&$ - \frac{1}{2}$&$\frac{1}{2\,\sqrt{3}}$&$\frac{1}{6}$&$-\frac{1}{3}$&$-\frac{1}{3}$\\
\hline
12&$ d_{R}^{c2} $&$\stackrel{03}{[-i]}\,\stackrel{12}{(-)}|
\stackrel{56}{(-)}\,\stackrel{78}{[-]}
||\stackrel{9 \;10}{[-]}\;\;\stackrel{11\;12}{(+)}\;\;\stackrel{13\;14}{[-]} $
&1&$-\frac{1}{2}$&0&
$-\frac{1}{2}$&$-\frac{1}{2}$&$\frac{1}{2\,\sqrt{3}}$&$\frac{1}{6}$&$-\frac{1}{3}$&$-\frac{1}{3}$\\
\hline
13&$d_{L}^{c2}$&$\stackrel{03}{[-i]}\,\stackrel{12}{[+]}|\stackrel{56}{(-)}\,\stackrel{78}{(+)}
||\stackrel{9 \;10}{[-]}\;\;\stackrel{11\;12}{(+)}\;\;\stackrel{13\;14}{[-]}$
&-1&$\frac{1}{2}$&
$-\frac{1}{2}$&0&$-\frac{1}{2}$&$\frac{1}{2\,\sqrt{3}}$&$\frac{1}{6}$&$\frac{1}{6}$&$-\frac{1}{3}$\\
\hline
14&$d_{L}^{c2} $&$ - \stackrel{03}{(+i)}\,\stackrel{12}{(-)}|\stackrel{56}{(-)}\,\stackrel{78}{(+)}
||\stackrel{9 \;10}{[-]}\;\;\stackrel{11\;12}{(+)}\;\;\stackrel{13\;14}{[-]} $&-1&$-\frac{1}{2}$&
$-\frac{1}{2}$&0&$-\frac{1}{2}$&$\frac{1}{2\,\sqrt{3}}$&$\frac{1}{6}$&$\frac{1}{6}$&$-\frac{1}{3}$\\
\hline
15&$ u_{L}^{c2}$&$ - \stackrel{03}{[-i]}\,\stackrel{12}{[+]}|\stackrel{56}{[+]}\,\stackrel{78}{[-]}
||\stackrel{9 \;10}{[-]}\;\;\stackrel{11\;12}{(+)}\;\;\stackrel{13\;14}{[-]}$ &-1&$\frac{1}{2}$&
$\frac{1}{2}$&0 &$-\frac{1}{2}$&$\frac{1}{2\,\sqrt{3}}$&$\frac{1}{6}$&$\frac{1}{6}$&$\frac{2}{3}$\\
\hline
16&$u_{L}^{c2}$&$\stackrel{03}{(+i)}\,\stackrel{12}{(-)}|\stackrel{56}{[+]}\,\stackrel{78}{[-]}
||\stackrel{9 \;10}{[-]}\;\;\stackrel{11\;12}{(+)}\;\;\stackrel{13\;14}{[-]}$&-1&$-\frac{1}{2}$&
$\frac{1}{2}$&0&$-\frac{1}{2}$&$\frac{1}{2\,\sqrt{3}}$&$\frac{1}{6}$&$\frac{1}{6}$&$\frac{2}{3}$\\
\hline\hline
17&$ u_{R}^{c3}$&$ \stackrel{03}{(+i)}\,\stackrel{12}{[+]}|
\stackrel{56}{[+]}\,\stackrel{78}{(+)}
||\stackrel{9 \;10}{[-]}\;\;\stackrel{11\;12}{[-]}\;\;\stackrel{13\;14}{(+)} $ &1&$\frac{1}{2}$&0&
$\frac{1}{2}$&$0$&$-\frac{1}{\sqrt{3}}$&$\frac{1}{6}$&$\frac{2}{3}$&$\frac{2}{3}$\\
\hline
18&$u_{R}^{c3}$&$\stackrel{03}{[-i]}\,\stackrel{12}{(-)}|\stackrel{56}{[+]}\,\stackrel{78}{(+)}
||\stackrel{9 \;10}{[-]}\;\;\stackrel{11\;12}{[-]}\;\;\stackrel{13\;14}{(+)}$&1&$-\frac{1}{2}$&0&
$\frac{1}{2}$&$0$&$-\frac{1}{\sqrt{3}}$&$\frac{1}{6}$&$\frac{2}{3}$&$\frac{2}{3}$\\
\hline
%$\cdots$&&&&&&&&&&\\
19&$d_{R}^{c3}$&$\stackrel{03}{(+i)}\,\stackrel{12}{[+]}|\stackrel{56}{(-)}\,\stackrel{78}{[-]}
||\stackrel{9 \;10}{[-]}\;\;\stackrel{11\;12}{[-]}\;\;\stackrel{13\;14}{(+)}$&1&$\frac{1}{2}$&0&
$-\frac{1}{2}$&$0$&$-\frac{1}{\sqrt{3}}$&$\frac{1}{6}$&$-\frac{1}{3}$&$-\frac{1}{3}$\\
\hline
20&$ d_{R}^{c3} $&$\stackrel{03}{[-i]}\,\stackrel{12}{(-)}|
\stackrel{56}{(-)}\,\stackrel{78}{[-]}
||\stackrel{9 \;10}{[-]}\;\;\stackrel{11\;12}{[-]}\;\;\stackrel{13\;14}{(+)} $&1&$-\frac{1}{2}$&0&
$-\frac{1}{2}$&$0$&$-\frac{1}{\sqrt{3}}$&$\frac{1}{6}$&$-\frac{1}{3}$&$-\frac{1}{3}$\\
\hline
21&$d_{L}^{c3}$&$\stackrel{03}{[-i]}\,\stackrel{12}{[+]}|\stackrel{56}{(-)}\,\stackrel{78}{(+)}
||\stackrel{9 \;10}{[-]}\;\;\stackrel{11\;12}{[-]}\;\;\stackrel{13\;14}{(+)}$&-1&$\frac{1}{2}$&
$-\frac{1}{2}$&0&$0$&$-\frac{1}{\sqrt{3}}$&$\frac{1}{6}$&$\frac{1}{6}$&$-\frac{1}{3}$\\
\hline
22&$d_{L}^{c3} $&$ - \stackrel{03}{(+i)}\,\stackrel{12}{(-)}|\stackrel{56}{(-)}\,\stackrel{78}{(+)}
||\stackrel{9 \;10}{[-]}\;\;\stackrel{11\;12}{[-]}\;\;\stackrel{13\;14}{(+)} $&-1&$-\frac{1}{2}$&
$-\frac{1}{2}$&0&$0$&$-\frac{1}{\sqrt{3}}$&$\frac{1}{6}$&$\frac{1}{6}$&$-\frac{1}{3}$\\
\hline
23&$ u_{L}^{c3}$&$ - \stackrel{03}{[-i]}\,\stackrel{12}{[+]}|\stackrel{56}{[+]}\,\stackrel{78}{[-]}
||\stackrel{9 \;10}{[-]}\;\;\stackrel{11\;12}{[-]}\;\;\stackrel{13\;14}{(+)}$ &-1&$\frac{1}{2}$&
$\frac{1}{2}$&0 &$0$&$-\frac{1}{\sqrt{3}}$&$\frac{1}{6}$&$\frac{1}{6}$&$\frac{2}{3}$\\
\hline
24&$u_{L}^{c3}$&$\stackrel{03}{(+i)}\,\stackrel{12}{(-)}|\stackrel{56}{[+]}\,\stackrel{78}{[-]}
||\stackrel{9 \;10}{[-]}\;\;\stackrel{11\;12}{[-]}\;\;\stackrel{13\;14}{(+)}$&-1&$-\frac{1}{2}$&
$\frac{1}{2}$&0&$0$&$-\frac{1}{\sqrt{3}}$&$\frac{1}{6}$&$\frac{1}{6}$&$\frac{2}{3}$\\
\hline\hline
25&$ \nu_{R}$&$ \stackrel{03}{(+i)}\,\stackrel{12}{[+]}|
\stackrel{56}{[+]}\,\stackrel{78}{(+)}
||\stackrel{9 \;10}{(+)}\;\;\stackrel{11\;12}{(+)}\;\;\stackrel{13\;14}{(+)} $ &1&$\frac{1}{2}$&0&
$\frac{1}{2}$&$0$&$0$&$-\frac{1}{2}$&$0$&$0$\\
\hline
26&$\nu_{R}$&$\stackrel{03}{[-i]}\,\stackrel{12}{(-)}|\stackrel{56}{[+]}\,\stackrel{78}{(+)}
||\stackrel{9 \;10}{(+)}\;\;\stackrel{11\;12}{(+)}\;\;\stackrel{13\;14}{(+)}$&1&$-\frac{1}{2}$&0&
$\frac{1}{2}$ &$0$&$0$&$-\frac{1}{2}$&$0$&$0$\\
\hline
27&$e_{R}$&$\stackrel{03}{(+i)}\,\stackrel{12}{[+]}|\stackrel{56}{(-)}\,\stackrel{78}{[-]}
||\stackrel{9 \;10}{(+)}\;\;\stackrel{11\;12}{(+)}\;\;\stackrel{13\;14}{(+)}$&1&$\frac{1}{2}$&0&
$-\frac{1}{2}$&$0$&$0$&$-\frac{1}{2}$&$-1$&$-1$\\
\hline
28&$ e_{R} $&$\stackrel{03}{[-i]}\,\stackrel{12}{(-)}|
\stackrel{56}{(-)}\,\stackrel{78}{[-]}
||\stackrel{9 \;10}{(+)}\;\;\stackrel{11\;12}{(+)}\;\;\stackrel{13\;14}{(+)} $&1&$-\frac{1}{2}$&0&
$-\frac{1}{2}$&$0$&$0$&$-\frac{1}{2}$&$-1$&$-1$\\
\hline
29&$e_{L}$&$\stackrel{03}{[-i]}\,\stackrel{12}{[+]}|\stackrel{56}{(-)}\,\stackrel{78}{(+)}
||\stackrel{9 \;10}{(+)}\;\;\stackrel{11\;12}{(+)}\;\;\stackrel{13\;14}{(+)}$&-1&$\frac{1}{2}$&
$-\frac{1}{2}$&0&$0$&$0$&$-\frac{1}{2}$&$-\frac{1}{2}$&$-1$\\
\hline
30&$e_{L} $&$ - \stackrel{03}{(+i)}\,\stackrel{12}{(-)}|\stackrel{56}{(-)}\,\stackrel{78}{(+)}
||\stackrel{9 \;10}{(+)}\;\;\stackrel{11\;12}{(+)}\;\;\stackrel{13\;14}{(+)} $&-1&$-\frac{1}{2}$&
$-\frac{1}{2}$&0&$0$&$0$&$-\frac{1}{2}$&$-\frac{1}{2}$&$-1$\\
\hline
31&$ \nu_{L}$&$ - \stackrel{03}{[-i]}\,\stackrel{12}{[+]}|\stackrel{56}{[+]}\,\stackrel{78}{[-]}
||\stackrel{9 \;10}{(+)}\;\;\stackrel{11\;12}{(+)}\;\;\stackrel{13\;14}{(+)}$ &-1&$\frac{1}{2}$&
$\frac{1}{2}$&0 &$0$&$0$&$-\frac{1}{2}$&$-\frac{1}{2}$&$0$\\
\hline
32&$\nu_{L}$&$\stackrel{03}{(+i)}\,\stackrel{12}{(-)}|\stackrel{56}{[+]}\,\stackrel{78}{[-]}
||\stackrel{9 \;10}{(+)}\;\;\stackrel{11\;12}{(+)}\;\;\stackrel{13\;14}{(+)}$&-1&$-\frac{1}{2}$&
$\frac{1}{2}$&0&$0$&$0$&$-\frac{1}{2}$&$-\frac{1}{2}$&$0$\\
\hline\hline
33&$ \bar{d}_{L}^{\bar{c1}}$&$ \stackrel{03}{[-i]}\,\stackrel{12}{[+]}|
\stackrel{56}{[+]}\,\stackrel{78}{(+)}
||\stackrel{9 \;10}{[-]}\;\;\stackrel{11\;12}{(+)}\;\;\stackrel{13\;14}{(+)} $ &-1&$\frac{1}{2}$&0&
$\frac{1}{2}$&$-\frac{1}{2}$&$-\frac{1}{2\,\sqrt{3}}$&$-\frac{1}{6}$&$\frac{1}{3}$&$\frac{1}{3}$\\
\hline
34&$\bar{d}_{L}^{\bar{c1}}$&$\stackrel{03}{(+i)}\,\stackrel{12}{(-)}|\stackrel{56}{[+]}\,\stackrel{78}{(+)}
||\stackrel{9 \;10}{[-]}\;\;\stackrel{11\;12}{(+)}\;\;\stackrel{13\;14}{(+)}$&-1&$-\frac{1}{2}$&0&
$\frac{1}{2}$&$-\frac{1}{2}$&$-\frac{1}{2\,\sqrt{3}}$&$-\frac{1}{6}$&$\frac{1}{3}$&$\frac{1}{3}$\\
\hline
35&$\bar{u}_{L}^{\bar{c1}}$&$ - \stackrel{03}{[-i]}\,\stackrel{12}{[+]}|\stackrel{56}{(-)}\,\stackrel{78}{[-]}
||\stackrel{9 \;10}{[-]}\;\;\stackrel{11\;12}{(+)}\;\;\stackrel{13\;14}{(+)}$&-1&$\frac{1}{2}$&0&
$-\frac{1}{2}$&$-\frac{1}{2}$&$-\frac{1}{2\,\sqrt{3}}$&$-\frac{1}{6}$&$-\frac{2}{3}$&$-\frac{2}{3}$\\
\hline
36&$ \bar{u}_{L}^{\bar{c1}} $&$ - \stackrel{03}{(+i)}\,\stackrel{12}{(-)}|
\stackrel{56}{(-)}\,\stackrel{78}{[-]}
||\stackrel{9 \;10}{[-]}\;\;\stackrel{11\;12}{(+)}\;\;\stackrel{13\;14}{(+)} $&-1&$-\frac{1}{2}$&0&
$-\frac{1}{2}$&$-\frac{1}{2}$&$-\frac{1}{2\,\sqrt{3}}$&$-\frac{1}{6}$&$-\frac{2}{3}$&$-\frac{2}{3}$\\
\hline
37&$\bar{d}_{R}^{\bar{c1}}$&$\stackrel{03}{(+i)}\,\stackrel{12}{[+]}|\stackrel{56}{[+]}\,\stackrel{78}{[-]}
||\stackrel{9 \;10}{[-]}\;\;\stackrel{11\;12}{(+)}\;\;\stackrel{13\;14}{(+)}$&1&$\frac{1}{2}$&
$\frac{1}{2}$&0&$-\frac{1}{2}$&$-\frac{1}{2\,\sqrt{3}}$&$-\frac{1}{6}$&$-\frac{1}{6}$&$\frac{1}{3}$\\
\hline
38&$\bar{d}_{R}^{\bar{c1}} $&$ - \stackrel{03}{[-i]}\,\stackrel{12}{(-)}|\stackrel{56}{[+]}\,\stackrel{78}{[-]}
||\stackrel{9 \;10}{[-]}\;\;\stackrel{11\;12}{(+)}\;\;\stackrel{13\;14}{(+)} $&1&$-\frac{1}{2}$&
$\frac{1}{2}$&0&$-\frac{1}{2}$&$-\frac{1}{2\,\sqrt{3}}$&$-\frac{1}{6}$&$-\frac{1}{6}$&$\frac{1}{3}$\\
\hline
39&$ \bar{u}_{R}^{\bar{c1}}$&$\stackrel{03}{(+i)}\,\stackrel{12}{[+]}|\stackrel{56}{(-)}\,\stackrel{78}{(+)}
||\stackrel{9 \;10}{[-]}\;\;\stackrel{11\;12}{(+)}\;\;\stackrel{13\;14}{(+)}$ &1&$\frac{1}{2}$&
$-\frac{1}{2}$&0 &$-\frac{1}{2}$&$-\frac{1}{2\,\sqrt{3}}$&$-\frac{1}{6}$&$-\frac{1}{6}$&$-\frac{2}{3}$\\
\hline
40&$\bar{u}_{R}^{\bar{c1}}$&$\stackrel{03}{[-i]}\,\stackrel{12}{(-)}|\stackrel{56}{(-)}\,\stackrel{78}{(+)}
||\stackrel{9 \;10}{[-]}\;\;\stackrel{11\;12}{(+)}\;\;\stackrel{13\;14}{(+)}$
%\stackrel{9 \;10}{[-]}\;\;\stackrel{11\;12}{[+]}\;\;\stackrel{13\;14}{[+]}$
&1&$-\frac{1}{2}$&
$-\frac{1}{2}$&0&$-\frac{1}{2}$&$-\frac{1}{2\,\sqrt{3}}$&$-\frac{1}{6}$&$-\frac{1}{6}$&$-\frac{2}{3}$\\
\hline\hline
41&$ \bar{d}_{L}^{\bar{c2}}$&$ \stackrel{03}{[-i]}\,\stackrel{12}{[+]}|
\stackrel{56}{[+]}\,\stackrel{78}{(+)}
||\stackrel{9 \;10}{(+)}\;\;\stackrel{11\;12}{[-]}\;\;\stackrel{13\;14}{(+)} $
&-1&$\frac{1}{2}$&0&
$\frac{1}{2}$&$\frac{1}{2}$&$-\frac{1}{2\,\sqrt{3}}$&$-\frac{1}{6}$&$\frac{1}{3}$&$\frac{1}{3}$\\
\hline
42&$\bar{d}_{L}^{\bar{c2}}$&$\stackrel{03}{(+i)}\,\stackrel{12}{(-)}|\stackrel{56}{[+]}\,\stackrel{78}{(+)}
||\stackrel{9 \;10}{(+)}\;\;\stackrel{11\;12}{[-]}\;\;\stackrel{13\;14}{(+)}$
&-1&$-\frac{1}{2}$&0&
$\frac{1}{2}$&$\frac{1}{2}$&$-\frac{1}{2\,\sqrt{3}}$&$-\frac{1}{6}$&$\frac{1}{3}$&$\frac{1}{3}$\\
\hline
43&$\bar{u}_{L}^{\bar{c2}}$&$ - \stackrel{03}{[-i]}\,\stackrel{12}{[+]}|\stackrel{56}{(-)}\,\stackrel{78}{[-]}
||\stackrel{9 \;10}{(+)}\;\;\stackrel{11\;12}{[-]}\;\;\stackrel{13\;14}{(+)}$
&-1&$\frac{1}{2}$&0&
$-\frac{1}{2}$&$\frac{1}{2}$&$-\frac{1}{2\,\sqrt{3}}$&$-\frac{1}{6}$&$-\frac{2}{3}$&$-\frac{2}{3}$\\
\hline
44&$ \bar{u}_{L}^{\bar{c2}} $&$ - \stackrel{03}{(+i)}\,\stackrel{12}{(-)}|
\stackrel{56}{(-)}\,\stackrel{78}{[-]}
||\stackrel{9 \;10}{(+)}\;\;\stackrel{11\;12}{[-]}\;\;\stackrel{13\;14}{(+)} $
&-1&$-\frac{1}{2}$&0&
$-\frac{1}{2}$&$\frac{1}{2}$&$-\frac{1}{2\,\sqrt{3}}$&$-\frac{1}{6}$&$-\frac{2}{3}$&$-\frac{2}{3}$\\
\hline
45&$\bar{d}_{R}^{\bar{c2}}$&$\stackrel{03}{(+i)}\,\stackrel{12}{[+]}|\stackrel{56}{[+]}\,\stackrel{78}{[-]}
||\stackrel{9 \;10}{(+)}\;\;\stackrel{11\;12}{[-]}\;\;\stackrel{13\;14}{(+)}$
&1&$\frac{1}{2}$&
$\frac{1}{2}$&0&$\frac{1}{2}$&$-\frac{1}{2\,\sqrt{3}}$&$-\frac{1}{6}$&$-\frac{1}{6}$&$\frac{1}{3}$\\
\hline
46&$\bar{d}_{R}^{\bar{c2}} $&$ - \stackrel{03}{[-i]}\,\stackrel{12}{(-)}|\stackrel{56}{[+]}\,\stackrel{78}{[-]}
||\stackrel{9 \;10}{(+)}\;\;\stackrel{11\;12}{[-]}\;\;\stackrel{13\;14}{(+)} $
&1&$-\frac{1}{2}$&
$\frac{1}{2}$&0&$\frac{1}{2}$&$-\frac{1}{2\,\sqrt{3}}$&$-\frac{1}{6}$&$-\frac{1}{6}$&$\frac{1}{3}$\\
\hline
47&$ \bar{u}_{R}^{\bar{c2}}$&$\stackrel{03}{(+i)}\,\stackrel{12}{[+]}|\stackrel{56}{(-)}\,\stackrel{78}{(+)}
||\stackrel{9 \;10}{(+)}\;\;\stackrel{11\;12}{[-]}\;\;\stackrel{13\;14}{(+)}$
 &1&$\frac{1}{2}$&
$-\frac{1}{2}$&0 &$\frac{1}{2}$&$-\frac{1}{2\,\sqrt{3}}$&$-\frac{1}{6}$&$-\frac{1}{6}$&$-\frac{2}{3}$\\
\hline
48&$\bar{u}_{R}^{\bar{c2}}$&$\stackrel{03}{[-i]}\,\stackrel{12}{(-)}|\stackrel{56}{(-)}\,\stackrel{78}{(+)}
||\stackrel{9 \;10}{(+)}\;\;\stackrel{11\;12}{[-]}\;\;\stackrel{13\;14}{(+)}$
&1&$-\frac{1}{2}$&
$-\frac{1}{2}$&0&$\frac{1}{2}$&$-\frac{1}{2\,\sqrt{3}}$&$-\frac{1}{6}$&$-\frac{1}{6}$&$-\frac{2}{3}$\\
\hline\hline
%\hline
%$\cdots$ &&&&&&&&&&& \\
%\hline\hline
49&$ \bar{d}_{L}^{\bar{c3}}$&$ \stackrel{03}{[-i]}\,\stackrel{12}{[+]}|
\stackrel{56}{[+]}\,\stackrel{78}{(+)}
||\stackrel{9 \;10}{(+)}\;\;\stackrel{11\;12}{(+)}\;\;\stackrel{13\;14}{[-]} $ &-1&$\frac{1}{2}$&0&
$\frac{1}{2}$&$0$&$\frac{1}{\sqrt{3}}$&$-\frac{1}{6}$&$\frac{1}{3}$&$\frac{1}{3}$\\
\hline
50&$\bar{d}_{L}^{\bar{c3}}$&$\stackrel{03}{(+i)}\,\stackrel{12}{(-)}|\stackrel{56}{[+]}\,\stackrel{78}{(+)}
||\stackrel{9 \;10}{(+)}\;\;\stackrel{11\;12}{(+)}\;\;\stackrel{13\;14}{[-]} $&-1&$-\frac{1}{2}$&0&
$\frac{1}{2}$&$0$&$\frac{1}{\sqrt{3}}$&$-\frac{1}{6}$&$\frac{1}{3}$&$\frac{1}{3}$\\
\hline
51&$\bar{u}_{L}^{\bar{c3}}$&$ - \stackrel{03}{[-i]}\,\stackrel{12}{[+]}|\stackrel{56}{(-)}\,\stackrel{78}{[-]}
||\stackrel{9 \;10}{(+)}\;\;\stackrel{11\;12}{(+)}\;\;\stackrel{13\;14}{[-]} $&-1&$\frac{1}{2}$&0&
$-\frac{1}{2}$&$0$&$\frac{1}{\sqrt{3}}$&$-\frac{1}{6}$&$-\frac{2}{3}$&$-\frac{2}{3}$\\
\hline
52&$ \bar{u}_{L}^{\bar{c3}} $&$ - \stackrel{03}{(+i)}\,\stackrel{12}{(-)}|
\stackrel{56}{(-)}\,\stackrel{78}{[-]}
||\stackrel{9 \;10}{(+)}\;\;\stackrel{11\;12}{(+)}\;\;\stackrel{13\;14}{[-]}  $&-1&$-\frac{1}{2}$&0&
$-\frac{1}{2}$&$0$&$\frac{1}{\sqrt{3}}$&$-\frac{1}{6}$&$-\frac{2}{3}$&$-\frac{2}{3}$\\
\hline
53&$\bar{d}_{R}^{\bar{c3}}$&$\stackrel{03}{(+i)}\,\stackrel{12}{[+]}|\stackrel{56}{[+]}\,\stackrel{78}{[-]}
||\stackrel{9 \;10}{(+)}\;\;\stackrel{11\;12}{(+)}\;\;\stackrel{13\;14}{[-]} $&1&$\frac{1}{2}$&
$\frac{1}{2}$&0&$0$&$\frac{1}{\sqrt{3}}$&$-\frac{1}{6}$&$-\frac{1}{6}$&$\frac{1}{3}$\\
\hline
54&$\bar{d}_{R}^{\bar{c3}} $&$ - \stackrel{03}{[-i]}\,\stackrel{12}{(-)}|\stackrel{56}{[+]}\,\stackrel{78}{[-]}
||\stackrel{9 \;10}{(+)}\;\;\stackrel{11\;12}{(+)}\;\;\stackrel{13\;14}{[-]} $&1&$-\frac{1}{2}$&
$\frac{1}{2}$&0&$0$&$\frac{1}{\sqrt{3}}$&$-\frac{1}{6}$&$-\frac{1}{6}$&$\frac{1}{3}$\\
\hline
55&$ \bar{u}_{R}^{\bar{c3}}$&$\stackrel{03}{(+i)}\,\stackrel{12}{[+]}|\stackrel{56}{(-)}\,\stackrel{78}{(+)}
||\stackrel{9 \;10}{(+)}\;\;\stackrel{11\;12}{(+)}\;\;\stackrel{13\;14}{[-]} $ &1&$\frac{1}{2}$&
$-\frac{1}{2}$&0 &$0$&$\frac{1}{\sqrt{3}}$&$-\frac{1}{6}$&$-\frac{1}{6}$&$-\frac{2}{3}$\\
\hline
56&$\bar{u}_{R}^{\bar{c3}}$&$\stackrel{03}{[-i]}\,\stackrel{12}{(-)}|\stackrel{56}{(-)}\,\stackrel{78}{(+)}
||\stackrel{9 \;10}{(+)}\;\;\stackrel{11\;12}{(+)}\;\;\stackrel{13\;14}{[-]} $&1&$-\frac{1}{2}$&
$-\frac{1}{2}$&0&$0$&$\frac{1}{\sqrt{3}}$&$-\frac{1}{6}$&$-\frac{1}{6}$&$-\frac{2}{3}$\\
\hline\hline
%$\cdots$ &&&&&&&&&&& \\
%\hline\hline
57&$ \bar{e}_{L}$&$ \stackrel{03}{[-i]}\,\stackrel{12}{[+]}|
\stackrel{56}{[+]}\,\stackrel{78}{(+)}
||\stackrel{9 \;10}{[-]}\;\;\stackrel{11\;12}{[-]}\;\;\stackrel{13\;14}{[-]} $ &-1&$\frac{1}{2}$&0&
$\frac{1}{2}$&$0$&$0$&$\frac{1}{2}$&$1$&$1$\\
\hline
58&$\bar{e}_{L}$&$\stackrel{03}{(+i)}\,\stackrel{12}{(-)}|\stackrel{56}{[+]}\,\stackrel{78}{(+)}
||\stackrel{9 \;10}{[-]}\;\;\stackrel{11\;12}{[-]}\;\;\stackrel{13\;14}{[-]}$&-1&$-\frac{1}{2}$&0&
$\frac{1}{2}$ &$0$&$0$&$\frac{1}{2}$&$1$&$1$\\
\hline
59&$\bar{\nu}_{L}$&$ - \stackrel{03}{[-i]}\,\stackrel{12}{[+]}|\stackrel{56}{(-)}\,\stackrel{78}{[-]}
||\stackrel{9 \;10}{[-]}\;\;\stackrel{11\;12}{[-]}\;\;\stackrel{13\;14}{[-]}$&-1&$\frac{1}{2}$&0&
$-\frac{1}{2}$&$0$&$0$&$\frac{1}{2}$&$0$&$0$\\
\hline
60&$ \bar{\nu}_{L} $&$ - \stackrel{03}{(+i)}\,\stackrel{12}{(-)}|
\stackrel{56}{(-)}\,\stackrel{78}{[-]}
||\stackrel{9 \;10}{[-]}\;\;\stackrel{11\;12}{[-]}\;\;\stackrel{13\;14}{[-]} $&-1&$-\frac{1}{2}$&0&
$-\frac{1}{2}$&$0$&$0$&$\frac{1}{2}$&$0$&$0$\\
\hline
61&$\bar{\nu}_{R}$&$\stackrel{03}{(+i)}\,\stackrel{12}{[+]}|\stackrel{56}{(-)}\,\stackrel{78}{(+)}
||\stackrel{9 \;10}{[-]}\;\;\stackrel{11\;12}{[-]}\;\;\stackrel{13\;14}{[-]}$&1&$\frac{1}{2}$&
$-\frac{1}{2}$&0&$0$&$0$&$\frac{1}{2}$&$\frac{1}{2}$&$0$\\
\hline
62&$\bar{\nu}_{R} $&$ - \stackrel{03}{[-i]}\,\stackrel{12}{(-)}|\stackrel{56}{(-)}\,\stackrel{78}{(+)}
||\stackrel{9 \;10}{[-]}\;\;\stackrel{11\;12}{[-]}\;\;\stackrel{13\;14}{[-]} $&1&$-\frac{1}{2}$&
$-\frac{1}{2}$&0&$0$&$0$&$\frac{1}{2}$&$\frac{1}{2}$&$0$\\
\hline
63&$ \bar{e}_{R}$&$\stackrel{03}{(+i)}\,\stackrel{12}{[+]}|\stackrel{56}{[+]}\,\stackrel{78}{[-]}
||\stackrel{9 \;10}{[-]}\;\;\stackrel{11\;12}{[-]}\;\;\stackrel{13\;14}{[-]}$ &1&$\frac{1}{2}$&
$\frac{1}{2}$&0 &$0$&$0$&$\frac{1}{2}$&$\frac{1}{2}$&$1$\\
\hline
64&$\bar{e}_{R}$&$\stackrel{03}{[-i]}\,\stackrel{12}{(-)}|\stackrel{56}{[+]}\,\stackrel{78}{[-]}
||\stackrel{9 \;10}{[-]}\;\;\stackrel{11\;12}{[-]}\;\;\stackrel{13\;14}{[-]}$&1&$-\frac{1}{2}$&
$\frac{1}{2}$&0&$0$&$0$&$\frac{1}{2}$&$\frac{1}{2}$&$1$\\
\hline
\end{supertabular}
\end{tiny}

\vspace {3mm}
\section{Grassmann and Clifford algebras}
\label{grassmannclifford}

This part is taken from Ref.~\cite{n2024NPB,nh2025Bled}, 
following Refs.~~\cite{norma93,nh2021RPPNP,n2014matterantimatter}.

In Grassmann $d$-dimensional space there are $d$ anti-commuting (operators)
$\theta^{a}$,
% $\{\theta^{a}, \theta^{b}\}_{+}=0$, $a=(0,1,2,3,5,..,d)$,
and $d$ anti-commuting operators which are derivatives with respect to $\theta^{a}$,
$\frac{\partial}{\partial \theta_{a}}$.
%$\{\frac{\partial}{\partial \theta_{a}}, \frac{\partial}{\partial \theta_{b}}\}_{+} =0$
%
\begin{eqnarray}
\label{thetaderanti0}
\{\theta^{a}, \theta^{b}\}_{+}=0\,, \, && \,
\{\frac{\partial}{\partial \theta_{a}}, \frac{\partial}{\partial \theta_{b}}\}_{+} =0\,,
\nonumber\\
\{\theta_{a},\frac{\partial}{\partial \theta_{b}}\}_{+} &=&\delta^b_{a}\,,
\;(a,b)=(0,1,2,3,5,\cdots,d)\,.
% ,\nonumber\\ \eta^{ab}&=&diag(1,-1,-1,\dots,-1)\,.
\end{eqnarray}
The choice%~\cite{nh2018}
\begin{eqnarray}
(\theta^{a})^{\dagger} &=& \eta^{a a} \frac{\partial}{\partial \theta_{a}}\,,\quad
{\rm leads \, to} \quad
(\frac{\partial}{\partial \theta_{a}})^{\dagger}= \eta^{a a} \theta^{a}\,,
\label{thetaderher0}
\end{eqnarray}
with $\eta^{a b}=diag\{1,-1,-1,\cdots,-1\}$.

$ \theta^{a}$ and $ \frac{\partial}{\partial \theta_{a}}$ are, up to the sign, 
Hermitian conjugate to each other. The identity is a self-adjoint member of 
the algebra.

In $d$-dimensional space, there are $2^d$ superposition of products of 
$\theta^{a}$, the Hermitian conjugated partners of which are the 
corresponding superposition of products of 
$\frac{\partial}{\partial \theta_{a}}$~\cite{nh2021RPPNP}. %,n2019PRD}.

We can make from $\theta^{a}$'s and their conjugate momenta
$p^{\theta a}= i \,\frac{\partial}{\partial \theta_{a}}$ two kinds of the 
operators, $\gamma^{a}$ and
$\tilde{\gamma}^{a}$~\cite{norma93},
\begin{eqnarray}
\label{clifftheta1}
\gamma^{a} &=& (\theta^{a} + \frac{\partial}{\partial \theta_{a}})\,, \quad
\tilde{\gamma}^{a} =i \,(\theta^{a} - \frac{\partial}{\partial \theta_{a}})\,,\nonumber\\
\theta^{a} &=&\frac{1}{2} \,(\gamma^{a} - i \tilde{\gamma}^{a})\,, \quad
\frac{\partial}{\partial \theta_{a}}= \frac{1}{2} \,(\gamma^{a} + i \tilde{\gamma}^{a})\,,
\nonumber\\
\end{eqnarray}
each offers $2^d$ superposition of products of $\gamma^{a}$ or 
$\tilde{\gamma}^{a}$~(\cite{nh2021RPPNP} and
references therein)
\begin{eqnarray}                                    % {gammatildeantiher0}
\label{gammatildeantiher0}
\{\gamma^{a}, \gamma^{b}\}_{+}&=&2 \eta^{a b}= \{\tilde{\gamma}^{a},
\tilde{\gamma}^{b}\}_{+}\,, \nonumber\\
\{\gamma^{a}, \tilde{\gamma}^{b}\}_{+}&=&0\,,\quad
(a,b)=(0,1,2,3,5,\cdots,d)\,, \nonumber\\
(\gamma^{a})^{\dagger} &=& \eta^{aa}\, \gamma^{a}\, , \quad
(\tilde{\gamma}^{a})^{\dagger} = \eta^{a a}\, \tilde{\gamma}^{a}\,.
\end{eqnarray}

The Grassmann algebra offers the description of the internal space of 
{\it anti-commuting integer spin second quantized fields} and of the 
{\it commuting integer spin second quantized fields}~\cite{nh2021RPPNP}. 

Both algebras, the
superposition of odd products of  $\gamma^a$'s or of $\tilde{\gamma}^a$'s,
offer the description of the second quantized half integer spins and charges
in the fundamental representations of the group~\cite{nh2021RPPNP}, 
Table~\ref{Table so13+1.} represents the internal spaces of one family 
of quarks and leptons and antiquarks and antileptons, the internal space 
of which is determined by odd products of  $\gamma^a$'s.

The superposition of even products of either $\gamma^a$'s or 
$\tilde{\gamma}^a$'s offer the description of the internal spaces of 
commuting second quantized boson fields with integer spins~\cite{n2023NPB,n2023MDPI,%
nh2025Bled}),
%)
manifesting from the point of the subgroups of the $SO(d-1,1)$ group, 
spins and charges in the adjoint representations.

\vspace{2mm}

There is so far observed only one kind of the anti-commuting half-integer spin 
second quantized fields. \\

The {\it postulate}, which determines how does $\tilde{\gamma}^{a}$
operate on $\gamma^a$, reduces the presentations of the two Clifford subalgebras, 
$\gamma^a$ and $\tilde{\gamma}^a$, to the one described by
$\gamma^a$~\cite{nh02,norma93,n2014matterantimatter}%,nh2018}
\begin{eqnarray}
\{\tilde{\gamma}^a B &=&(-)^B\, i \, B \gamma^a\}\, |\psi_{oc}>\,,
\label{tildegammareduced0}
\end{eqnarray}
with $(-)^B = -1$, if $B$ is (a function of) odd products of $\gamma^a$'s, 
otherwise $(-)^B = 1$~\cite{nh02}, the vacuum state $|\psi_{oc}>$ is 
defined in Eq.~(\ref{vaccliffodd}) of Subsect.~\ref{basisvectors}.

\vspace{2mm}

After the postulate of Eq.~(\ref{tildegammareduced0}) the  vector space of 
$\gamma^a$'s are chosen to describe the internal space of fermions, while 
$\tilde{\gamma}^{a}$'s  are used to determine the family quantum numbers 
of the fermion fields.

\section{Some useful relations}
\label{usefulrelations}
%

% EXPLAIN FOR WHAT THIS IS USEFUL! {grassmannclifford}

 In this appendix some useful relations, needed in this paper in 
 Sects.~(\ref{basisvectors}, %\ref{bosons13+1and5+1}, 
 \ref{13+1representation}) are presented, taken from 
 Ref.~\cite{n2024NPB}, App.~E.

One can find if taking into account Eq.~(\ref{gammatildeantiher0})
\begin{small}
\begin{eqnarray}
\label{usefulrel0}
%{\rm it \; then} &&\; {\rm follows}\nonumber\\
\gamma^a \stackrel{ab}{(k)}&=& \eta^{aa}\stackrel{ab}{[-k]},\; \quad
\gamma^b \stackrel{ab}{(k)}= -ik \stackrel{ab}{[-k]}, \; \quad % \nonumber\\
\gamma^a \stackrel{ab}{[k]}= \stackrel{ab}{(-k)},\;\quad \;\;
\gamma^b \stackrel{ab}{[k]}= -ik \eta^{aa} \stackrel{ab}{(-k)}\,,\nonumber\\
\tilde{\gamma^a} \stackrel{ab}{(k)} &=& - i\eta^{aa}\stackrel{ab}{[k]},\quad
\tilde{\gamma^b} \stackrel{ab}{(k)} = - k \stackrel{ab}{[k]}, \;\qquad \,%\nonumber\\
\tilde{\gamma^a} \stackrel{ab}{[k]} = \;\;i\stackrel{ab}{(k)},\; \quad
\tilde{\gamma^b} \stackrel{ab}{[k]} = -k \eta^{aa} \stackrel{ab}{(k)}\,,
\nonumber\\ %
\stackrel{ab}{(k)}\stackrel{ab}{(-k)}& =& \eta^{aa} \stackrel{ab}{[k]}\,,\quad \;%\qquad \qquad
\stackrel{ab}{(-k)}\stackrel{ab}{(k)} = \eta^{aa} \stackrel{ab}{[-k]}\,,\quad\;
\stackrel{ab}{(k)}\stackrel{ab}{[k]} =0\,,\quad \quad\,
\stackrel{ab}{(k)}\stackrel{ab}{[-k]} =
\stackrel{ab}{(k)}\,,\quad %\qquad \qquad
\nonumber\\ %%
\stackrel{ab}{(-k)}\stackrel{ab}{[k]} &=& \stackrel{ab}{(-k)}\,,\quad \quad\quad
\stackrel{ab}{[k]}\stackrel{ab}{(k)}= \stackrel{ab}{(k)}\,,
%\stackrel{ab}{[k]}\stackrel{ab}{(k)}= \stackrel{ab}{(k)}\,, \quad \quad \quad
% \stackrel{ab}{(k)}\stackrel{ab}{[-k]} = \stackrel{ab}{(k)}\,,
\quad \quad \quad \;
\stackrel{ab}{[k]}\stackrel{ab}{(-k)} =0\,,\quad \quad\,
\stackrel{ab}{[k]}\stackrel{ab}{[-k]} =0\,,\quad
\nonumber\\
\stackrel{ab}{(k)}^{\dagger} &=& \eta^{aa}\stackrel{ab}{(-k)}\,,\quad
(\stackrel{ab}{(k)})^2 =0\,, \quad \stackrel{ab}{(k)}\stackrel{ab}{(-k)}
=\eta^{aa}\stackrel{ab}{[k]}\,,\nonumber\\
% \stackrel{ab}{[k]}:&=&
% \frac{1}{2}(1+ \frac{i}{k} \gamma^a \gamma^b)\,,\quad \;\,
\stackrel{ab}{[k]}^{\dagger} &=& \,\stackrel{ab}{[k]}\,, \quad \quad \quad \quad
(\stackrel{ab}{[k]})^2 = \stackrel{ab}{[k]}\,,
\quad \stackrel{ab}{[k]}\stackrel{ab}{[-k]}=0\,.
 \end{eqnarray}
 \end{small} 
\begin{small}
\begin{eqnarray}
\stackrel{ab}{\tilde{(k)}} \stackrel{ab}{(k)}&=& 0\,,\quad
\stackrel{ab}{\tilde{(k)}} \stackrel{ab}{(-k)}=-i \eta^{aa}\stackrel{ab}{[-k]}
\,,\quad
\stackrel{ab}{\widetilde{(-k)}} \stackrel{ab}{(k)}=-i \eta^{aa}\stackrel{ab}{[k]}
\,,\quad
\stackrel{ab}{\tilde{(k)}} \stackrel{ab}{[k]}= i  \stackrel{ab}{(k)}
\,, \quad \nonumber\\
%\stackrel{ab}{\widetilde{(-k)}}\stackrel{ab}{\tilde{[k]}}&=& 0\,,\quad 
%\stackrel{ab}{\widetilde{(-k)}}\stackrel{ab}{\widetilde{[-k]}}= i  \stackrel{ab}{\widetilde{(-k)}}\,,
%
\stackrel{ab}{\tilde{(k)}} \stackrel{ab}{[-k]}&=&0
\,,\quad
\stackrel{ab}{\widetilde{(-k)}} \stackrel{ab}{[k]}=0\,,
\quad
\stackrel{ab}{\widetilde{(-k)}} \stackrel{ab}{[-k]}=i  \stackrel{ab}{(-k)}\,,
\quad \stackrel{ab}{\tilde{[k]}} \stackrel{ab}{(k)}=\stackrel{ab}{(k)}\,,\nonumber\\
\stackrel{ab}{\tilde{[k]}} \stackrel{ab}{(-k)}&=&0\,, \quad
 \stackrel{ab}{\widetilde{[k]}} \stackrel{ab}{[k]}=0\,,\quad % CHECK
 \stackrel{ab}{\widetilde{[-k]}} \stackrel{ab}{[k]}=
\stackrel{ab}{[k]}\,,\quad
\stackrel{ab}{\tilde{[k]}}\stackrel{ab}{[-k]}=\stackrel{ab}{[-k]}\,,
\label{graficcliff0}
 \end{eqnarray}
\end{small}
%

% EXPLAIN!!!

% Let us present the odd ``basis vectors'' and their Hermitian conjugate partners
% for $d=(3+1)$. The odd ``basis vectors'' appear in two families, each family has 
% two members.

%%%%%%%%%%%%%%%%%%%%

%
\begin{small}
\begin{eqnarray}
\label{graficfollow1}
S^{ac}\stackrel{ab}{(k)}\stackrel{cd}{(k)} &=& -\frac{i}{2} \eta^{aa} \eta^{cc} 
\stackrel{ab}{[-k]}\stackrel{cd}{[-k]}\,, \quad 
S^{ac}\stackrel{ab}{[k]}\stackrel{cd}{[k]} = 
\frac{i}{2} \stackrel{ab}{(-k)}\stackrel{cd}{(-k)}\,,\nonumber\\
S^{ac}\stackrel{ab}{(k)}\stackrel{cd}{[k]} &=& -\frac{i}{2} \eta^{aa}  
\stackrel{ab}{[-k]}\stackrel{cd}{(-k)}\,, \quad
S^{ac}\stackrel{ab}{[k]}\stackrel{cd}{(k)} = \frac{i}{2} \eta^{cc}  
\stackrel{ab}{(-k)}\stackrel{cd}{[-k]}\,, \nonumber\\
\tilde{S}^{ac} \stackrel{ab}{(k)}\stackrel{cd}{(k)} &=& \frac{i}{2} \eta^{aa} \eta^{cc} 
\stackrel{ab}{[k]}\stackrel{cd}{[k]}\,, \quad 
\tilde{S}^{ac}\stackrel{ab}{[k]}\stackrel{cd}{[k]} = 
-\frac{i}{2} \stackrel{ab}{(k)}\stackrel{cd}{(k)}\,,\nonumber\\
\tilde{S}^{ac}\stackrel{ab}{(k)}\stackrel{cd}{[k]} &=& -\frac{i}{2} \eta^{aa}  
\stackrel{ab}{[k]}\stackrel{cd}{(k)}\,, \quad
\tilde{S}^{ac}\stackrel{ab}{[k]}\stackrel{cd}{(k)} = \frac{i}{2} \eta^{cc}  
\stackrel{ab}{(k)}\stackrel{cd}{[k]}\,. 
\end{eqnarray}
\end{small}

% {graficfollow1}{plusminus}

The reader can calculate all the quantum numbers  of Table~\ref{Table so13+1.},
 App.~\ref{13+1representation},
if taking into account
the generators of the two $SU(2)$  ($\subset SO(3,1)$ $\subset SO(7,1) \subset SO(13,1)$) groups, describing  spins of fermions and  the corresponding family quantum numbers
\begin{eqnarray}
\label{so1+3}
&&\vec{N}_{\pm}(= \vec{N}_{(L,R)}): = \,\frac{1}{2} (S^{23}\pm i S^{01},
S^{31}\pm i S^{02}, S^{12}\pm i S^{03} )\,,\nonumber\\
% \nonumber\\ &&N^{\pm}_{+}      = N^{1}_{+} \pm i \,N^{2}_{+} = 
%  - \stackrel{03}{(\mp i)} \stackrel{12}{(\pm )}\,, \quad N^{\pm}_{-}= 
% N^{1}_{-} \pm  i\,N^{2}_{-} =    \stackrel{03}{(\pm i)} \stackrel{12}{(\pm )}\,.
&&\vec{\tilde{N}}_{\pm}(=\vec{\tilde{N}}_{(L,R)}): =
 \,\frac{1}{2} (\tilde{S}^{23}\pm i \tilde{S}^{01}\,,
\tilde{S}^{31}\pm i \tilde{S}^{02}, \tilde{S}^{12}\pm i \tilde{S}^{03} )\,,
\end{eqnarray}
the generators of the two $SU(2)$ ($SU(2)$ $\subset SO(4)$ $\subset SO(7,1) 
\subset SO(13,1)$) groups, describing  the weak charge, $\vec{\tau}^{1}$, and
the second kind of the weak charge, $\vec{\tau}^{2}$,  of fermions and 
the corresponding family quantum numbers
% in the {\it spin-charge-family } theory
%
 \begin{eqnarray}
 \label{so42}
 \vec{\tau}^{1}:&=&\frac{1}{2} (S^{58}-  S^{67}, \,S^{57} + S^{68}, \,S^{56}-  S^{78} )\,,%\nonumber\\
 \quad
 \vec{\tau}^{2}:= \frac{1}{2} (S^{58}+  S^{67}, \,S^{57} - S^{68}, \,S^{56}+  S^{78} )\,,
 \nonumber\\
% \tau^{1\pm}         &=& (\mp)\, \stackrel{56}{(\pm )} \stackrel{78}{(\mp )} \,, \qquad   
% \qquad \qquad  \qquad \qquad \;\;
% \tau^{2\mp}=            (\mp)\, \stackrel{56}{(\mp )} \stackrel{78}{(\mp )} \,,
 \vec{\tilde{\tau}}^{1}:&=&\frac{1}{2} (\tilde{S}^{58}-  \tilde{S}^{67}, \,\tilde{S}^{57} + 
 \tilde{S}^{68}, \,\tilde{S}^{56}-  \tilde{S}^{78} )\,, \quad 
 \vec{\tilde{\tau}}^{2}:=\frac{1}{2} (\tilde{S}^{58}+  \tilde{S}^{67}, \,\tilde{S}^{57} - 
 \tilde{S}^{68}, \,\tilde{S}^{56}+  \tilde{S}^{78} ),\,\,\;\;
 \end{eqnarray}
and the generators of $SU(3)$ and $U(1)$ subgroups of $SO(6)$ 
$\subset SO(13,1)$, describing  the colour charge and the ''fermion'' 
charge  of fermions as well as the corresponding family quantum 
number $\tilde{\tau}^4$
%in the {\it spin-charge-family } theory
%
 \begin{eqnarray}
 \label{so64}
 \vec{\tau}^{3}: = &&\frac{1}{2} \,\{  S^{9\;12} - S^{10\;11} \,,
  S^{9\;11} + S^{10\;12} ,\, S^{9\;10} - S^{11\;12}\, ,  %\nonumber\\
  S^{9\;14} -  S^{10\;13} ,\,  \nonumber\\
  && S^{9\;13} + S^{10\;14} \,,  S^{11\;14} -  S^{12\;13}\,, %\nonumber\\
  S^{11\;13} +  S^{12\;14} ,\,  \frac{1}{\sqrt{3}} ( S^{9\;10} + S^{11\;12} - 
 2 S^{13\;14})\}\,,\nonumber\\
 \tau^{4}: = &&-\frac{1}{3}(S^{9\;10} + S^{11\;12} + S^{13\;14})\,,\;\;\nonumber\\
 \tilde{\tau}^{4}: = &&-\frac{1}{3}(\tilde{S}^{9\;10} + \tilde{S}^{11\;12} + \tilde{S}^{13\;14})\,.
 \end{eqnarray}
The (chosen) Cartan subalgebra operators, determining the commuting 
operators in the above equations, is presented in Eq.~(\ref{cartangrasscliff}). 

The  hypercharge $Y$ and the electromagnetic charge $Q$ and the corresponding family  quantum numbers then follows as
 \begin{eqnarray}
 \label{YQY'Q'andtilde}
 Y:= \tau^{4} + \tau^{23}\,,\;\; Q: =  \tau^{13} + Y\,,\;\; 
 Y':= -\tau^{4}\tan^2\vartheta_2 + \tau^{23}\,, %\;\;
 \;\; Q':= -Y \tan^2\vartheta_1 + \tau^{13} \,,&&\nonumber\\
  \tilde{Y}:= \tilde{\tau}^{4} + \tilde{\tau}^{23}\,,\,\;\tilde{Q}:= 
  \tilde{Y} + \tilde{\tau}^{13}\,,\;\;
   \tilde{Y'}:= -\tilde{\tau}^{4} 
  \tan^2 \vartheta_2 + \tilde{\tau}^{23}\,,\;
  \;\; \tilde{Q'}= -\tilde{Y} \tan^2 \vartheta_1 
  + \tilde{\tau}^{13}\,. &&\,
 % \,,%\nonumber\\
%%  %{\cal Y}:&=& {\cal \tau}^{4} + {\cal \tau}^{23}\,,\;\; {\cal Y'}:=
%%% -{\cal \tau}^{4}\tan^2 \vartheta_2 + 
 %% %{\cal \tau}^{23}\,,\;\;
 %% %{\cal Q}: =  {\cal \tau}^{13} + {\cal Y}\,,\;\; {\cal Q'}:=
 %%% -{\cal Y} \tan^2 \vartheta_1 + %\cal \tau}^{13}\,
 %% %\nonumber\\
  \end{eqnarray}
 %
% are presented.
 \begin{small}
Below are some of the above expressions written in terms of  nilpotents and projectors
 \begin{eqnarray}
\label{plusminus}
 N^{\pm}_{+}         &=& N^{1}_{+} \pm i \,N^{2}_{+} = 
 - \stackrel{03}{(\mp i)} \stackrel{12}{(\pm )}\,, \quad N^{\pm}_{-}= 
 N^{1}_{-} \pm  i\,N^{2}_{-} = 
   \stackrel{03}{(\pm i)} \stackrel{12}{(\pm )}\,,
%   N^{\pm}_{+}      &=& N^{1}_{+} \pm i \,N^{2}_{+} = 
 % - \stackrel{03}{(\mp i)} \stackrel{12}{(\pm )}\,, \quad N^{\pm}_{-}= 
 % N^{1}_{-} \pm  i\,N^{2}_{-} = 
%   \stackrel{03}{(\pm i)} \stackrel{12}{(\pm )}\,,
\nonumber\\
 \tilde{N}^{\pm}_{+} &=& - \stackrel{03}{\tilde{(\mp i)}} 
 \stackrel{12}{\tilde{(\pm )}}\,, \quad 
 \tilde{N}^{\pm}_{-}= %\tilde{N}^{1}_{-} \pm i\,\tilde{N}^{2}_{-} = 
   \stackrel{03} {\tilde{(\pm i)}} \stackrel{12} {\tilde{(\pm )}}\,,\nonumber\\ 
 \tau^{1\pm}         &=& (\mp)\, \stackrel{56}{(\pm )} \stackrel{78}{(\mp )} \,, \quad   
 \tau^{2\mp}=            (\mp)\, \stackrel{56}{(\mp )} \stackrel{78}{(\mp )} \,,\nonumber\\ 
 \tilde{\tau}^{1\pm} &=& (\mp)\, \stackrel{56}{\tilde{(\pm )}} 
 \stackrel{78}{\tilde{(\mp )}}\,,\quad   
 \tilde{\tau}^{2\mp}= (\mp)\, \stackrel{56}{\tilde{(\mp )}} \stackrel{78}{\tilde{(\mp )}}\,.
 \end{eqnarray}
\end{small}

 For fermions, the operator of handedness $\Gamma^d$ is determined as follows:
  \begin{small}
\begin{eqnarray}
\label{Gamma}
%\begin{displaymath}
 \Gamma^{(d)}= \prod_a (\sqrt{\eta^{aa}} \gamma^a)  \cdot \left \{ \begin{array}{l l}
 (i)^{\frac{d}{2}} \,, &\rm{ for\, d \,even}\,,\\
 (i)^{\frac{d-1}{2}}\,,&\rm{for \, d \,odd}\,.
  \end{array} \right.
% \end{displaymath}
 \end{eqnarray}
 \end{small}
\section{Odd and even ``basis vectors'' in $(5+1)$-dimensional space}
\label{oddevend}
%
%
 % THINK ABOUT ADD MORE FROM n2024NPB\\
 
In this appendix, the even and odd ``basis vectors'' are presented 
for the choice $d=(5+1)$, needed in Subsect.~(\ref{statesactivein3+1}). 
The presentation follows the paper~\cite{n2023NPB}.

Table~\ref{Table Clifffourplet.} presents $2^{d=6}$ ``eigenvectors" 
of the Cartan subalgebra members, Eq.~(\ref{cartangrasscliff}), of the 
odd and even ``basis vectors'' which are the superposition of odd,  
${\hat b}^{m \dagger}_f$, appearing in $2^{\frac{d=6}{2}-1}=4$
families with $2^{\frac{d=6}{2}-1}=4$ members each, and their 
Hermitian conjugate partners, $({\hat b}^{m \dagger}_f)^{\dagger}$, 
$16$, and of $2^{\frac{d=6}{2}-1} \times 2^{\frac{d=6}{2}-1}= 16$ 
even, ${}^{I}{\cal A}^{m}_f$, 
and of $16$ even, ${}^{II}{\cal A}^{m}_f$, products of $\gamma^{a}$'s, 
helpful in Sect.~(\ref{statesactivein3+1}). Table~\ref{Table Clifffourplet.} 
is presented in several papers~(\cite{n2023NPB,nh2021RPPNP}, and 
references therein). 

Odd and even ``basis vectors'' are presented as products of nilpotents
and projectors, Eqs.~(\ref{calsab}, \ref{usefulrel0}). The odd ``basis 
vectors'' are products of odd number of nilpotents, one or three, the rest 
are projectors, two or zero; the even ``basis vectors'' are products of 
even number of nilpotents, zero or two, the rest are projectors, three 
or one.

As explained in Eqs.~(\ref{AIbbdagger},~\ref{AIIbdaggerb}), even 
``basis vectors'' can be expressed as algebraic products of odd 
``basis vectors'' and their Hermitian conjugated partners: 
${}^{I}{\hat{\cal A}}^{m \dagger}_{f}=\hat{b}^{m' \dagger}_{f `} *_A 
(\hat{b}^{m'' \dagger}_{f `})^{\dagger}$\,,
while 
$ {}^{II}{\hat{\cal A}}^{m \dagger}_{f}=
(\hat{b}^{m' \dagger}_{f `})^{\dagger} *_A 
\hat{b}^{m' \dagger}_{f `'}\,. $

The ``basis vectors'' ${}^{I}{\hat{\cal A}}^{m \dagger}_{f}$ 
are presented in Tables~(\ref{transverseCliff basis5+1even I.},~
\ref{S120Cliff basis5+1even I.}), expressed with the odd ``basis vectors''
of the first group of Table~\ref{Table Clifffourplet.}, and the first family, 
and their Hermitian conjugated partners. Any of the families generates 
the same  ${}^{I}{\hat{\cal A}}^{m \dagger}_{f}$. The first, 
Table~\ref{transverseCliff basis5+1even I.}, collects the eight members 
with ${\cal S}^{12}=\pm 1$, the second, Table~
\ref{S120Cliff basis5+1even I.}, collects the remaining eight 
members with ${\cal S}^{12}=0$. 
 
 The ``basis vectors'' ${}^{II}{\hat{\cal A}}^{m \dagger}_{f}$ 
are presented in Tables~(\ref{transverseCliff basis5+1even II.},~
\ref{S120Cliff basis5+1even II.}), expressed with the odd ``basis 
vectors'' of the first group of Table~\ref{Table Clifffourplet.}, and 
the first members of any family, and with their Hermitian conjugated 
partners. Any of the family members generates the same
 ${}^{II}{\hat{\cal A}}^{m \dagger}_{f}$. The first 
 Table~\ref{transverseCliff basis5+1even II.} collects the eight 
members with
${\cal S}^{12}=\pm 1$, the second one, 
Table~\ref{S120Cliff basis5+1even I.}, collects the remaining eight 
members with ${\cal S}^{12}=0$.

 %
%******\\02.04.2026\\

% G  
\begin{table*}
\begin{small}
%\begin{center}
\caption{\label{Table Clifffourplet.}  This table, taken from~\cite{n2023NPB}, 
represents for the internal space $d=(5+1)$ $2^d=64$ ``eigenvectors" of 
the Cartan subalgebra, Eq.~(\ref{cartangrasscliff}), members of the odd and 
even ``basis vectors'' which are the superposition of odd and even products 
of $\gamma^{a}$'s  in $d=(5+1)$-dimensional internal space. Table is 
divided into four groups. The first group, $odd \,I$, is (chosen) to represent 
``basis vectors", ${\hat b}^{m \dagger}_f$, appearing in 
$2^{\frac{d}{2}-1}=4$ ``families" ($f=1,2,3,4$), each ''family'' having 
$2^{\frac{d}{2}-1}=4$ ``family'' members ($m=1,2,3,4$).
The second group, $odd\,II$, contains Hermitian conjugate partners of the 
first group for each ``family'' separately, ${\hat b}^{m}_f=$
$({\hat b}^{m \dagger}_f)^{\dagger}$. The $odd \,I$ or $odd \,II$ are 
products of an odd number of nilpotents (one or three) and projectors (two 
or none).
The ``family" quantum numbers of ${\hat b}^{m \dagger}_f$, that is the 
eigenvalues of $(\tilde{S}^{03}, \tilde{S}^{12},\tilde{S}^{56})$, appear 
for the first {\it odd I } group, and the two last {\it even I} and {\it even II}
groups above each ``family", the quantum numbers of the ``family'' 
members $(S^{03}, S^{12}, S^{56})$ are 
%in these three cases
written in the last three columns. % before the last two columns.
For the Hermitian conjugated partners of {\it odd I}, presented in the group 
{\it odd II}, the quantum numbers $(S^{03}, S^{12}, S^{56})$ are 
presented above each group of the Hermitian conjugate partners, the 
last three columns %before the last two
tell eigenvalues of $(\tilde{S}^{03}, \tilde{S}^{12},\tilde{S}^{56})$.
Each of the two groups with the even number of $\gamma^a$'s,
{\it even \,I} and {\it even \,II},  has their Hermitian conjugated partners 
within its group.
The quantum numbers $f$, that is the eigenvalues of $(\tilde{S}^{03}, 
\tilde{S}^{12},\tilde{S}^{56})$, are written above each column of
four members, the quantum numbers of the members, $(S^{03}, 
S^{12}, S^{56})$, are written in the last three columns. The quantum 
numbers of $({\cal {\bf S}}^{03}, {\cal {\bf S}}^{12}, {\cal {\bf S}}^{56})$, 
for the even two groups, the {\it even \,I} and {\it even \,II},  
are ${\cal {\bf S}}^{ab}$ $= S^{ab} + \tilde{S}^{ab} $,which are for all projectors, according to Eq.~(\ref{calsab})	, equal zero and for nilpotents equal
to $\pm i$ or  $\pm 1$.
 \vspace{2mm}}
 \end{small}
\begin{tiny}
\begin{center}
  \begin{tabular}{|c|c|c|c|c|c|r|r|r|}%r|r|}
\hline
$ $&$$&$ $&$ $&$ $&&$$&$$&$$\\% &&\\
$''basis\, vectors'' $&$m$&$ f=1$&$ f=2 $&$ f=3 $&
$ f=4 $&$$&$$&$$\\ %&$$&$$\\
$(\tilde{S}^{03}, \tilde{S}^{12}, \tilde{S}^{56})$&$\rightarrow$&$(\frac{i}{2},- \frac{1}{2},-\frac{1}{2})$&$(-\frac{i}{2},-\frac{1}{2},\frac{1}{2})$&
$(-\frac{i}{2},\frac{1}{2},-\frac{1}{2})$&$(\frac{i}{2},\frac{1}{2},\frac{1}{2})$&$S^{03}$
 &$S^{12}$&$S^{56}$\\ %&$\Gamma^{(5+1)}$&$\Gamma^{(3+1)}$\\
%$(\tilde{S}^{03}, \tilde{S}^{12}, \tilde{S}^{56})$&$\rightarrow$& 
%$\stackrel{03}{\;\,}\;\;\,\stackrel{12}{\;\,}\;\;\,\stackrel{56}{\;\,}$&
%$\stackrel{03}{\;\,}\;\;\,\stackrel{12}{\;\,}\;\;\,\stackrel{56}{\;\,}$&
%$\stackrel{03}{\;\,}\;\;\,\stackrel{12}{\;\,}\;\;\,\stackrel{56}{\;\,}$&
%$\stackrel{03}{\;\,}\;\;\,\stackrel{12}{\;\,}\;\;\,\stackrel{56}{\;\,}$&
%&&&&\\
\hline
$ $&$$&$ $&$ $&$ $&&$$&$$&$$\\ %&&\\
$odd \,I\; {\hat b}^{m \dagger}_f$&$1$& 
$\stackrel{03}{(+i)}\stackrel{12}{[+]}\stackrel{56}{[+]}$&
                        $\stackrel{03}{[+i]}\stackrel{12}{[+]}\stackrel{56}{(+)}$ & 
                        $\stackrel{03}{[+i]}\stackrel{12}{(+)}\stackrel{56}{[+]}$ &  
                        $\stackrel{03}{(+i)}\stackrel{12}{(+)}\stackrel{56}{(+)}$ &
                        $\frac{i}{2}$&$\frac{1}{2}$&$\frac{1}{2}$\\ %&$1$&$1$\\
$$&$2$&    $[-i](-)[+] $ & $(-i)(-)(+) $ & $(-i)[-][+] $ & $[-i][-](+) $ &$-\frac{i}{2}$&
$-\frac{1}{2}$&$\frac{1}{2}$\\ %&$1$&$1$\\
$$&$3$&    $[-i] [+](-)$ & $(-i)[+][-] $ & $(-i)(+)(-) $ & $[-i](+)[-] $&$-\frac{i}{2}$&
$\frac{1}{2}$&$-\frac{1}{2}$\\ %&$1$&$-1$ \\
$$&$4$&    $(+i)(-)(-)$ & $[+i](-)[-] $ & $[+i][-](-) $ & $(+i)[-][-]$&$\frac{i}{2}$&
$-\frac{1}{2}$&$-\frac{1}{2}$\\ %&$1$&$-1$ \\
\hline
$ $&$$&$ $&$ $&$ $&&$$&$$&$$\\ %&&\\
$(S^{03}, S^{12}, S^{56})$&$\rightarrow$&$(-\frac{i}{2}, \frac{1}{2},\frac{1}{2})$&
$(\frac{i}{2},\frac{1}{2},-\frac{1}{2})$&
$(\frac{i}{2},- \frac{1}{2},\frac{1}{2})$&$(-\frac{i}{2},-\frac{1}{2},-\frac{1}{2})$&
$\tilde{S}^{03}$
&$\tilde{S}^{12}$&$\tilde{S}^{56}$\\ %&$\Gamma^{(5+1)}$&$\tilde{\Gamma}^{(3+1)}$\\
%$ $&$$&$ $&$ $&$ $&&$$&$$&$$&&\\
&&
$\stackrel{03}{\;\,}\;\;\,\stackrel{12}{\;\,}\;\;\,\stackrel{56}{\;\,}$&
$\stackrel{03}{\;\,}\;\;\,\stackrel{12}{\;\,}\;\;\,\stackrel{56}{\;\,}$&
$\stackrel{03}{\;\,}\;\;\,\stackrel{12}{\;\,}\;\;\,\stackrel{56}{\;\,}$&
$\stackrel{03}{\;\,}\;\;\,\stackrel{12}{\;\,}\;\;\,\stackrel{56}{\;\,}$&
&&\\% & %$\Gamma^{(5+1)}$ &$$\\
\hline
$ $&$$&$ $&$ $&$ $&&$$&$$&$$\\ %&&\\
$odd\,II\; {\hat b}^{m}_f$&$1$ &$(-i)[+][+]$ & $[+i][+](-)$ & $[+i](-)[+]$ & $(-i)(-)(-)$&
$-\frac{i}{2}$&$-\frac{1}{2}$&$-\frac{1}{2}$\\ %&$-1$&$1$ \\
$$&$2$&$[-i](+)[+]$ & $(+i)(+)(-)$ & $(+i)[-][+]$ & $[-i][-](-)$&
$\frac{i}{2}$&$\frac{1}{2}$&$-\frac{1}{2}$\\ % &$-1$&$1$ \\
$$&$3$&$[-i][+](+)$ & $(+i)[+][-]$ & $(+i)(-)(+)$ & $[-i](-)[-]$&
$\frac{i}{2}$&$-\frac{1}{2}$&$\frac{1}{2}$\\ %5 &$-1$&$-1$ \\
$$&$4$&$(-i)(+)(+)$ & $[+i](+)[-]$ & $[+i][-](+)$ & $(-i)[-][-]$&
$-\frac{i}{2}$&$\frac{1}{2}$&$\frac{1}{2}$\\ %&$-1$&$-1$ \\
\hline
&&&&&&&&\\ %&&\\
\hline
$ $&$$&$ $&$ $&$ $&&$$&$$&$$\\ %&&\\
$(\tilde{S}^{03}, \tilde{S}^{12}, \tilde{S}^{56})$&$\rightarrow$&
$(-\frac{i}{2},\frac{1}{2},\frac{1}{2})$&$(\frac{i}{2},-\frac{1}{2},\frac{1}{2})$&
$(-\frac{i}{2},-\frac{1}{2},-\frac{1}{2})$&$(\frac{i}{2},\frac{1}{2},-\frac{1}{2})$&
%$ $&
$S^{03}$&$S^{12}$&$S^{56}$\\ %&$\Gamma^{(5+1)}$&$\Gamma^{(3+1)}$\\ 
%&&
%$(-\frac{i}{2},\frac{1}{2},\frac{1}{2})$&$(\frac{i}{2},-\frac{1}{2},\frac{1}{2})$&
%$(-\frac{i}{2},-\frac{1}{2},-\frac{1}{2})$&$(\frac{i}{2},\frac{1}{2},-\frac{1}{2})$&
%&&&&\\
&& 
$\stackrel{03}{\;\,}\;\;\,\stackrel{12}{\;\,}\;\;\,\stackrel{56}{\;\,}$&
$\stackrel{03}{\;\,}\;\;\,\stackrel{12}{\;\,}\;\;\,\stackrel{56}{\;\,}$&

$\stackrel{03}{\;\,}\;\;\,\stackrel{12}{\;\,}\;\;\,\stackrel{56}{\;\,}$&
$\stackrel{03}{\;\,}\;\;\,\stackrel{12}{\;\,}\;\;\,\stackrel{56}{\;\,}$&
&&\\ %&&\\
\hline
$ $&$$&$ $&$ $&$ $&&$$&$$&$$\\ % &\\
$even\,I \; {}^{I}{\cal A}^{m}_f$&$1$&$[+i](+)(+) $ & $(+i)[+](+) $ & $[+i][+][+] $ & $(+i)(+)[+] $ &$\frac{i}{2}$&
$\frac{1}{2}$&$\frac{1}{2}$\\ %&$1$&$1$ \\
$$&$2$&$(-i)[-](+) $ & $[-i](-)(+) $ & $(-i)(-)[+] $ & $[-i][-][+] $ &$-\frac{i}{2}$&
$-\frac{1}{2}$&$\frac{1}{2}$\\ % &$1$&$1$ \\
$$&$3$&$(-i)(+)[-] $ & $[-i][+][-] $ & $(-i)[+](-) $ & $[-i](+)(-) $&$-\frac{i}{2}$&
$\frac{1}{2}$&$-\frac{1}{2}$\\ % &$1$&$-1$ \\
$$&$4$&$[+i][-][-] $ & $(+i)(-)[-] $ & $[+i](-)(-) $ & $(+i)[-](-) $&$\frac{i}{2}$&
$-\frac{1}{2}$&$-\frac{1}{2}$\\ % &$1$&$-1$ \\ 
\hline
$ $&$$&$ $&$ $&$ $&&$$&$$&$$\\ %&&\\
$(\tilde{S}^{03}, \tilde{S}^{12}, \tilde{S}^{56})$&$\rightarrow$&
$(\frac{i}{2},\frac{1}{2},\frac{1}{2})$&$(-\frac{i}{2},-\frac{1}{2},\frac{1}{2})$&
$(\frac{i}{2},-\frac{1}{2},-\frac{1}{2})$&$(-\frac{i}{2},\frac{1}{2},-\frac{1}{2})$&
%$ even\, II$&$m$&$ $&$$&$ $&$$&
$S^{03}$&$S^{12}$&$S^{56}$\\ %&$\Gamma^{(5+1)}$&$\Gamma^{(3+1)}$\\
%&&$(\frac{i}{2},\frac{1}{2},\frac{1}{2})$&$(-\frac{i}{2},-\frac{1}{2},\frac{1}{2})$&
%$(\frac{i}{2},-\frac{1}{2},-\frac{1}{2})$&$(-\frac{i}{2},\frac{1}{2},-\frac{1}{2})$&
%&&&&\\
&& 
$\stackrel{03}{\;\,}\;\;\,\stackrel{12}{\;\,}\;\;\,\stackrel{56}{\;\,}$&
$\stackrel{03}{\;\,}\;\;\,\stackrel{12}{\;\,}\;\;\,\stackrel{56}{\;\,}$&
$\stackrel{03}{\;\,}\;\;\,\stackrel{12}{\;\,}\;\;\,\stackrel{56}{\;\,}$&
$\stackrel{03}{\;\,}\;\;\,\stackrel{12}{\;\,}\;\;\,\stackrel{56}{\;\,}$&
&&\\ %&&\\
\hline
$ $&$$&$ $&$ $&$ $&&$$&$$&$$\\ %&&\\
$even\,II \; {}^{II}{\cal A}^{m}_f$&$1$& $[-i](+)(+) $ & $(-i)[+](+) $ & $[-i][+][+] $ & 
$(-i)(+)[+] $ &$-\frac{i}{2}$&
$\frac{1}{2}$&$\frac{1}{2}$\\ %&$-1$&$-1$ \\ 
$$&$2$&    $(+i)[-](+) $ & $[+i](-)(+) $ & $(+i)(-)[+] $ & $[+i][-][+] $ &$\frac{i}{2}$&
$-\frac{1}{2}$&$\frac{1}{2}$ \\ % &$-1$&$-1$ \\
$$&$3$&    $(+i)(+)[-] $ & $[+i][+][-] $ & $(+i)[+](-) $ & $[+i](+)(-) $&$\frac{i}{2}$&
$\frac{1}{2}$&$-\frac{1}{2}$\\ % &$-1$&$1$ \\
$$&$4$&    $[-i][-][-] $ & $(-i)(-)[-] $ & $[-i](-)(-) $ & $(-i)[-](-) $&$-\frac{i}{2}$&
$-\frac{1}{2}$&$-\frac{1}{2}$\\ %&$-1$&$1$ \\
\hline
 \end{tabular}
\end{center}
\end{tiny}
\end{table*}
%

%******\\02.04.2026\\

%%%%%%%%%%%%%%%%%%%%%%%%%%%
\begin{table}
\begin{tiny}
% \begin{center}
%\begin{minipage}[t]{16.5 cm}
%\begin{small}
\caption{% CHANGE!!
The even ``basis vectors'' ${}^{I}{\hat{\cal A}}^{m \dagger}_{f}$,
belonging to transverse momentum in internal space, ${\cal S}^{12}=$
$1$, the first half of ${}^{I}{\hat{\cal A}}^{m \dagger}_{f}$, and
${\cal S}^{12}= -1$, the
second half of ${}^{I}{\hat{\cal A}}^{m \dagger}_{f}$, for $d=(5+1)$, are
presented as algebraic products of the $f=1$ family ``basis vectors''
$\hat{b}^{m' \dagger}_{1}$ and their Hermitian conjugated partners
($\hat{b}^{m'' \dagger}_{1})^{\dagger}$: $\hat{b}^{m' \dagger}_{1} *_{A}$
($\hat{b}^{m'' \dagger}_{1})^{\dagger}$. Two
${}^{I}{\hat{\cal A}}^{m \dagger}_{f}$ which are the Hermitian conjugated
partners are marked with the same symbol ($\star \star$, $\ddagger$, $\otimes$,
$\odot \odot$).
The even ``basis vectors'' ${}^{I}{\hat{\cal A}}^{m \dagger}_{f}$
are products of one projector and two nilpotents, the odd ``basis
vectors'' and their Hermitian conjugated partners are products of one nilpotent
and two projectors or of three nilpotents. The even and odd
objects are eigenvectors of all the corresponding Cartan subalgebra members,
Eq.~(\ref{cartangrasscliff}). There are $\frac{1}{2} \times 2^{\frac{6}{2}-1}
\times 2^{\frac{6}{2}-1}$ algebraic products of $\hat{b}^{m' \dagger}_{1} *_{A}$
($\hat{b}^{m'' \dagger}_{1})^{\dagger}$ with ${\cal S}^{12}$ equal
to $ 1$ or $-1$ . The rest $8$ of $16$ members
present ${}^{I}{\hat{\cal A}}^{m \dagger}_{f}$ with ${\cal S}^{12}=0$.
The members $\hat{b}^{m' \dagger}_{f}$ together with their Hermitian
conjugated partners of each of the four families, $f=(1,2,3,4)$, offer the
same ${}^{I}{\hat{\cal A}}^{m \dagger}_{f}$ with ${\cal S}^{12}=\pm1$
as the ones presented in this table.
(And equivalently for ${\cal S}^{12}=0$.) Table is taken from
Ref.~\cite{nh2025Bled}.
\vspace{3mm}}
\label{transverseCliff basis5+1even I.} % \label{Cliff basis5+1even I.}
%\label{Clifford even basis5+1 I.}\ref{Cliff basis5+1even II.}
 %
 %\begin{tiny}
 \begin{center}
 \begin{tabular}{ |c| c| c c|}
 \hline
 $$&$$&$$&$$\\
 %$$&$$&$$&$$&$$&$$&$$&$$&$$&$$%&$$&$$
%$\, f $&
${\cal S}^{12} $&$symbol$&${}^{I}\hat{\cal A}^{m \dagger}_f=$
&$\hat{b}^{m' \dagger}_{f `} *_A (\hat{b}^{m'' \dagger}_{f `})^{\dagger}$\\
%$$&$$&$$&$$\\
\\
\hline
%
% 1
$$&$$&$$&$$\\
$1$&$\star \star$&${}^{I}\hat{\cal A}^{1 \dagger}_1=$&
$\hat{b}^{1 \dagger}_{1} *_{A} (\hat{b}^{4 \dagger}_{1})^{\dagger}$\\
$$&$$&$$&$$\\
$ $&$$ &$\stackrel{03}{[+i]}\,\stackrel{12}{(+)} \stackrel{56}{(+)}$&
$\stackrel{03}{(+i)}\,\stackrel{12}{[+]} \stackrel{56}{[+]} *_{A} 
\stackrel{03}{(-i)}\,\stackrel{12}{(+)} \stackrel{56}{(+)}$\\
\hline
% 2
$$&$$&$$&$$\\
$1$&$\ddagger$&${}^{I}\hat{\cal A}^{3 \dagger}_1=$&
$\hat{b}^{3 \dagger}_{1} *_{A} (\hat{b}^{4 \dagger}_{1})^{\dagger}$\\
$$&$$&$$&$$\\
$ $&$$ &$\stackrel{03}{(-i)}\,\stackrel{12}{(+)} \stackrel{56}{[-]}$&
$\stackrel{03}{[-i]}\,\stackrel{12}{[+]} \stackrel{56}{(-)} *_{A} 
\stackrel{03}{(-i)}\,\stackrel{12}{(+)} \stackrel{56}{(+)}$\\
\hline
% 3
$$&$$&$$&$$\\
$1$&$\odot \odot$&${}^{I}\hat{\cal A}^{1 \dagger}_4=$&
$\hat{b}^{1 \dagger}_{1} *_{A} (\hat{b}^{2 \dagger}_{1})^{\dagger}$\\
$$&$$&$$&$$\\
$ $&$$ &$\stackrel{03}{(+i)}\,\stackrel{12}{(+)} \stackrel{56}{[+]}$&
$\stackrel{03}{(+i)}\,\stackrel{12}{[+]} \stackrel{56}{[+]} *_{A} 
\stackrel{03}{[-i]}\,\stackrel{12}{(+)} \stackrel{56}{[+]}$\\
\hline
% 4
$$&$$&$$&$$\\
$1$&$\otimes$&${}^{I}\hat{\cal A}^{3 \dagger}_4=$&
$\hat{b}^{3 \dagger}_{1} *_{A} (\hat{b}^{2 \dagger}_{1})^{\dagger}$\\
$$&$$&$$&$$\\
$ $&$$ &$\stackrel{03}{[-i]}\,\stackrel{12}{(+)} \stackrel{56}{(-)}$&
$\stackrel{03}{[-i]}\,\stackrel{12}{[+]} \stackrel{56}{(-)} *_{A} 
\stackrel{03}{[-i]}\,\stackrel{12}{(+)} \stackrel{56}{[+]}$\\
\hline
\hline
% 1
$$&$$&$$&$$\\
$-1$&$\otimes$&${}^{I}\hat{\cal A}^{2 \dagger}_2=$&
$\hat{b}^{2 \dagger}_{1} *_{A} (\hat{b}^{3 \dagger}_{1})^{\dagger}$\\
$$&$$&$$&$$\\
$ $&$$ &$\stackrel{03}{[-i]}\,\stackrel{12}{(-)} \stackrel{56}{(+)}$&
$\stackrel{03}{[-i]}\,\stackrel{12}{(-)} \stackrel{56}{[+]} *_{A} 
\stackrel{03}{[-i]}\,\stackrel{12}{[+]} \stackrel{56}{(+)}$\\
\hline
% 2
$$&$$&$$&$$\\
$-1$&$\ddagger$&${}^{I}\hat{\cal A}^{4 \dagger}_2=$&
$\hat{b}^{4 \dagger}_{1} *_{A} (\hat{b}^{3 \dagger}_{1})^{\dagger}$\\
$$&$$&$$&$$\\
$ $&$$ &$\stackrel{03}{(+i)}\,\stackrel{12}{(-)} \stackrel{56}{[-]}$&
$\stackrel{03}{(+i)}\,\stackrel{12}{(-)} \stackrel{56}{(-)} *_{A} 
\stackrel{03}{[-i]}\,\stackrel{12}{[+]} \stackrel{56}{(+)}$\\
\hline
% 3
$$&$$&$$&$$\\
$-1$&$\odot \odot$&${}^{I}\hat{\cal A}^{2 \dagger}_3=$&
$\hat{b}^{2 \dagger}_{1} *_{A} (\hat{b}^{1 \dagger}_{1})^{\dagger}$\\
$$&$$&$$&$$\\
$ $&$$ &$\stackrel{03}{(-i)}\,\stackrel{12}{(-)} \stackrel{56}{[+]}$&
$\stackrel{03}{[-i]}\,\stackrel{12}{(-)} \stackrel{56}{[+]} *_{A} 
\stackrel{03}{(-i)}\,\stackrel{12}{[+]} \stackrel{56}{[+]}$\\
\hline
% 4
$$&$$&$$&$$\\
$-1$&$\star \star$&${}^{I}\hat{\cal A}^{4 \dagger}_3=$&
$\hat{b}^{4 \dagger}_{1} *_{A} (\hat{b}^{1 \dagger}_{1})^{\dagger}$\\
$$&$$&$$&$$\\
$ $&$$ &$\stackrel{03}{[+i]}\,\stackrel{12}{(-)} \stackrel{56}{(-)}$&
$\stackrel{03}{(+i)}\,\stackrel{12}{(-)} \stackrel{56}{(-)} *_{A} 
\stackrel{03}{(-i)}\,\stackrel{12}{[+]} \stackrel{56}{[+]}$\\
\hline
 \end{tabular}
 \end{center}
\end{tiny}
%\end{small}
\end{table}
%\end{document}
%
%*
% G Done!
%CHECKED 11.04.2024.
\begin{table}
\begin{tiny}
% \begin{center}
%\begin{minipage}[t]{16.5 cm}
%\begin{small}
\caption{
The even ``basis vectors'' ${}^{I}{\hat{\cal A}}^{m \dagger}_{f}$,
belonging to zero momentum in internal space, ${\cal S}^{12}=$ $0$,
for $d=(5+1)$, are presented as algebraic products of the $f=1$ family
``basis vectors'' $\hat{b}^{m' \dagger}_{1}$ and their Hermitian conjugated
partners ($\hat{b}^{m'' \dagger}_{1})^{\dagger}$:
$\hat{b}^{m' \dagger}_{1} *_{A}$ ($\hat{b}^{m'' \dagger}_{1})^{\dagger}$.
The two ${}^{I}{\hat{\cal A}}^{m \dagger}_{f}$ which are Hermitian
conjugated partners, are marked with the same symbol (either $\bigtriangleup$ or $\bullet$). The symbol $\bigcirc$ presents selfadjoint members, four of eight.
The even ``basis vectors'' ${}^{I}{\hat{\cal A}}^{m \dagger}_{f}$
are products of one projector and two nilpotents or three projectors (they are
self-adjoint), the odd ``basis
vectors'' and their Hermitian conjugated partners are products of one nilpotent
and two projectors or of three nilpotents. The even and odd
objects are eigenvectors of all the corresponding Cartan subalgebra members,
Eq.~(\ref{cartangrasscliff}). There are $\frac{1}{2} \times 2^{\frac{6}{2}-1}
\times 2^{\frac{6}{2}-1}$ algebraic products of $\hat{b}^{m' \dagger}_{1}
*_{A}$ ($\hat{b}^{m'' \dagger}_{1})^{\dagger}$. The rest $8$ of $16$
members have ${}^{I}{\hat{\cal A}}^{m \dagger}_{f}$ with
${\cal S}^{12}=+ 1$ (four) and with ${\cal S}^{12}=- 1$ (four), present
in Table~\ref{transverseCliff basis5+1even I.}.
The members $\hat{b}^{m' \dagger}_{f}$ together with their Hermitian
conjugated partners of each of the four families, $f=(1,2,3,4)$, offer the
same ${}^{I}{\hat{\cal A}}^{m \dagger}_{f}$ with ${\cal S}^{12}=0$
as the ones presented in this table. The table is taken from
Ref.~\cite{nh2025Bled}.
\vspace{3mm}}
\label{S120Cliff basis5+1even I.} 
% \label{Cliff basis5+1even I.} 
% \label{Cliff basis5+1even I.}
%\label{Clifford even basis5+1 I.}\ref{Cliff basis5+1even II.}
 %
 %\begin{tiny}
 \begin{center}
 \begin{tabular}{ |c| c| c c|}
 \hline
 $$&$$&$$&$$\\
 %$$&$$&$$&$$&$$&$$&$$&$$&$$&$$%&$$&$$
%$\, f $&
${\cal S}^{12} $&$symbol$&${}^{I}\hat{\cal A}^{m \dagger}_f=$
&$\hat{b}^{m' \dagger}_{f `} *_A (\hat{b}^{m'' \dagger}_{f `})^{\dagger}$\\
%$$&$$&$$&$$\\
\\
\hline
%
% 1
$$&$$&$$&$$\\
$0$&$\bigtriangleup$&${}^{I}\hat{\cal A}^{2 \dagger}_1=$&
$\hat{b}^{2 \dagger}_{1} *_{A} (\hat{b}^{4 \dagger}_{1})^{\dagger}$\\
$$&$$&$$&$$\\
$ $&$$ &$\stackrel{03}{(-i)}\,\stackrel{12}{[-]} \stackrel{56}{(+)}$&
$\stackrel{03}{[-i]}\,\stackrel{12}{(-)} \stackrel{56}{[+]} *_{A} 
\stackrel{03}{(-i)}\,\stackrel{12}{(+)} \stackrel{56}{(+)}$\\
\hline
% 2
$$&$$&$$&$$\\
$0$&$\bigcirc$&${}^{I}\hat{\cal A}^{4 \dagger}_1=$&
$\hat{b}^{4 \dagger}_{1} *_{A} (\hat{b}^{4 \dagger}_{1})^{\dagger}$\\
$$&$$&$$&$$\\
$ $&$$ &$\stackrel{03}{[+i]}\,\stackrel{12}{[-]} \stackrel{56}{[-]}$&
$\stackrel{03}{(+i)}\,\stackrel{12}{(-)} \stackrel{56}{(-)} *_{A} 
\stackrel{03}{(-i)}\,\stackrel{12}{(+)} \stackrel{56}{(+)}$\\
\hline
% 3
$$&$$&$$&$$\\
$0$&$\bullet$&${}^{I}\hat{\cal A}^{1 \dagger}_2=$&
$\hat{b}^{1 \dagger}_{1} *_{A} (\hat{b}^{3 \dagger}_{1})^{\dagger}$\\
$$&$$&$$&$$\\
$ $&$$ &$\stackrel{03}{(+i)}\,\stackrel{12}{[+]} \stackrel{56}{(+)}$&
$\stackrel{03}{(+i)}\,\stackrel{12}{[+]} \stackrel{56}{[+]} *_{A} 
\stackrel{03}{[-i]}\,\stackrel{12}{[+]} \stackrel{56}{(+)}$\\
\hline
% 4
$$&$$&$$&$$\\
$0$&$\bigcirc$&${}^{I}\hat{\cal A}^{3 \dagger}_2=$&
$\hat{b}^{3 \dagger}_{1} *_{A} (\hat{b}^{3 \dagger}_{1})^{\dagger}$\\
$$&$$&$$&$$\\
$ $&$$ &$\stackrel{03}{[-i]}\,\stackrel{12}{[+]} \stackrel{56}{[-]}$&
$\stackrel{03}{[-i]}\,\stackrel{12}{[+]} \stackrel{56}{(-)} *_{A} 
\stackrel{03}{[-i]}\,\stackrel{12}{[+]} \stackrel{56}{(+)}$\\
\hline
\hline
% 1
$$&$$&$$&$$\\
$0$&$\bigcirc$&${}^{I}\hat{\cal A}^{1 \dagger}_3=$&
$\hat{b}^{1 \dagger}_{1} *_{A} (\hat{b}^{1 \dagger}_{1})^{\dagger}$\\
$$&$$&$$&$$\\
$ $&$$ &$\stackrel{03}{[+i]}\,\stackrel{12}{[+]} \stackrel{56}{[+]}$&
$\stackrel{03}{(+i)}\,\stackrel{12}{[+]} \stackrel{56}{[+]} *_{A} 
\stackrel{03}{(-i)}\,\stackrel{12}{[+]} \stackrel{56}{[+]}$\\
\hline
% 2
$$&$$&$$&$$\\
$0$&$\bullet$&${}^{I}\hat{\cal A}^{3 \dagger}_3=$&
$\hat{b}^{3 \dagger}_{1} *_{A} (\hat{b}^{1 \dagger}_{1})^{\dagger}$\\
$$&$$&$$&$$\\
$ $&$$ &$\stackrel{03}{(-i)}\,\stackrel{12}{[+]} \stackrel{56}{(-)}$&
$\stackrel{03}{[-i]}\,\stackrel{12}{[+]} \stackrel{56}{(-)} *_{A} 
\stackrel{03}{(-i)}\,\stackrel{12}{[+]} \stackrel{56}{[+]}$\\
\hline
% 3
$$&$$&$$&$$\\
$0$&$\bigcirc$&${}^{I}\hat{\cal A}^{2 \dagger}_4=$&
$\hat{b}^{2 \dagger}_{1} *_{A} (\hat{b}^{2 \dagger}_{1})^{\dagger}$\\
$$&$$&$$&$$\\
$ $&$$ &$\stackrel{03}{[-i]}\,\stackrel{12}{[-]} \stackrel{56}{[+]}$&
$\stackrel{03}{[-i]}\,\stackrel{12}{(-)} \stackrel{56}{[+]} *_{A} 
\stackrel{03}{[-i]}\,\stackrel{12}{(+)} \stackrel{56}{[+]}$\\
\hline
% 4
$$&$$&$$&$$\\
$0$&$\bigtriangleup$&${}^{I}\hat{\cal A}^{4 \dagger}_4=$&
$\hat{b}^{4 \dagger}_{1} *_{A} (\hat{b}^{2 \dagger}_{1})^{\dagger}$\\
$$&$$&$$&$$\\
$ $&$$ &$\stackrel{03}{(+i)}\,\stackrel{12}{[-]} \stackrel{56}{(-)}$&
$\stackrel{03}{(+i)}\,\stackrel{12}{(-)} \stackrel{56}{(-)} *_{A} 
\stackrel{03}{[-i]}\,\stackrel{12}{(+)} \stackrel{56}{[+]}$\\
\hline
 \end{tabular}
 \end{center}
\end{tiny}
%\end{small}
\end{table}
\begin{table}
\begin{tiny}           
\caption{
The even ``basis vectors'' ${}^{II}{\hat{\cal A}}^{m \dagger}_{f}$,
belonging to transverse momentum in internal space, ${\cal S}^{12}=$ $1$,
the first half ${}^{II}{\hat{\cal A}}^{m \dagger}_{f}$, and
${\cal S}^{12}=-1$, the
second half ${}^{II}{\hat{\cal A}}^{m \dagger}_{f}$, for $d=(5+1)$, are
presented as algebraic products of the first, $m=1$, member of ``basis
vectors'' $\hat{b}^{m'=1 \dagger}_{f '}$ and the Hermitian conjugated
partners ($\hat{b}^{m' =1 \dagger}_{f ''})^{\dagger}$. Two
${}^{II}{\hat{\cal A}}^{m \dagger}_{f}$ which are the Hermitian
conjugated partners are marked with the same symbol.
The even ``basis vectors'' ${}^{II}{\hat{\cal A}}^{m \dagger}_{f}$
are products of one projector and two nilpotents, the odd ``basis
vectors'' and the Hermitian conjugated partners are products of one nilpotent
and two projectors or of three nilpotents. Even and odd
objects are eigenvectors of the corresponding Cartan subalgebra members,
Eq.~(\ref{cartangrasscliff}). There are $2^{\frac{6}{2}-1}\times
2^{\frac{6}{2}-1}$ algebraic products of
$(\hat{b}^{m' \dagger}_{f `})^{\dagger}$ and
$\hat{b}^{m' \dagger}_{f ``}$, $f `$ and $f ``$ run over all
four families. The rest of the $16$ members present
${}^{II}{\hat{\cal A}}^{m \dagger}_{f}$ with ${\cal S}^{12}=0$.
The members $(\hat{b}^{m' \dagger}_{f `})^{\dagger}$ together with
$\hat{b}^{m' \dagger}_{f `'}$, $m'=(1,2,3,4)$, offer the
same ${}^{II}{\hat{\cal A}}^{m \dagger}_{f}$ with
${\cal S}^{12}=\pm1$ as the ones presented in this table.
(And equivalently for ${\cal S}^{12}=0$.) The table is taken from
Ref.~\cite{nh2025Bled}.
\vspace{3mm}}
\label{transverseCliff basis5+1even II.} % \label{Cliff basis5+1even I.}
%\label{Clifford even basis5+1 I.}\ref{Cliff basis5+1even II.}
 %
 %\begin{tiny}
 \begin{center}
 \begin{tabular}{ |c| c| c c|}
 \hline
 $$&$$&$$&$$\\
 %$$&$$&$$&$$&$$&$$&$$&$$&$$&$$%&$$&$$
%$\, f $&
${\cal S}^{12} $&$symbol$&${}^{II}\hat{\cal A}^{m \dagger}_f=$
&$(\hat{b}^{1 \dagger}_{f `})^{\dagger} *_A \hat{b}^{1 \dagger}_{f ``}$\\
%$$&$$&$$&$$\\
\\
\hline
%
% CORRECT 
% 1
$$&$$&$$&$$\\
$1$&$\star \star$&${}^{II}\hat{\cal A}^{1 \dagger}_1=$&
$(\hat{b}^{1 \dagger}_{1})^{\dagger} *_{A} \hat{b}^{1 \dagger}_{4}$\\
$$&$$&$$&$$\\
$ $&$$ &$\stackrel{03}{[-i]}\,\stackrel{12}{(+)} \stackrel{56}{(+)}$&
$\stackrel{03}{(-i)}\,\stackrel{12}{[+]} \stackrel{56}{[+]} *_{A} 
\stackrel{03}{(+i)}\,\stackrel{12}{(+)} \stackrel{56}{(+)}$\\
\hline
% 2
$$&$$&$$&$$\\
$1$&$\odot \odot$&${}^{II}\hat{\cal A}^{3 \dagger}_1=$&
$(\hat{b}^{1 \dagger}_{2})^{\dagger} *_{A} \hat{b}^{1\dagger}_{4}$\\
$$&$$&$$&$$\\
$ $&$$ &$\stackrel{03}{(+i)}\,\stackrel{12}{(+)} \stackrel{56}{[-]}$&
$\stackrel{03}{[+i]}\,\stackrel{12}{[+]} \stackrel{56}{(-)} *_{A} 
\stackrel{03}{(+i)}\,\stackrel{12}{(+)} \stackrel{56}{(+)}$\\
\hline
% 3
$$&$$&$$&$$\\
$1$&$\ddagger$&${}^{II}\hat{\cal A}^{1 \dagger}_4=$&
$(\hat{b}^{1 \dagger}_{1})^{\dagger} *_{A} \hat{b}^{1 \dagger}_{3}$\\
$$&$$&$$&$$\\
$ $&$$ &$\stackrel{03}{(-i)}\,\stackrel{12}{(+)} \stackrel{56}{[+]}$&
$\stackrel{03}{(-i)}\,\stackrel{12}{[+]} \stackrel{56}{[+]} *_{A} 
\stackrel{03}{[+i]}\,\stackrel{12}{(+)} \stackrel{56}{[+]}$\\
\hline
% 4
$$&$$&$$&$$\\
$1$&$\otimes$&${}^{II}\hat{\cal A}^{3 \dagger}_4=$&
$(\hat{b}^{1 \dagger}_{2})^{\dagger} *_{A} \hat{b}^{1 \dagger}_{3}$\\
$$&$$&$$&$$\\
$ $&$$ &$\stackrel{03}{[+i]}\,\stackrel{12}{(+)} \stackrel{56}{(-)}$&
$\stackrel{03}{[+i]}\,\stackrel{12}{[+]} \stackrel{56}{(-)} *_{A} 
\stackrel{03}{[+i]}\,\stackrel{12}{(+)} \stackrel{56}{[+]}$\\
\hline
\hline
% 1
$$&$$&$$&$$\\
$-1$&$\otimes$&${}^{II}\hat{\cal A}^{2 \dagger}_2=$&
$(\hat{b}^{1 \dagger}_{3})^{\dagger} *_{A} \hat{b}^{1 \dagger}_{2}$\\
$$&$$&$$&$$\\
$ $&$$ &$\stackrel{03}{[+i]}\,\stackrel{12}{(-)} \stackrel{56}{(+)}$&
$\stackrel{03}{[+i]}\,\stackrel{12}{(-)} \stackrel{56}{[+]} *_{A} 
\stackrel{03}{[+i]}\,\stackrel{12}{[+]} \stackrel{56}{(+)}$\\
\hline
% 2
$$&$$&$$&$$\\
% HERE CORRECTED $\otimes \otimes$ into $\odot \odot$
$-1$&$\odot \odot$&${}^{II}\hat{\cal A}^{4 \dagger}_2=$&
$(\hat{b}^{1 \dagger}_{4})^{\dagger} *_{A} \hat{b}^{1 \dagger}_{2}$\\
$$&$$&$$&$$\\
$ $&$$ &$\stackrel{03}{(-i)}\,\stackrel{12}{(-)} \stackrel{56}{[-]}$&
$\stackrel{03}{(-i)}\,\stackrel{12}{(-)} \stackrel{56}{(-)} *_{A} 
\stackrel{03}{[+i]}\,\stackrel{12}{[+]} \stackrel{56}{(+)}$\\
\hline
% 3
$$&$$&$$&$$\\
$-1$&$\ddagger$&${}^{II} \hat{\cal A}^{2 \dagger}_3=$&
$(\hat{b}^{1 \dagger}_{3})^{\dagger} *_{A} \hat{b}^{1 \dagger}_{1}$\\
$$&$$&$$&$$\\
$ $&$$ &$\stackrel{03}{(+i)}\,\stackrel{12}{(-)} \stackrel{56}{[+]}$&
$\stackrel{03}{[+i]}\,\stackrel{12}{(-)} \stackrel{56}{[+]} *_{A} 
\stackrel{03}{(+i)}\,\stackrel{12}{[+]} \stackrel{56}{[+]}$\\
\hline
% 4
$$&$$&$$&$$\\
$-1$&$\star \star$&${}^{II}\hat{\cal A}^{4 \dagger}_3=$&
$(\hat{b}^{1 \dagger}_{4})^{\dagger} *_{A} \hat{b}^{1 \dagger}_{1}$\\
$$&$$&$$&$$\\
$ $&$$ &$\stackrel{03}{[-i]}\,\stackrel{12}{(-)} \stackrel{56}{(-)}$&
$\stackrel{03}{(-i)}\,\stackrel{12}{(-)} \stackrel{56}{(-)} *_{A} 
\stackrel{03}{(+i)}\,\stackrel{12}{[+]} \stackrel{56}{[+]}$\\
\hline
 \end{tabular}
 \end{center}
\end{tiny}
%\end{small}
\end{table}
%\end{document}
%

%
\begin{table}
\begin{tiny}
% \begin{center}
%\begin{minipage}[t]{16.5 cm}
%\begin{small}
\caption{
The even ``basis vectors'' ${}^{II}{\hat{\cal A}}^{m \dagger}_{f}$,
belonging to %transverse momentum in internal space, 
${\cal S}^{12}=0$ in internal space,  for $d=(5+1)$, are 
presented as algebraic products of the first, $m=1$, member of ``basis 
vectors'' $\hat{b}^{m'=1 \dagger}_{f '}$ and the Hermitian conjugated 
partners ($\hat{b}^{m' =1 \dagger}_{f ''})^{\dagger}$. The Hermitian 
conjugated partners  of two ${}^{II}{\hat{\cal A}}^{m \dagger}_{f}$ are 
marked with the same symbol (either $\bigtriangleup$ or $\bullet$). The
symbol $\bigcirc$ presents four selfadjoint members.
The even ``basis vectors''  ${}^{II}{\hat{\cal A}}^{m \dagger}_{f}$
are the products of one projector and two nilpotents, or of three projectors
(they are self adjoint), the odd ``basis 
vectors'' and the Hermitian conjugated partners are products of one nilpotent 
and two projectors or of three nilpotents. Even and odd
objects are eigenvectors of all the corresponding Cartan subalgebra members, 
Eq.~(\ref{cartangrasscliff}).  There are $\frac{1}{2}\times 2^{\frac{6}{2}-1}
\times 2^{\frac{6}{2}-1}$ algebraic products of $
(\hat{b}^{m'\dagger}_{f `})^{\dagger} *_{A}$  
$\hat{b}^{m' \dagger}_{f ``}$, $f `$ and $f ``$ run over all
four families. The rest of $16$ members present
${}^{II}{\hat{\cal A}}^{m \dagger}_{f}$ with ${\cal S}^{12}=\pm 1$.
The members $(\hat{b}^{m' \dagger}_{f `})^{\dagger}$ together with
$\hat{b}^{m' \dagger}_{f ``}$ $m'=(1,2,3,4)$, offer the same 
 ${}^{II}{\hat{\cal A}}^{m \dagger}_{f}$, all with ${\cal S}^{12}=0$.
 Table is taken from ~\cite{nh2025Bled}.
\vspace{3mm}}
\label{S120Cliff basis5+1even II.} % \label{Cliff basis5+1even I.}
%\label{Clifford even basis5+1 I.}\ref{Cliff basis5+1even II.}
 %
 %\begin{tiny}
 \begin{center}
 \begin{tabular}{ |c| c| c c|}
 \hline
 $$&$$&$$&$$\\
 %$$&$$&$$&$$&$$&$$&$$&$$&$$&$$%&$$&$$
%$\, f $&
${\cal S}^{12} $&$symbol$&${}^{II}\hat{\cal A}^{m \dagger}_f=$
&$(\hat{b}^{1 \dagger}_{f `})^{\dagger} *_A \hat{b}^{1 \dagger}_{f ``}$\\
%$$&$$&$$&$$\\
\\
\hline
%
%CORRECT
% 1
$$&$$&$$&$$\\
$0$&$\bigtriangleup$&${}^{II}\hat{\cal A}^{2 \dagger}_1=$&
$(\hat{b}^{1 \dagger}_{3})^{\dagger} *_{A} \hat{b}^{1 \dagger}_{4}$\\
$$&$$&$$&$$\\
$ $&$$ &$\stackrel{03}{(+i)}\,\stackrel{12}{[-]} \stackrel{56}{(+)}$&
$\stackrel{03}{[+i]}\,\stackrel{12}{(-)} \stackrel{56}{[+]} *_{A} 
\stackrel{03}{(+i)}\,\stackrel{12}{(+)} \stackrel{56}{(+)}$\\
\hline
% 2
$$&$$&$$&$$\\
$0$&$\bigcirc$&${}^{II}\hat{\cal A}^{4 \dagger}_1=$&
$(\hat{b}^{1 \dagger}_{4})^{\dagger} *_{A} \hat{b}^{1\dagger}_{4}$\\
$$&$$&$$&$$\\
$ $&$$ &$\stackrel{03}{[-i]}\,\stackrel{12}{[-]} \stackrel{56}{[-]}$&
$\stackrel{03}{(-i)}\,\stackrel{12}{(-)} \stackrel{56}{(-)} *_{A} 
\stackrel{03}{(+i)}\,\stackrel{12}{(+)} \stackrel{56}{(+)}$\\
\hline
% 3
$$&$$&$$&$$\\
$0$&$\bullet$&${}^{II}\hat{\cal A}^{1 \dagger}_2=$&
$(\hat{b}^{1 \dagger}_{1})^{\dagger} *_{A} \hat{b}^{1 \dagger}_{2}$\\
$$&$$&$$&$$\\
$ $&$$ &$\stackrel{03}{(-i)}\,\stackrel{12}{[+]} \stackrel{56}{(+)}$&
$\stackrel{03}{(-i)}\,\stackrel{12}{[+]} \stackrel{56}{[+]} *_{A} 
\stackrel{03}{[+i]}\,\stackrel{12}{[+]} \stackrel{56}{(+)}$\\
\hline
% 4
$$&$$&$$&$$\\
$0$&$\bigcirc$&${}^{II}\hat{\cal A}^{3 \dagger}_2=$&
$(\hat{b}^{1 \dagger}_{2})^{\dagger} *_{A} \hat{b}^{1 \dagger}_{2}$\\
$$&$$&$$&$$\\
$ $&$$ &$\stackrel{03}{[+i]}\,\stackrel{12}{[+]} \stackrel{56}{[-]}$&
$\stackrel{03}{[+i]}\,\stackrel{12}{[+]} \stackrel{56}{(-)} *_{A} 
\stackrel{03}{[+i]}\,\stackrel{12}{[+]} \stackrel{56}{(+)}$\\
\hline
\hline
% 1
$$&$$&$$&$$\\
$0$&$\bigcirc$&${}^{II}\hat{\cal A}^{1 \dagger}_3=$&
$(\hat{b}^{1 \dagger}_{1})^{\dagger} *_{A} \hat{b}^{1 \dagger}_{1}$\\
$$&$$&$$&$$\\
$ $&$$ &$\stackrel{03}{[-i]}\,\stackrel{12}{[+]} \stackrel{56}{[+]}$&
$\stackrel{03}{(-i)}\,\stackrel{12}{([+]} \stackrel{56}{[+]} *_{A} 
\stackrel{03}{(+i)}\,\stackrel{12}{[+]} \stackrel{56}{[+]}$\\
\hline
% 2
$$&$$&$$&$$\\
$0$&$\bullet$&${}^{II}\hat{\cal A}^{3 \dagger}_3=$&
$(\hat{b}^{1 \dagger}_{2})^{\dagger} *_{A} \hat{b}^{1 \dagger}_{1}$\\
$$&$$&$$&$$\\
$ $&$$ &$\stackrel{03}{(+i)}\,\stackrel{12}{[+]} \stackrel{56}{(-)}$&
$\stackrel{03}{[+i]}\,\stackrel{12}{[+]} \stackrel{56}{(-)} *_{A} 
\stackrel{03}{(+i)}\,\stackrel{12}{[+]} \stackrel{56}{[+]}$\\
\hline
% 3
$$&$$&$$&$$\\
$0$&$\bigcirc$&${}^{II} \hat{\cal A}^{2 \dagger}_4=$&
$(\hat{b}^{1 \dagger}_{3})^{\dagger} *_{A} \hat{b}^{1 \dagger}_{3}$\\
$$&$$&$$&$$\\
$ $&$$ &$\stackrel{03}{[+i]}\,\stackrel{12}{[-]} \stackrel{56}{[+]}$&
$\stackrel{03}{[+i]}\,\stackrel{12}{(-)} \stackrel{56}{[+]} *_{A} 
\stackrel{03}{[+i]}\,\stackrel{12}{(+)} \stackrel{56}{[+]}$\\
\hline
% 4
$$&$$&$$&$$\\
$0$&$\bigtriangleup$&${}^{II}\hat{\cal A}^{4 \dagger}_4=$&
$(\hat{b}^{1 \dagger}_{4})^{\dagger} *_{A} \hat{b}^{1 \dagger}_{3}$\\
$$&$$&$$&$$\\
$ $&$$ &$\stackrel{03}{(-i)}\,\stackrel{12}{[-]} \stackrel{56}{(-)}$&
$\stackrel{03}{(-i)}\,\stackrel{12}{(-)} \stackrel{56}{(-)} *_{A} 
\stackrel{03}{[+i]}\,\stackrel{12}{(+)} \stackrel{56}{[+]}$\\
\hline
 \end{tabular}
 \end{center}
\end{tiny}
%\end{small}
\end{table}

%%%%%%%%%%%%%%%%%%%%%%%%%%%
\vspace {3mm}
 
\section{Acknowledgement}
%\begin{acknowledgements}
 The author thanks Department of Physics, FMF, University of Ljubljana, 
Society of  Mathematicians, Physicists and Astronomers of Slovenia for supporting  
the research on the  {\it spin-charge-family} theory, and
 Matja\v z Breskvar of  Beyond Semiconductor for donations, in particular for 
sponsoring the annual workshops entitled  "What comes beyond the standard
 models" at Bled, in which the ideas and realizations, presented in this paper,
 were discussed. The author thanks Holger Beck Nielsen for fruitful discussions.
 
%
% {norma92,norma93,norma95,pikanorma2005,
% nh02,nd2017,2020PartIPartII,nh2021RPPNP,{nh2017FDP},%
% n2024Bled,gmdn2008,gn2009,n2014matterantimatter,JMP2013,%
% gn2014,nm2015}


\begin{thebibliography}{}
\bibitem{n2024NPB}
   N. S. Manko\v c Bor\v stnik,
 "Can the ``basis vectors'', describing the internal spaces of fermion and boson fields with the Clifford odd 
(for fermion) and Clifford even (for boson) objects, explain interactions among fields, with gravitons 
included?" % NPB-D-24-00256  
[arxiv: 2407.09482].
\end{thebibliography}
\end{document}